\documentstyle[12pt,epsf]{article}
\def\la{\mathrel{\mathpalette\fun <}}

\def\fun#1#2{\lower3.6pt\vbox{\baselineskip0pt\lineskip.9pt
\ialign{$\mathsurround=0pt#1\hfil##\hfil$\crcr#2\crcr\sim\crcr}}}
\title{Electroweak radiative corrections in Z boson decays}
\author{M I Vysotsky, V A Novikov, L B Okun and A N Rozanov
\thanks{CPPM, Marseille, France}\\
Institute of Theoretical and Experimental Physics,\\
117259 Moscow, Russia}

\date{}

\begin{document}

\maketitle
\newpage
\begin{center}
{\bf Contents.}
\end{center}
\vspace{3mm}

\begin{enumerate}
\item {\bf Introduction.}\\
 New theories, new symmetries, new particles, new phenomena.\\
 $W$ and $Z$ boson `factories'. \\
 What's the point of studying loop corrections?.
\item {\bf Brief history of electroweak radiative corrections.} \\
 Muon and neutron decays. \\
 Main relations of electroweak theory. \\
 Traditional parametrization of corrections to the $\mu$-decay
 and the running $\alpha$.\\
 Deep inelastic neutrino scattering by nucleons. \\
 Other processes involving neutral currents.
\item {\bf On optimal parametrization of the theory and
 the choice of the Born approximation.} \\
 Traditional choice of the main parameters. \\
 Optimal choice of the main parameters. \\
 $Z$ boson decays. Amplitudes and widths. \\
 Asymmetries. \\
 The Born approximation for hadronless observables.
\item {\bf One-loop corrections to hadronless observables.} \\
 Four types of Feynman diagrams. \\
 Asymptotic limit at $m_t^2 \gg m_Z^2$. \\
 The functions $V_m(t,h)$, $V_A(t,h)$ and $V_R(t,h)$.\\
 The corrections $\delta V_i(t,h)$. \\
 Accidental (?) compensation and the mass of the $t$-quark.\\
 How to calculate $V_i$? `Five steps'.
\item {\bf Hadronic decays of the ${\bf{Z}}$ boson.} \\
 The leading quarks and hadrons.\\
 Decays to pairs of light quarks. \\
 Decays to a pair $b\bar{b}$.
\item {\bf Comparison of theoretical results
with  experimental data.} \\
 LEPTOP code. \\
 General fit.  \item {\bf
Conclusions.}\\
Achievements.\\
Problems.\\
Prospects.\\
\item {\bf Appendices.}\\
Appendix A. Feynman rules in electroweak theory.\\
Appendix B. Relation between $\bar{\alpha}$ and
$\alpha(0)$. \\
Appendix C. Summary of the results for
$\bar{\alpha}$. \\
Appendix D. How $\alpha_W(q^2)$ and
$\alpha_Z(q^2)$ `crawl'. \\
Appendix E. Relation between
$\bar{\alpha}$, $G_{\mu}$, $m_Z$ and the bare quantities.\\
Appendix
F. The radiators $R_{Aq}$ and $R_{Vq}$.\\
Appendix G. Derivation of formulas for the asymmetries.\\
Appendix H. Corrections proportional to $m_t^2$.\\
Appendix I. Explicit form of the functions $T_i(t)$ and $H_i(h)$. \\
Appendix J. The contribution of heavy fermions to the self-energy
of the vector bosons. \\
Appendix K. The contribution of light fermions to self-energy
of the vector bosons. \\
Appendix L. The contribution
of vector and scalar bosons to the self-energy of the vector
 bosons. \\
 Appendix M. The vertex parts of $F_{Af}$ and $F_{Vf}$ and
the constants $C_i$. \\
Appendix N. The functions $\phi(t)$ and
$\delta\phi(t)$ in the $Z\to b\bar{b}$ decay. \\
Appendix O. The
$\delta_2 V_i$ corrections. \\
Appendix P. The $\delta_5 V_i$
corrections. \\
\\ {\bf Bibliography.}\\

\vspace{2mm}
\end{enumerate}

\newpage

\newpage
\section{Introduction.}

\begin{center}
{\bf New theories, new symmetries, new particles, new phenomena.}
\end{center}
\vspace{3mm}

The creation of the unified electroweak theory at the end of the
1960s \cite{1}, \cite{2} and of the quantum chromodynamics at
the beginning of the 1970s \cite{3} has dramatically changed the
entire picture of elementary particle physics.
Its foundation changed to gauge symmetries: electroweak symmetry
$SU(2)_L \times U(1)$ and the color symmetry $SU(3)_c$. It
became clear that the gauge symmetries determine the dynamics of
the fundamental physical processes in which the key players are
the gauge vector bosons---the well-familiar photon and
a host of new particles: $W^+$, $W^-$, $Z$ bosons and eight gluons
which differ from one another in color charge. Even though the
Higgs condensate, filling the entire space, remains
enigmatic in the electroweak theory, while the problem of confinement
is still unsolved in chromodynamics, the two theories are
nevertheless so inseparable from modern physics that they were
given the name of the Minimal Standard Model (MSM). We assume in
this review that the reader is familiar with the basics of the
MSM (see, for example, the monographs \cite{15}).

The concept of quarks has undergone dramatic expansion in the
process of creation of the MSM. In \cite{1} the electroweak theory
was suggested for leptons (electron and electron neutrino). The
subsequent inclusion of quarks into the theory led to the
hypothesis that in addition to the three quarks known at the
time---$u$, $d$ and $s$---there exists the fourth
quark, $c$. According to \cite{2}, if $d$- and $s$-quarks are
analogues, respectively, of $e$ and $\mu$, then the mutually orthogonal
combinations $u\cos\theta_c +c\sin\theta_c$ and
$-u\sin\theta_c + c\cos\theta_c$, where
$\theta_c$ is the Cabibbo angle, must constitute the analogues of
$\nu_e$ and $\nu_{\mu}$ \cite{4}.

One of the important consequences of the electroweak theory
was the prediction of the weak neutral currents. According to
the theory, the neutral weak currents must be diagonal;
in other words, neutral currents changing quark flavor (FCNC)
are forbidden. This explained the
absence of such decays as $K^0 \to \bar{e}e$, $K^0 \to \bar{\mu}\mu$
and $K^+ \to \pi^+\bar{e}e$. Since there is no neutral current $\bar{s}d$
in the Lagrangian, these decays cannot occur in the tree
approximation: they require loops with virtual $W$-bosons.
This is also true for the transitions $K^0$---$\bar{K}^0$ that are
responsible for the mass difference between the $K^0_L$ and
$K^0_S$-mesons.

Diagonal neutral currents were discovered in reactions with neutrino
beams \cite{13}, in rotation of the polarization plane of a
laser beam in bismuth vapor \cite{1013}, and in the
scattering of polarized electrons by deuterons \cite{2013}.

The charmed quark $c$ was discovered in 1974 \cite{5}. Even
before that, Kobayashi and Maskawa \cite{6} conjectured that in
addition to two generations of leptons and quarks,
$(\nu_e, e, u, d)$, $(\nu_{\mu}, \mu, c, s)$, there must
exist the third generation $(\nu_{\tau}, \tau, t, b)$. The $2\times 2$
Cabibbo matrix for two generations,
$$
\left(\begin{array}{rl} \cos\theta_c & \sin\theta_c \\ -\sin\theta_c &
\cos\theta_c
\end{array} \right)
$$
is replaced in the case of three generations
by a $3\times 3$ unitary matrix that in its most general form
contains three angles and one phase; the phase is nonzero
if the CP-invariance is violated. This is how the mechanism of
CP violation at the level of quark currents was proposed.

The $\tau$-lepton \cite{7} and the $b$-quark \cite{8} were
discovered experimentally in mid-1970s. The heaviest
fermion---the $t$-quark---was discovered only two decades later
\cite{9} \cite{10}. As for the mechanism of CP-invariance
violation, it remains unknown even now.

The renormalizability of the electroweak theory and quantum
chromodynamics \cite{11} and the property of asymptotic freedom
in QCD \cite{12} opened a wide field for reliable computations
based on perturbation theory. On the basis of such computations
in the tree-diagram approximation, it was possible to predict such
qualitatively novel phenomena as quark and gluon jets; using the
data on neutral currents, it proved possible to perform
preliminary computations of the masses and partial widths of the
$W$ and $Z$ bosons even before their actual discoveries.

\begin{center}
\vspace{3mm}
{\bf ${\bf{W}}$ and ${\bf{Z}}$ boson `factories'.}
\vspace{3mm}
\end{center}

To test the predictions of the electroweak theory,
proton--antiproton colliders were built at the beginning of
the 1980s in Europe (at CERN) and then in the USA (at FNAL).
The discovery of the $W$ and $Z$ bosons \cite{14}
provided spectacular confirmation of the tree-diagram calculations 
\cite{15} and made feasible and urgent  precision tests
of the electroweak theory with loops included. (When speaking
about the tree (Born) approximation and loops, we always mean the
corresponding Feynman diagrams.)

A unique object for such tests was the $Z$ boson. To carry out
a precision study of its properties, electron--positron
colliders were built at the end of the 1980s: LEP1 at CERN and
SLC
at SLAC. Electron and positron in these colliders collide at the
center-of-mass energy equal to the $Z$ boson mass. This results in
the resonance creation and decay of a $Z$ boson (see figure~1).
\begin{figure}
\epsfxsize=200pt
\begin{center}
\parbox{\epsfxsize}{\epsfbox{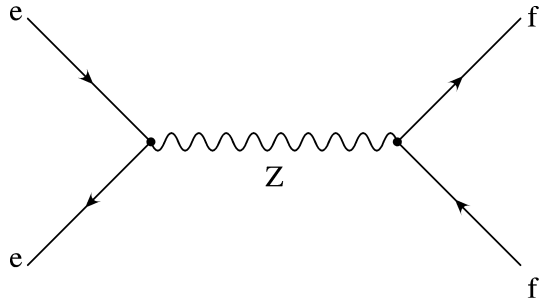}}
\end{center}
\caption{The $Z$ boson as a resonance in $e^+e^-$ annihilations. 
A fermion--antifermion pair in the $Z$-boson decay is denoted by 
$f\bar{f}$, where $f$ can be either a lepton $(e,\mu, \tau,
\nu_e, \nu_{\mu}, \nu_{\tau})$, or a quark $(u, d, s, c,
b)$. In this last case a quark-antiquark pair typically transforms, owing 
to interactions with gluons, to a multi-hadron state. 
The outgoing arrow in this and subsequent diagrams corresponds
to the emission of the fermion ($f$) and to absorption of the 
antifermion ($e^+$); an incoming arrow denotes the emission of
the antifermion ($\bar{f}$) and the absorption of the fermion ($e^-$).  }
\end{figure}
The LEP1 completed its operations in October 1995; about 20
million $Z$ bosons were detected in the four detectors
of this collider (ALEPH, DELPHI, L3, OPAL). The total number of
$Z$ bosons recorded at the SLC by the sole SLD detector was
approximately $10^5$; however, since the colliding electrons are
longitudinally polarized, it was possible to study the
dependence of the annihilation cross section $e^+e^-$ of the $Z$
boson on the sign of beam polarization. As a result, the SLC
proved its competitiveness even with substantially lower
statistics. The statistical and systematic accuracy achieved in
the study of the properties of the $Z$-boson are of the order of
$10^{-5}$ for the $Z$ boson mass and of the order of several
thousandths for the observables that characterize its
decays.

\begin{center}
\vspace{3mm}
{\bf What's the point of studying loop corrections?}
\vspace{3mm}
\end{center}

A natural question is: why do we need to compare the
experimental data and the loop corrections of the electroweak
theory? We need it mostly to gather data on the not yet
discovered particles. For instance, even before the $t$-quark
was discovered on the Tevatron by CDF \cite{9} and D0 \cite{10}
collaborations, its mass was predicted  by analyzing the
radiative loop corrections and the LEP1 and SLC data \cite{16}. The
main loop involving the virtual $t$ and $\bar{t}$-quarks is shown in
figure 2a.
\begin{figure}
\epsfxsize=400pt
\begin{center}
\parbox{\epsfxsize}{\epsfbox{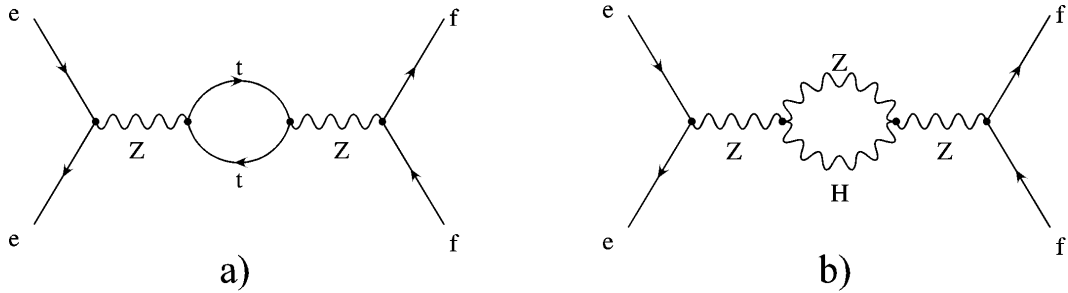}}
\end{center}
\caption{ Contribution of the $t\bar{t}$ to the $Z$-boson propagator 
(a). 
Contribution of the loop with virtual $Z$-boson and a higgs 
to the $Z$-boson propagator (b). }
\end{figure}
The scalar Higgs boson (or simply the higgs) had not been found
yet. In the minimal version of the theory, the so-called Minimal
Standard Model (MSM), there is a single higgs---a neutral particle
whose mass is not fixed by the model. In the Minimal
Supersymmetric Standard Model (MSSM) we have three neutral and
two charged higgses. The lightest of the neutral higgses must
not be heavier than 135 GeV \cite{1900}, \cite{1901}. The
simplest diagram involving a virtual higgs is shown in figure 2b.

When planning experiments on LEP1 and SLC, people had great
expectations that precision measurements would detect pronounced
deviations from the predictions of the standard model and would
thereby unambiguously point to the reality of some sort of `new
physics'. In fact, even though some discrepancies with the MSM
were found, they  go beyond the three standard deviations
only in
a single case (that of the decay of a $Z$ boson to a $b\bar{b}$-pair). If
these discrepancies are not caused by some sort of systematic
error, they may indicate (see Conclusions) the existence of
the relatively light ($\sim 100$ GeV) squarks and gluino---the
supersymmetric partners of quarks and gluons, respectively.  %

\section{Brief history of electroweak radiative corrections.}

The pioneer calculations of electroweak corrections in MSM were
performed in the 1970s, long before the discovery of the $W$ and
$Z$ bosons. The calculations were devoted to the muon decay and the
$\beta$-decays of the neutron and nuclei and to deep inelastic
processes. In connection with the construction of LEP and SLC,
a number of teams of theorists carried out detailed calculations
of the required radiative corrections. These calculations were
discussed and compared at special workshops and meetings. The
result of this work was the publication of two so-called `CERN
yellow reports' \cite{2000}, \cite{2001}, which, together with
the `yellow report' \cite{16}, became the `must' books for
experimenters and theoreticians studying the $Z$ boson.

\begin{center}
\vspace{3mm}
{\bf Muon and neutron decays.}
\vspace{3mm}
\end{center}

Sirlin \cite{17} calculated the radiative corrections to the muon decay
due to one-loop Feynman diagrams (see figure~3).
\begin{figure}
\epsfxsize=400pt
\begin{center}
\parbox{\epsfxsize}{\epsfbox{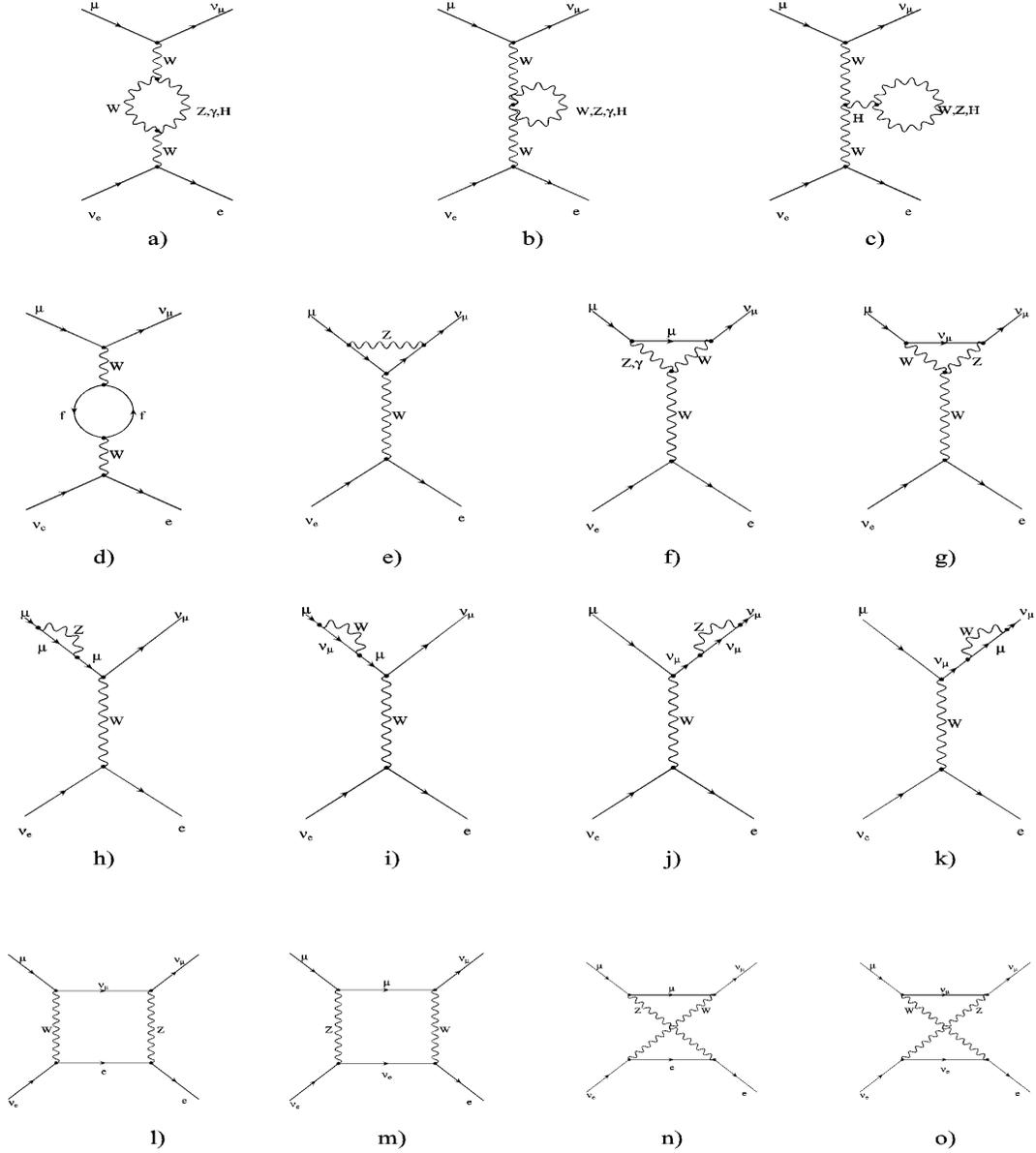}}
\end{center}
\caption{ One-loop diagrams in muon decay.  Loops in the
$W$-boson propagator (a), (b), (c), (d); in the $W$-vertex
(e), (f) and (g) and in external fermion lines (h), (i), (j),
(k) (similar diagrams for $e$ and $\bar{\nu}_e$ are assumed), as
well as square-type diagrams (l), (m), (n), (o).  }
\end{figure}
We must emphasize that the purely electromagnetic correction to
the muon decay due to the exchange of virtual photons and the emission
of real photons was calculated even earlier \cite{18},
for the pointlike four-fermion interaction, i.e.\ without taking
the $W$-boson into account (see figure 4). 
\begin{figure}
\epsfxsize=400pt
\begin{center}
\parbox{\epsfxsize}{\epsfbox{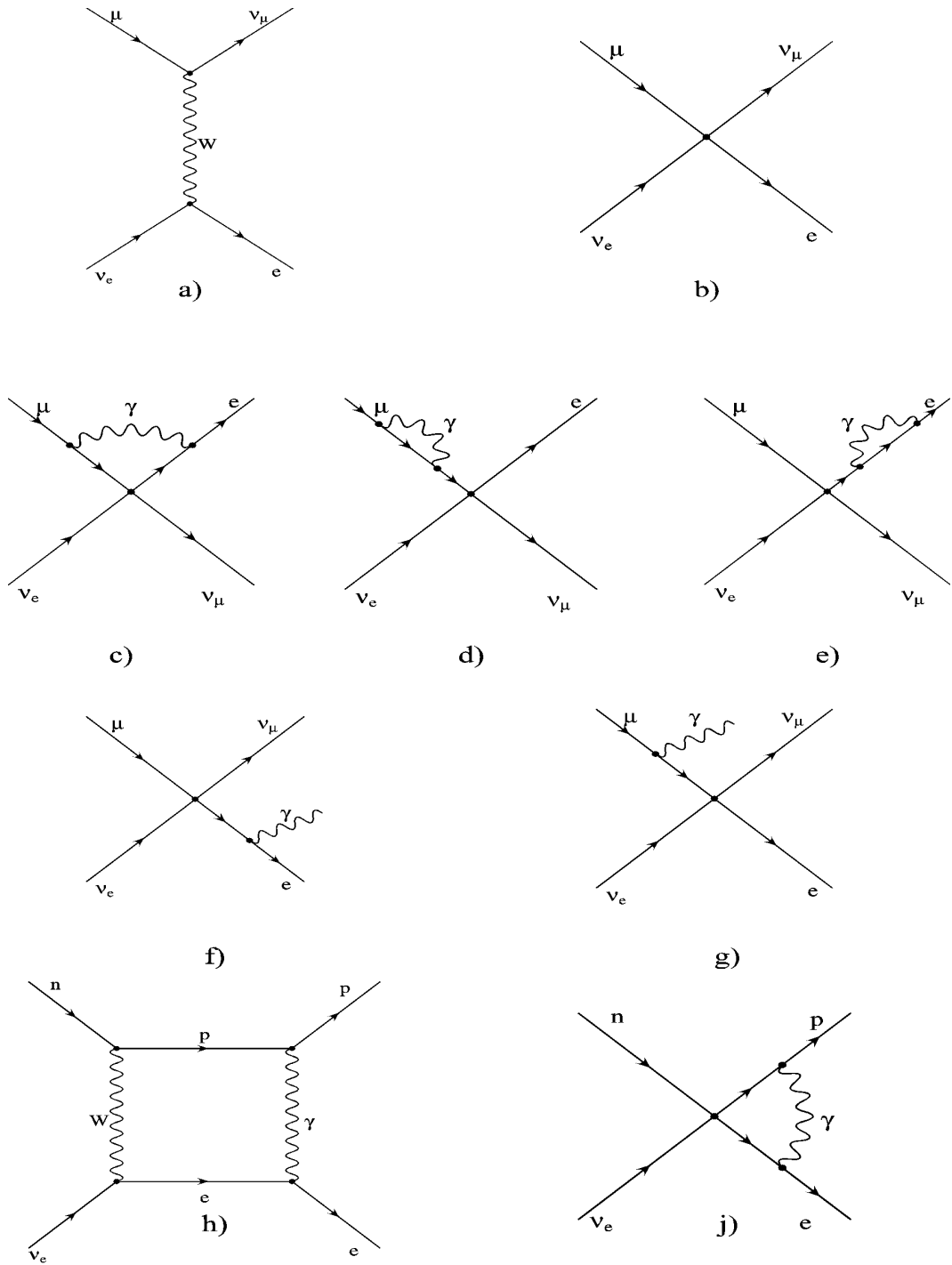}}
\end{center}
\caption{ Muon decay in tree-diagram approximation (a) 
and in local four-fermion approximation (b). 
 Electromagnetic corrections to muon decay in the local
approximation (c) -(g).  
 Electromagnetic corrections to the $\beta$-decay of the neutron:
in tree-diagram approximation (h), in four-fermion approximation (j). }
\end{figure}
It was found that the correction is finite: it contained no divergences. The
 four-fermion interaction constant $G_{\mu}$, extracted from
the muon lifetime $\tau_{\mu}$,
\begin{equation}
\frac{1}{\tau_{\mu}}=\Gamma_{\mu} = \frac{G_{\mu}^2}{192\pi^3}m^5_{\mu}
f(\frac{m^2_e}{m^2_{\mu}})(1-\frac{\alpha}{2\pi}(\pi^2
-\frac{25}{4}))
\label{1}
\end{equation}
where $f(x) = 1-8x +8x^3 -x^4 -12x^2\ln x$, already includes
this electromagnetic correction proportional to $\alpha$. 
$ G_{\mu}=1.16639(2)10^{-5} GeV^{-2} .$ The
finiteness of the purely electromagnetic correction in the muon
decay is caused by the V--A nature of the interacting charged currents
$\bar{\nu}_{\mu}\mu$ and $\bar{e}\nu_e$.

In the neutron decay, the purely electromagnetic correction to
the four-fermion interaction (figure 4(j)) diverges
logarithmically. In view of the $W$-boson propagator (figure
4(h)), the logarithmic divergence is cut off at the $W$-boson
mass.

This correction to the vector
vertex in the leading logarithmic approximation, calculated in \cite{19},
is given by the factor
\begin{equation}
1+ \frac{3\alpha}{2\pi} \ln \frac{m_W}{m_p} \;\; ,
\label{2}
\end{equation}
where $m_p$ is the proton mass. Numerically, its value is of the
order of 1.7\%. Only after this correction is taken into
account, does $\cos^2\theta_c$
extracted from the nuclear $\beta$-decay become equal to
$1-\sin^2\theta_c$, where $\sin^2\theta_c$ is found from the
decays of strange particles ($\theta_c$ is the Cabibbo angle).
Although the correction we discuss now contains $m_W$ in the
logarithmic term, it is essentially electromagnetic and not electroweak,
since it is insensitive to details of the electroweak theory
at short distances, in contrast to, say, electroweak corrections to the muon
decay (figure 3).

Calculations of electroweak corrections to the muon decay
show that the main contribution, exceeding all
others, is caused by the vacuum polarization of the photon
(figure 5a).
\begin{figure}
\epsfxsize=400pt
\begin{center}
\parbox{\epsfxsize}{\epsfbox{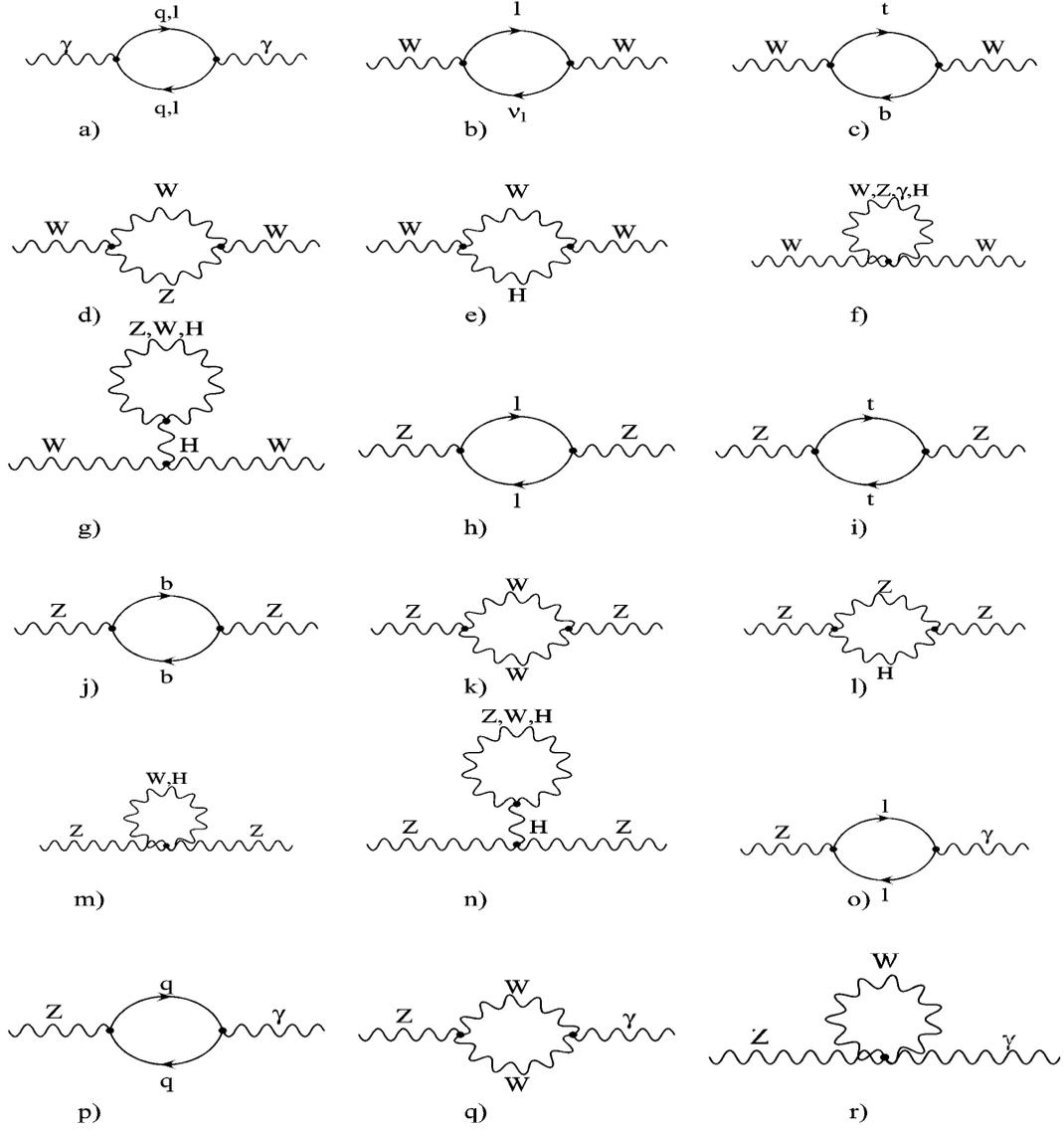}}
\end{center}
\caption{ Photon polarization of the vacuum, resulting in the
logarithmic running of the electromagnetic charge 
$e$ and the `fine structure constant' $\alpha \equiv
\frac{e^2}{4\pi}$, as a function of $q^2$, where $q$ is the 4-momentum 
of the photon (a). 
Some of the diagram that contribute to the proper 
energy of the $W$-boson (b)-(g).
Some diagrams that contribute to the proper energy of the
$Z$-boson (h)-(n). 
Some of the diagram that contribute to the $Z \leftrightarrow
\gamma$ transition (o)-(r). }
\end{figure}
At the first glance, this correction should not have emerged in
the $\mu$-decay in the one-loop approximation: it is not there
among the loops in figure 3. It does appear, however, when
$G_{\mu}$ is expressed in terms of the fine structure constant
$\alpha$ and the masses of the $W$ and $Z$ bosons.

\begin{center}
\vspace{3mm}
{\bf Main relations of electroweak theory.}
\vspace{3mm}
\end{center}

It is well known \cite{15} that in the Born approximation we
have (see figure 4a)
\begin{equation}
G_{\mu} =\frac{g^2}{4\sqrt{2}m^2_W} \;\; ,
\label{3}
\end{equation}
where $m_W$ is the $W$ boson mass and $g$ is its coupling constant
to the charged current.

On the other hand, we have in the same approximation
\begin{equation}
m_W=\frac{1}{2}g\eta \;\; ,
\label{4}
\end{equation}
where $\eta$ is the vacuum
expectation value of the higgs field.

Likewise,
\begin{equation}
m_Z = \frac{1}{2} f\eta \;\; ,
\label{5}
\end{equation}
where $m_Z$ is the $Z$ boson mass and $f$ is its coupling constant
to the neutral left-handed current.

Therefore,
\begin{equation}
\frac{m_W}{m_Z} =\frac{g}{f} \;\; .
\label{6}
\end{equation}
If we introduce the famous Weinberg angle \cite{1}, it becomes
obvious that the two definitions are equally valid in the
Born approximation:
$$
\cos\theta_W =\frac{m_W}{m_Z}  ,
$$
\begin{equation}
\cos\theta_W=\frac{g}{f}  .
\label{7}
\end{equation}
It is well known \cite{1} that the angle $\theta_W$
in the electroweak theory
defines the relation between  the electric charge
$e$ and the weak charge $g$:
\begin{equation}
e=g \sin \theta_W \;\; .
\label{8}
\end{equation}
In the Born approximation, therefore,
\begin{equation}
G_{\mu}=\frac{g^2}{4\sqrt{2}m^2_W}=\frac{1}{\sqrt{2}\eta^2} =
\frac{\pi\alpha}{\sqrt{2}m^2_W\sin^2\theta_W} =
\frac{\pi\alpha}{\sqrt{2}m^2_Z\sin^2\theta_W\cos^2\theta_W} \;\; .
\label{9}
\end{equation}
If we now take into account the electroweak corrections,
we find
$$
\frac{m_W}{m_Z}\neq \frac{g}{f} \;\; .
$$

\begin{center}
{\bf Traditional parametrization of corrections to the $\mu$-decay
and the running $\alpha$.}
\end{center}

\vspace{3mm}

Sirlin's definitions \cite{17} are widely used in the literature
(see the review by Langacker and Erler
\cite{29} and references therein);
according to them
\begin{equation}
s_W^2 \equiv \sin^2\theta_W\equiv
1-c^2_W \equiv 1-\cos^2\theta_W = 1-\frac{m^2_W}{m^2_Z} \;\; ,
\label{10}
\end{equation}
\begin{equation}
G_{\mu} =\frac{\pi\alpha}{\sqrt{2}m^2_W s^2_W(1-\Delta r)} \;\; ,
\label{11}
\end{equation}
where
$$
\Delta r = \Delta r_{em}+\Delta r_{ew}
$$
includes both the truly electroweak correction $\Delta r_{ew}$
for loops in figure 3, and the purely electromagnetic correction
$\Delta r_{em}$ due to $\alpha$ running from $q^2=0$ to $q^2 \sim m^2_W,
m^2_Z$. This correction arises because
\begin{equation}
\alpha\equiv\alpha(q^2=0) =1/137.035985(61)
\label{12}
\end{equation}
is defined for $q^2 =0$, while typical momenta of
virtual particles in the electroweak loop are of the order of
the intermediate boson masses. It is convenient to denote
\begin{equation}
\bar{\alpha} \equiv \alpha(m^2_Z)=\frac{\alpha}{1-\Delta r_{em}}
\equiv \frac{\alpha}{1 - \delta\alpha} \;\; .
\label{13}
\end{equation}
$\Delta r_{em}$ was calculated in a number of papers \cite{20};
the necessary formulas are given in Appendices B and C.
The contribution of the lepton loops to $\Delta r_{em}$ is
described by the expression
\begin{equation}
\delta\alpha^l \equiv
\Delta r^l_{em} = \frac{\alpha}{3\pi}\sum_l [\ln\frac{m^2_Z}{m^2_l} -
\frac{5}{3}] = 0.03141 \;\; ,
\label{14}
\end{equation}
where $l=e, \mu, \tau$. The contribution of quark loops
cannot be calculated theoretically because the quark mass in the
logarithm has no rigorous theoretical definition.
This reflects our ignorance of strong interactions at large distances
(small momenta).
The quark
(hadron) part $\Delta r_{em}^h$ is therefore calculated by
substituting into the dispersion relation the experimental data
on the cross section of the $e^+e^-$-annihilation into hadrons:
\begin{equation}
\delta \alpha^h \equiv \Delta r^h_{em} = \frac{m^2_Z}{4\pi^2 \alpha}
\int^{\infty}_{4 m^2_{\pi}} \frac{ds}{m^2_Z - s}
\sigma^h_{e^+e^-}\;\; ,
\label{1155}
\end{equation}
where $\sigma^h_{e^+e^-}$ is the cross section of the $e^+e^-$ annihilation
into hadrons via one virtual photon.

In this review we make use of the recent result reported by
Jegerlehner and Eidelman \cite{20}:
$\Delta r^h_{em} = 0.02799(66)$, so that
\begin{equation}
\delta\alpha = \delta\alpha^l + \delta\alpha^h \equiv
\Delta r_{em}=\Delta r_{em}^l +
\Delta r_{em}^h = 0.05940(66) \;\; .
\label{15}
\end{equation}
As follows from (\ref{13}) and (\ref{15}),
\begin{equation}
\bar{\alpha} = [128.896(90)]^{-1} \;\; .
\label{1166}
\end{equation}
A summary of results of various calculations of $\bar{\alpha}$
is given in Appendix C.
Following tradition, the contributions of the $t$-quark loop
and the $W$-boson loop are not included into $\alpha(m_Z)$.
In the leading approximation in $1/m_t^2$, the contribution
of the $t$-quark loop is
\begin{equation}
\Delta
r^t_{em}=-\frac{\alpha}{\pi}\frac{4}{45}(\frac{m_Z}{m_t})^2 =
-0.00005 \;\;{\rm for}\;\; m_t =180 {\rm GeV}\;\; ,
\label{16}
\end{equation}
and the exact formula (see equation (\ref{555})) corresponds to
$$
\Delta r^t_{em} = - 0.00006 \;\; {\rm for}\;\; m_t = 180 {\rm
GeV}\;\; .
$$
The contribution of the $W$-boson loop is gauge-dependent.
In the 't Hooft--Feynman gauge it is $\Delta r_{em}^W = 0.00050$
(see (\ref{554})).

\begin{center}
{\bf Deep inelastic neutrino scattering by nucleons.}
\end{center}

\vspace{3mm}

A predominant part of theoretical work on electroweak
corrections prior to the discovery of the $W$- and $Z$-bosons was
devoted to calculating the neutrino--electron \cite{2400} and especially
nucleon--electron \cite{2401} interaction cross sections.
The reason for this is that after the discovery of
neutral currents the quantity $s^2_W
\equiv 1-\frac{m^2_W}{m^2_Z}$ was extracted precisely from a
comparison of the cross section of neutral currents (NC) and
the charged currents (CC). While  a $W$-boson interacts with a
charged current of $V-A$ type, for example,
\begin{equation}
\frac{g}{2\sqrt{2}}
W_{\alpha} \bar{u}(\gamma_{\alpha} + \gamma_{\alpha}\gamma_5)d\;\;,
\label{17}
\end{equation}
the $Z$-boson interacts with neutral currents that have a
more complex form,
\begin{equation}
\frac{f}{2} Z_{\alpha}
\bar{\psi}_f [T_3^f\gamma_{\alpha}\gamma_5 +(T_3^f -2Q^f s^2_W)
\gamma_{\alpha}]\psi_f \;\; ,
\label{18}
\end{equation}
where $T_3^f$ is the third projection of the weak isotopic spin
of the left-hand component of the fermion $f$
(quark or lepton), $Q^f$ is its charge,
and $\psi_f$ is the Dirac spinor describing it.
In the tree approximation, the  NC/CC cross
section ratio for purely axial interactions
(and isoscalar target) equals 1, since in
this approximation the quantity
\begin{equation}
\rho = \frac{f^2}{g^2}\frac{m^2_W}{m^2_Z}
\label{111}
\end{equation}
equals unity. With the vector current taken into account,
the ratio of NC and CC is a function of $s^2_W$. Measurements
of this ratio gave $s^2_W\approx 0.23$, which thus made it
possible to predict the masses of the $W$ and $Z$ bosons
(from the formula for the muon decay (\ref{9})).
More accurate measurements of NC/CC made it possible to improve
the accuracy of $s^2_W$ to such an extent that it became
necessary to take into account the electroweak radiative
corrections both in $s^2_W$ and in $\rho$, which now, with the
corrections taken into account, is not equal to unity any more.

Veltman was the first to point out \cite{21} that if $m_t/m_Z\gg 1$,
the main correction to $\rho$ and $s^2_W$ is caused by the
violation of the electroweak isotopic invariance by the masses
of the $t$- and $b$-quarks in loops of
self-energies of the $Z$ and $W$-bosons. To find $\rho$ in the
limit $m_t/m_Z \gg 1$, it is sufficient to consider these loops
neglecting the momentum of $W$ and $Z$ bosons $q$ in comparison
with the masses of the $W$ and $Z$, i.e.\ for $q^2 =0$.
Elementary calculation of the loops indicated above yields
(see Appendix H):
\begin{equation}
\rho =1+\frac{3\alpha_Z}{16\pi}(\frac{m_t}{m_Z})^2 =
1+\frac{3\alpha_W}{16\pi}(\frac{m_t}{m_W})^2 =
1+\frac{3G_{\mu}m^2_t}{8\sqrt{2}\pi^2} \;\; .
\label{19}
\end{equation}

Here and hereafter we denote
\begin{equation}
\alpha_Z = \frac{f^2}{4\pi}\;\; , \;\; \alpha_W =
\frac{g^2}{4\pi}\;\; .
\label{112}
\end{equation}

Since in real life $m_t/m_Z \simeq 2$, the sum of the remaining,
non-leading corrections is found to be comparable
to the correction proportional to $m_t^2$ (see below).

At the present moment, the quantity $s_W^2$ extracted from the
data on deep inelastic $\nu N$-scattering is determined as
$0.2260(48)$ (the global fit of the data from the collaborations CDHS
\cite{24} and CHARM \cite{25}), or $0.2218(59)$ (collaboration CCFR)
\cite{26}. The accuracy of these data is poorer than found by the
direct measurement of the $W$ boson mass $m_W$ and of the ratio
$m_W/m_Z$ by the collaborations UA2 in CERN \cite{27} and
CDF at Tevatron \cite{28}. According to PDG \cite{29},
the fitted quantity is $m_W = 80.22(26)$, which corresponds to
$s_W^2 = 0.2264(25)$. Note that according to the most recent data
\cite{029}, the measurement accuracy is even higher for
$m_W$: $m_W = 80.26(16) GeV$. For this reason we will not
discuss further the deep inelastic scattering of the neutrino.
Moreover, additional assumptions on the effective mass of the $c$-quark
and on the accuracy of the QCD corrections are necessary for the
interpretation of these experiments.

\begin{center}
{\bf Other processes involving neutral currents.}
\end{center}

\vspace{3mm}

This is true to even greater degree for the parity violation
in eD-scattering \cite{2013}, which, in addition, provides a considerably less
accurate $s_W^2 = 0.216(17)$.

The measurement accuracy of the $\nu_{\mu} e$- and
$\bar{\nu}_{\mu} e$-scattering even in the highest-accuracy
experiment  CHARM II \cite{30} is not sufficient for revealing
the genuine electroweak corrections. However, after an analysis in
the Born approximation, this experiment has demonstrated
for the first time that the interaction constant of the
current $\bar{\nu}_{\mu} \nu_{\mu}$ with the $Z$ boson
is in satisfactory agreement with the theory (see \cite{31}, \cite{32}).

The experiment on measuring parity violation in cesium
$^{133}$Cs$_{55}$ \cite{33} is also insufficiently sensitive, at the
accuracy achieved. Here the effect is produced by the interaction
of the nucleon vector current with the  electron axial current.
Since the characteristic momentum of electron in an atom is
small in comparison with nuclear dimensions, all nucleons of a
nucleus `function' coherently and the nucleus is characterized
by an aggregate weak charge $Q_W$, which is experimentally found
to be $Q_W^{exp} = -71.0 \pm 1.8$, while the theoretically
anticipated value is $-72.9 \pm 0.1$. A spectacular
property of $Q_W$ is the fact that owing to an accidental
cancellation of protons' and neutrons' contributions, it is
practically independent of $m_t$. On the contrary, $Q_W$ is very
sensitive to the contribution of neutral bosons $Z'$ and
$Z''$ heavier than the $Z$-bosons (if they exist).

The best object for testing the electroweak theory at the loop
level is therefore the $Z$ boson with which this review is
predominantly connected.
\section{On optimal parametrization of the theory and
 the choice of the Born approximation.}

The electroweak theory is in many ways different from
electrodynamics, in particular in the diversity of particle
interactions that must be taken into account when considering
any effect in the loop approximation. The parametrization of QED
is straightforward: fundamental quantities are the electron
mass and charge, which are known with very high accuracy. It is
therefore natural to express all theoretical predictions of QED
in terms of $\alpha$ and $m_e$.

\begin{center}
{\bf Traditional choice of the main parameters.}
\end{center}

\vspace{3mm}
The choice of the main parameters in electroweak theory is not
equally obvious. Historically, those selected  were
$G_{\mu}$ as an experimentally measured with best accuracy weak
decay constant,
$s_W^2 \equiv 1 - m_W^2/m_Z^2$, since $W$ and $Z$
bosons were not yet directly observed at that time, while the
value of $s_W^2$ was known from experiments with neutral
currents, and finally, $\alpha$. This parametrization of the
1970s proved to be surprisingly long-lived; the loop
parameters $\Delta r$ and $\rho$ connected with it are widely
used in the literature and are very likely to survive beyond the
end of this century.

In fact, this parametrization is far from being
optimal because $m_W$ (and thus $s_W$ as well) is measured
experimentally at much poorer accuracy than $m_Z$:
\begin{equation}
m_W = 80.26(16) {\rm GeV}
\label{20}
\end{equation}
\begin{equation}
m_Z = 91.1884(22) {\rm GeV}
\label{21}
\end{equation}
As a result, $s_W^2$ is extracted by fitting the loop formulas
for various  observables. This extraction inevitably
requires that we fix the values of the $t$-quark mass and the higgs
mass. Another drawback of this parametrization is that the
quantity $\alpha$, despite its superior accuracy, is not
directly related to the electroweak loops, which
are characterized by a quantity $\bar{\alpha}$, and these we know
with much less impressive accuracy. As a result, the
purely electromagnetic correction $\Delta r_{em}$
is not separated from the genuinely electroweak corrections, thus
blurring the interpretation of the experimental data.

\begin{center}
{\bf Optimal choice of the main parameters.}
\end{center}

\vspace{3mm}
As follows from the above remarks, the currently optimal
parametrization is the one based on $G_{\mu}$, $m_Z$ and
$\bar{\alpha}$. With this parametrization it is convenient to
introduce the weak angle $\theta$, defined (by analogy to
equation (\ref{9})) by the relation
\begin{equation}
G_{\mu} =
\frac{\pi\bar{\alpha}}{\sqrt{2} m_Z^2 s^2 c^2}\;\;,
\label{22}
\end{equation}
where $s^2\equiv \sin^2 \theta$, $c^2\equiv \cos^2 \theta$.

As follows from equation (\ref{22}),
$$
\sin^2 2\theta = \frac{4\pi \bar{\alpha}}{\sqrt{2}G_{\mu} m_Z^2} =
0.71078(50)\;\; ,
$$
\begin{equation}
s^2 = 0.23110(23) \;\; ,
\label{255}
\end{equation}
$$
c = 0.87687(13) \;\; .
$$

The angle $\theta$ was introduced in mid-1980s \cite{34}.
However, its consistent use began only after the publication of
\cite{35}.

Using $\theta$ instead of $\theta_W$ automatically takes into
account the running of $\alpha$ and makes it possible to
concentrate on the genuinely electroweak
corrections. Using $m_Z$ instead of $s_W$ allows one to
explicitly single out the dependence on $m_t$ and $m_H$ for each
electroweak observable.

Note that a different definition of the $Z$ boson mass
$\overline{m}_Z$ is known in the literature, related to a
different parametrization of the
shape of the $Z$ boson peak \cite{10W}. This mass $\overline{m}_Z$
is smaller than $m_Z$ by approximately 30 MeV. In this review, we
consistently use only $m_Z$, following the summary reports of LEP
collaborations \cite{4400}.

Let us show how the parametrization in terms of $G_{\mu}$, $m_Z$
and $s$ is applied to the decay amplitudes of Z boson and to
the ratio $m_W/m_Z$.

\begin{center}
{\bf ${\bf{Z}}$ boson decays. Amplitudes and widths.}
\end{center}

\vspace{3mm}
In correspondence with equation (\ref{18}), we rewrite the amplitude
of the Z boson decay into a fermion--antifermion pair $f\bar{f}$ in
the form
\begin{equation}
M(Z \to
f\bar{f}) = \frac{1}{2} \bar{f} Z_{\alpha} \bar{\psi}_f
(\gamma_{\alpha} g_{V f} + \gamma_{\alpha} \gamma_5 g_{Af}) \psi_f
\;\; .
\label{266}
\end{equation}
Here by definition $\bar{f}$ is the value of the coupling constant
$f$ in the Born approximation,
\begin{equation}
\bar{f}^2 = 4\sqrt{2} G_{\mu} m^2_Z =
0.54866(4)
\label{23}
\end{equation}
(the use of the same letter $f$ to denote both the fermion and
the coupling constant cannot lead to confusion, to such an extent
are these objects different). The high accuracy with which the
numerical value of $\bar{f}$ is known comes from the fact that
$\bar{f}$ is independent of $\bar{\alpha}$.
All electroweak radiative corrections are `hidden' in
dimensionless constants $g_{V f}$ and $g_{Af}$.
These coefficients do not include the
contribution of the  final state interactions  due to the
exchange of gluons (for quarks) and photons (for quarks and
leptons). The   final state
interactions  have nothing in
common with the electroweak corrections and must be taken into
account  as separate
factors in the expressions for the decay rates.
 These factors are sometimes
known as `radiators', since they cover not only exchange of photons
and gluons but also their emission.

Radiators are trivially unities in the case of decay to any of
the neutrino pairs $\nu_e\bar{\nu}_e$, $\nu_{\mu}\bar{\nu}_{\mu}$,
$\nu_{\tau}\bar{\nu}_{\tau}$ and therefore
\begin{equation}
\Gamma_{\nu} = \Gamma(Z \to \nu\bar{\nu}) = 4\Gamma_0(g^2_{A\nu} +
g^2_{V_{\nu}})\;\;,
\label{24}
\end{equation}
where $\Gamma_0$ is the so-called standard width:
\begin{equation}
\Gamma_0 = \frac{G_{\mu} m^3_Z}{24\sqrt{2}\pi} = 82.944(6) {\rm
MeV}  .
\label{25}
\end{equation}
If neutrino masses are assumed to be negligible, then
\begin{equation}
g_{A\nu} = g_{V_{\nu}} \equiv g_{\nu}
\label{26}
\end{equation}
so that
\begin{equation}
\Gamma_{\nu} = 8\Gamma_0 g^2_{\nu} .
\label{27}
\end{equation}

For decays to any of the pairs of charged leptons $l\bar{l}$
we have \cite{16}
\begin{equation}
\Gamma_l \equiv \Gamma(Z \to l\bar{l}) = 4\Gamma_0 [g^2_{V l}(1 +
\frac{3\bar{\alpha}}{4\pi}) + g^2_{Al}(1 +
\frac{3\bar{\alpha}}{4\pi} -
6\frac{m^2_l}{m^2_Z})]\;\; ,
\label{28}
\end{equation}
where the QED correction is taken into account only to the
lowest approximation in $\bar{\alpha}$; we neglect terms on the
order of $(\bar{\alpha}/\pi)^2 \sim 10^{-6}$. The term
proportional to $m^2_l$ is negligible  for $l = e, \mu$ and
must be included only for
$l = \tau$ ($m^2_{\tau}/m^2_Z = 3.8 \cdot 10^{-4}$).

For decays to any of the five pairs of quarks $q\bar{q}$ we have
\begin{equation}
\Gamma_q \equiv \Gamma(Z \to q\bar{q}) = 12 \cdot \Gamma_0 [g^2_{Aq}
R_{Aq} + g^2_{V q} R_{V q}] .
\label{29}
\end{equation}
Here the factor 3, additional in comparison with leptons, takes
into account the three colors of each quark. The `radiators' in
the first approximation are identical for the vector and the axial-vector
interactions,
\begin{equation}
R_{V_q} = R_{Aq} = 1 + \frac{\hat{\alpha}_s}{\pi}\;\; ,
\label{30}
\end{equation}
where $\hat{\alpha}_s$ is the constant of interaction of gluons
with quarks at $q^2 = m^2_Z$. There are different conventions
for the choice of $\hat{\alpha}_s$. In calculations of decays
of the $Z$-boson, it is quite typical to determine
$\hat{\alpha}_s$ using the so-called modified minimal
subtraction scheme, ${\overline{MS}}$ (see at the end of Appendix
A). The numerical value of $\hat{\alpha}_s$, found from $Z$-boson
decays, is of the order of 0.12. For additional details on the
value of $\hat{\alpha}_s$ and for more accurate expressions
for radiators see Appendix F. Here we only remark that vast
literature is devoted to radiator calculations; they are
calculated using perturbation theory up to terms
$(\hat{\alpha}_s/\pi)^3$, see \cite{A}, \cite{B},
\cite{C}, \cite{D}. The full hadron width is, to the accuracy
of very small corrections, the sum of widths of five quark
channels:
\begin{equation}
\Gamma_h = \Gamma_u + \Gamma_d + \Gamma_s + \Gamma_c + \Gamma_b\;\; .
\label{31}
\end{equation}
The full width of the $Z$ boson is given by the obvious
expression:
\begin{equation}
\Gamma_Z
= \Gamma_h + \Gamma_e + \Gamma_{\mu} + \Gamma_{\tau} +
3\Gamma_{\nu}\;\; .
\label{32}
\end{equation}
The annihilation cross section $e^+e^-$ into hadrons at the
$Z$-peak is given by the Breit--Wigner formula:
\begin{equation}
\sigma_h = \frac{12\pi}{M^2_Z} \frac{\Gamma_e\Gamma_h}{\Gamma^2_Z}
\;\; .
\label{33}
\end{equation}
Finally, the following notation for the ratio of partial widths
is widely used:
\begin{equation}
R_b = \frac{\Gamma_b}{\Gamma_h}\;\; , \;\; R_c =
\frac{\Gamma_c}{\Gamma_h}\;\; , \;\; R_l =
\frac{\Gamma_h}{\Gamma_l}\;\; .
\label{34}
\end{equation}
(Note that in contrast to $R_b$ and $R_c$, $\Gamma_h$ in $R_l$
is in the numerator.)

\begin{center}
{\bf Asymmetries.}
\end{center}

\vspace{3mm}
In addition to the total and partial widths of $Z$ boson decays,
experimentalists also measure effects due to parity
non-conservation, i.e. the interference of the vector and
axial--vector currents. For pairs of light quarks $(u, d, s, c)$
and leptons we determine the quantity
\begin{equation}
A_f = \frac{2 g_{Af} g_{V f}}{g^2_{Af} + g^2_{V f}}\;\; .
\label{35}
\end{equation}
For the pair $b\bar{b}$ (see Appendix G),
\begin{equation}
A_b = \frac{2g_{Ab} g_{V b}}{v^2 g^2_{Ab} + \frac{1}{2}(3 - v^2)
g^2_{V b}} \cdot v \;\; ,
\label{36}
\end{equation}
where $v$ is the velocity of the $b$-quark (in units of $c$):
\begin{equation}
v = \sqrt{1 - \frac{4\hat{m}^2_b}{m^2_Z}}\;\; .
\label{37}
\end{equation}
Here $\hat{m}_b$ is the value of the `running mass' of the
$b$-quark with momentum $m_Z$, calculated in  the scheme
${\overline{MS}}$ \cite{E}. The forward--backward charge asymmetry
in the decay to $f\bar{f}$ equals (see Appendix G)
\begin{equation}
A^f_{FB} \equiv \frac{N_F - N_B}{N_F + N_B} = \frac{3}{4} A_e A_f\;\;
,
\label{38}
\end{equation}
where $A_e$ refers to the creation of a $Z$ boson in $e^+e^-$-annihilations,
and $A_f$ refers to its decay into $f\bar{f}$.

The longitudinal polarization of the $\tau$-lepton in the
$Z \to \tau\bar{\tau}$ decay is
\begin{equation}
P_{\tau} = -A_{\tau}\;\;.
\label{39}
\end{equation}

If, however, we measure polarization as a function of the angle
$\theta$ between the momentum of a $\tau^-$ and the direction of
the electron beam, this permits the determination of not only
$A_{\tau}$ but $A_e$ as well:
\begin{equation}
P_{\tau}(\cos \theta) =
- \frac{A_{\tau}(1 + \cos^2 \theta) + A_e 2\cos \theta }
{(1 + \cos^2 \theta) + A_{\tau} A_e 2\cos \theta} \;\; .
\label{455}
\end{equation}

The polarization of $P_{\tau}$ is found from $P_{\tau} (\cos \theta)$
by separately
integrating the numerator and the denominator in (\ref{455})
over the total solid angle.

The relative difference between total cross sections in the
$Z$ peak for the left- and right-polarized electron that collide
with non-polarized positrons (this quantity is measured at the SLC
collider) is
\begin{equation}
A_{LR} \equiv \frac{\sigma_L - \sigma_R}{\sigma_L + \sigma_R} =
A_e\;\;.
\label{40}
\end{equation}

The measurement of the asymmetries outlined above allows one to
experimentally determine the quantities $g_{V f}/g_{Af}$, since
these asymmetries are caused by the interference of the
vector and axial--vector currents. In their turn,
measurements of the widths $\Gamma_f$, $\Gamma_h$, $\Gamma_Z$
mostly permit the experimental determination of
$g_{Af}$, since $|g_{V q}|^2 < |g_{Aq}|^2$ for quarks, and
for leptons $|g_{Vl}|^2 \ll |g_{Al}|^2$. As for
$\Gamma_q$ and $\Gamma_h$, getting these quantities allows one
to find $\hat{\alpha}_s$.

\begin{center}
{\bf The Born approximation for hadronless observables.}
\end{center}

\vspace{3mm}

Before discussing the loop electroweak corrections, let us consider
expressions for $m_W/m_Z$, $g_{Af}$ and $g_{V f}/g_{Af}$ in the so-called
$\bar{\alpha}$-Born approximation. Using the
angle $\theta$ introduced earlier, its $\sin \theta \equiv s$ and
$\cos \theta \equiv c$, we automatically take into account the
purely electromagnetic correction due to the running of $\alpha$. It
is easily shown that in the $\bar{\alpha}$-Born approximation
\begin{equation}
(m_W/m_Z)^B = c
\label{41}
\end{equation}
\begin{equation}
g_{Af}^B = T_{3f}
\label{42}
\end{equation}
\begin{equation}
(g_{V f}/g_{Af})^B = 1 - 4|Q_f|s^2
\label{43}
\end{equation}
It is of interest to compare the Born values with their experimental
values. Table 1 presents this comparison for the so-called
`hadronless' observables (here and below the experimental
results are taken  from \cite{4400}.)

\vspace{5mm}

\begin{center}
{\bf Table 1}

\vspace{5mm}

\begin{tabular}{|l|l|l|}
\hline
Observable & Experiment & $\bar{\alpha}$-Born \\ \hline \hline
$m_W/m_Z$ & 0.8802(18) & 0.8769(1) \\ \hline
$m_W$ (GeV) & 80.26(16) & 79.96(2) \\ \hline
$s^2_W$ & 0.2253(31) & 0.2311(2) \\ \hline
$g_{Al}$ & -0.5011(4) & -0.5000(0) \\ \hline
$\Gamma_l$ (MeV) & 83.93(14) & 83.57(2) \\ \hline
$g_{Vl}/g_{Al}$ & 0.0756(14) & 0.0756(9) \\ \hline
$s^2_l$ & 0.2311(4) & 0.2311(2) \\ \hline
\end{tabular}
\end{center}
For the reader's convenience, the table lists different
representations of the same observables known in the
literature. Thus, according to widely used definitions,
\begin{equation}
s^2_W = 1 - m^2_W/m^2_Z\;\; ,
\label{44}
\end{equation}
\begin{equation}
s^2_l \equiv s^2_{eff} \equiv \sin^2
\theta_{eff}^{lept} \equiv \frac{1}{4}(1 -g_{V l}/g_{Al})\;\; .
\label{45}
\end{equation}
The experimental value of $s^2_l$ in the table is the average of
two numbers, 0.2316(5) (LEP) and 0.2305(5) (SLC). It is assumed
in the table that the lepton universality holds,
thus  the lepton
decay data have been averaged over a number of observables.

Table 1 shows that the $\bar{\alpha}$-Born approximation
provides good description of the experimental data.
( An agreement is found for the hadron decays of
$Z$-bosons as well). An especially (and unexpectedly!) good
agreement, and the ensuing smallness of radiative corrections, is
found for $g_{Vl}/g_{Al}$.  The anomalous smallness of true
electroweak corrections was first pointed out in 1992 \cite{35}, when
the $\bar{\alpha}$-Born approximation was applied for the first time.
(Before that it was hidden in the shadow of the large contribution of
purely electromagnetic running of $\alpha$, that was not separated
from truly electroweak corrections).  The LEP1 data of 1992 were not
sufficiently accurate to allow detecting them.
Even the data presented to the Marseille conference in summer 1993
were not, as pointed out in a number of reports \cite{45},
sufficiently accurate for the detection of radiative corrections.
At the Glasgow conference in summer 1994 corrections were
detectable at the level of $2.3 \sigma$ for $g_{Al}$, $1.5 \sigma$
for $m_W$ and $1 \sigma$ for $g_{Vl}/g_{Al}$ \cite{46}. Note that the
difference between most recent experimental values $s^2_W =
0.2253(31)$ and $s^2_{l}= 0.2311(4)$ is a
$2\sigma$ manifestation of
electroweak radiative corrections, which is independent of either the
choice of the Born approximation or of the choice of calculation
scheme.

\section{One-loop corrections to hadronless observables.}
The fact that the experimentally observed electroweak corrections
are small suggests that one-loop approximation
would be sufficient for describing them \cite{35}. This idea is
supported by a thorough evaluation \cite{16} of
theoretical uncertainties contributed by higher-order perturbation
theory in electroweak interaction. There exists essentially a single
two-loop
diagram that should be taken into account. It was calculated
in \cite{50}, and we will discuss its contribution and take it
into account below (see formula (\ref{64})).

\begin{center}
{\bf Four types of Feynman diagrams.}
\end{center}

\vspace{3mm}
Four types of Feynman diagrams contribute to electroweak corrections
for the observables of interest to us here, $m_W/m_Z,
g_{Al}, g_{Vl}/g_{Al}$:
\begin{enumerate}
\item{Self-energy loops for $W$ and $Z$ bosons with
virtual $\nu, l, q, H, W$ and $Z$ in loops. Examples of some of
these diagrams are shown in figures 5b--5n.}
\item{Loops of charged particles that result in $Z \leftrightarrow
\gamma$ transitions (figures 5o--5r).}
\item{Vertex triangles with virtual leptons and a $W$ or $Z$
boson (figures 6a--6c).}
\begin{figure}
\epsfxsize=500pt
\begin{center}
\parbox{\epsfxsize}{\epsfbox{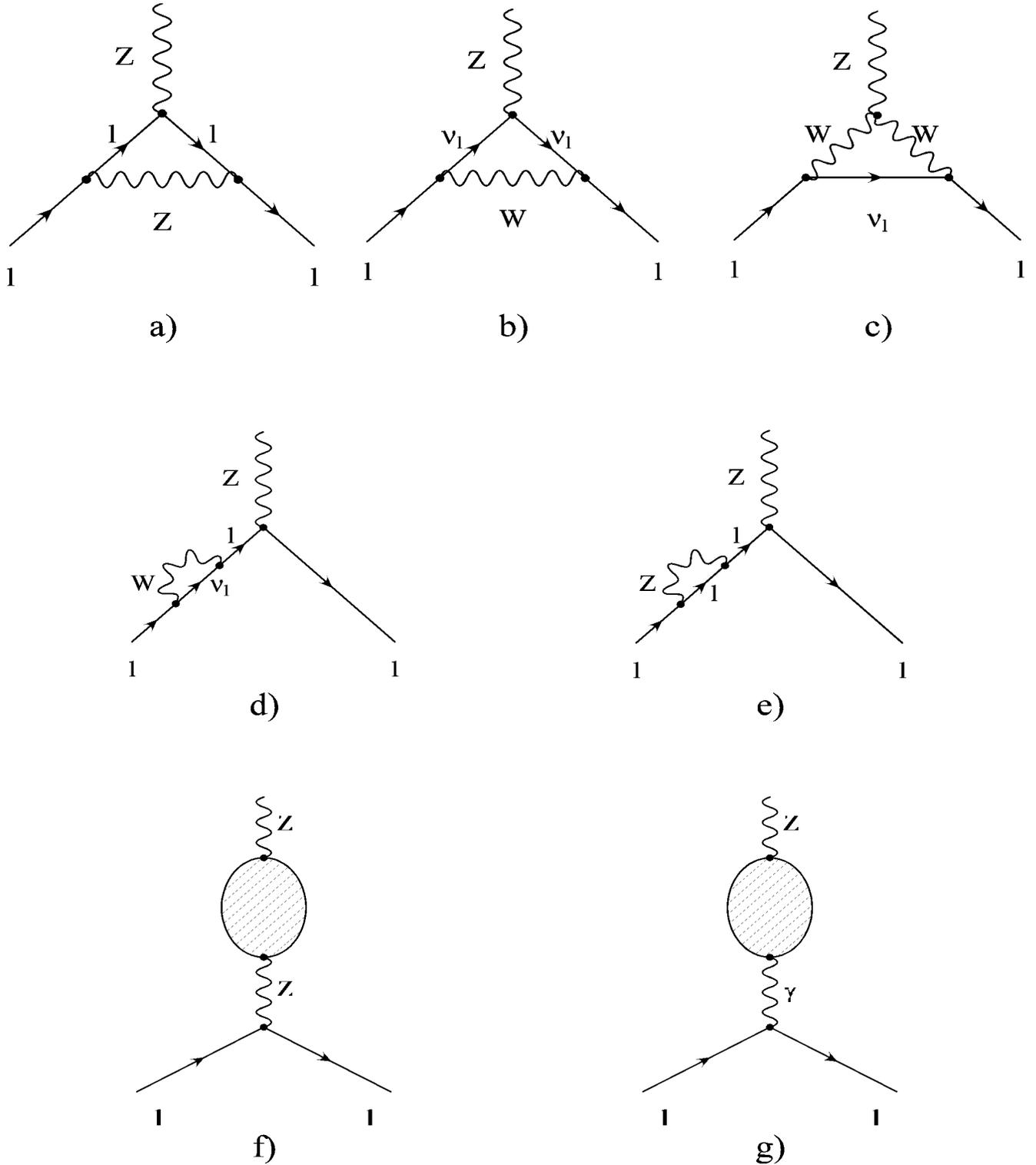}}
\end{center}
\caption{Vertex triangular diagrams in the $Z \to
l\bar{l}$ decay (a), (b), (c). 
Loops that renormalize the lepton wavefunctions in the $Z \to l\bar{l}$
decay. 
(Of course,antilepton have similar loops.)  (d), (e). 
Types of diagrams that renormalize the $Z$ boson wavefunction 
in the $Z \to l\bar{l}$ decay  (f), (g). 
Virtual particles in the loops are those that we were discussing above  }
\end{figure}
\item{Electroweak corrections to lepton wavefunctions
(figures 6d--6e).}
\end{enumerate}
It must be emphasized that loops shown in figures
5h--5n contribute not only to the $m_Z$ mass and, consequently,
to the $m_W/m_Z$ ratio but also to the $Z$ boson decay to
$l\bar{l}$, to which $Z \leftrightarrow \gamma$ transitions also
contribute (figures 5o--5r). This occurs owing to the diagrams
of the type of figures 6f--6g which give corrections to the
$Z$-boson wavefunction.

Obviously, electroweak corrections to $m_W/m_Z, g_{Al}$ and
$g_{Vl}/g_{Al}$ are dimensionless and thus can be expressed in
terms of $\bar{\alpha}, c, s$ and the dimensionless
parameters
$$
t =(\frac{m_t}{m_Z})^2\;\; , \;\; h = (\frac{m_H}{m_Z})^2\;\; ,
$$
where $m_t$ is the mass of the $t$-quark and $m_H$ is the higgs
mass. (We neglect the masses of leptons and all quarks except $t$.)

\begin{center}
{\bf The asymptotic limit at ${\bf{m^2_t \gg m^2_Z}}$.\\
The functions ${\bf{V_m(t,h), \;\; V_A(t,h)}}$ and ${\bf{V_R(t,h)}}$.}
\end{center}

\vspace{3mm}
It is convenient to split the calculation of corrections into a
number of stages and begin with calculating the asymptotic
limit for $t \gg 1$.

 According to the reasons mentioned above (see equation (\ref{19}),
 the main contribution comes from diagrams that contain $t$- and
 $b$-quarks (5c and 5i,j). A simple calculation (see Appendix H)
 gives the following result for the sum of the Born and loop
terms:
 \begin{equation}
 m_W/m_Z = c +
 \frac{3c}{32\pi s^2(c^2 - s^2)} \bar{\alpha}t\;\; .
 \label{46}
 \end{equation}
 \begin{equation}
 g_{Al} = -\frac{1}{2} - \frac{3}{64\pi s^2c^2} \bar{\alpha} t\;\; ,
 \label{47}
 \end{equation}
 \begin{equation}
R \equiv g_{Vl}/g_{Al} = 1 - 4s^2 + \frac{3}{4\pi(c^2 - s^2)}
\bar{\alpha} t\;\; ,
\label{48}
\end{equation}
\begin{equation}
g_{\nu} = \frac{1}{2} + \frac{3}{64\pi s^2 c^2} \bar{\alpha} t\;\;
.
\label{49}
\end{equation}

\begin{center}

{\bf The functions ${\bf{V_m(t,h), \;\; V_A(t,h)}}$ and ${\bf{V_R(t,h)}}$.}
\end{center}

\vspace{3mm}

If we now switch from the asymptotic case of $t \gg 1$ to the
case of $t \sim 1$, then, first, the change in the contribution
of the diagrams 5c, 5i and 5j can be written in the form
\begin{equation}
t \to t + T_i(t)\;\; ,
\label{50}
\end{equation}
where the index $i = m, A, R $, $\nu$ for $m_W/m_Z,\;\;
g_{Al},\;\; R \equiv g_{Vl}/g_{Al}$ and $g_{\nu}$, respectively. 

The functions $T_i$ are relatively simple combinations of
algebraic and logarithmic functions. They are listed in explicit
form in Appendix I. Their numerical values for a range of
values of $m_t$ are given in Table 2. The functions $T_i(t)$
thus describe the contribution of the quark doublet $t,b$
to $m_W/m_Z$, $g_A$, $R = g_{Vl}/g_{Al}$ and
$g_{\nu}$. If, however, we now take into account
the contributions of the remaining virtual particles, then the
result can be given in the form
\begin{equation}
t \to V_i(t,h) = t + T_i(t) + H_i(h) + C_i + \delta V_i(t,h)\;\; .
\label{51}
\end{equation}

\newpage
\begin{center}
{\bf Table 2}
\vspace{2mm}

\begin{tabular}{|r|r|r|r|r|}
\hline
$m_t$ & $t$ & $T_m$ & $T_A$ & $T_R$ \\
(GeV) & & & & \\ \hline
0 & 0 & -0.188 & 0.875 & 0.444 \\
10 & 0.012 & 0.192 & 0.934 & 0.038 \\
20 & 0.048 & -0.256 & 0.955 & -0.015 \\
30 & 0.108 & -0.430 & 0.812 & -0.305 \\
40 & 0.192 & -0.753 & 0.403 & -0.959 \\
50 & 0.301 & -0.985 & 0.111 & -0.748 \\
60 & 0.433 & -0.931 & 0.327 & -0.412 \\
70 & 0.589 & -0.688 & 0.390 & -0.250 \\
80 & 0.770 & -0.317 & 0.421 & -0.143 \\
90 & 0.974 & -0.080 & 0.440 & -0.061 \\
100 & 1.203 & 0.084 & 0.451 & 0.006 \\
110 & 1.455 & 0.214 & 0.460 & 0.062 \\
120 & 1.732 & 0.323 & 0.465 & 0.111 \\
130 & 2.032 & 0.418 & 0.470 & 0.154 \\
140 & 2.357 & 0.503 & 0.473 & 0.193 \\
150 & 2.706 & 0.579 & 0.476 & 0.228 \\
160 & 3.079 & 0.649 & 0.478 & 0.261 \\
170 & 3.476 & 0.713 & 0.480 & 0.291 \\
180 & 3.896 & 0.772 & 0.481 & 0.319 \\
190 & 4.341 & 0.828 & 0.483 & 0.345 \\
200 & 4.810 & 0.880 & 0.484 & 0.370 \\
210 & 5.303 & 0.929 & 0.485 & 0.393 \\
220 & 5.821 & 0.975 & 0.485 & 0.415 \\
230 & 6.362 & 1.019 & 0.486 & 0.436 \\
240 & 6.927 & 1.061 & 0.487 & 0.456 \\
250 & 7.516 & 1.101 & 0.487 & 0.475 \\
260 & 8.130 & 1.139 & 0.487 & 0.493 \\
270 & 8.767 & 1.176 & 0.488 & 0.510 \\
280 & 9.428 & 1.211 & 0.488 & 0.527 \\
290 & 10.114 & 1.245 & 0.489 & 0.543 \\
300 & 10.823 & 1.277 & 0.489 & 0.559 \\ \hline
\end{tabular}
\end{center}
\vspace{3mm}

Here $H_i(h)$ contain the contribution of the virtual vector and higgs
bosons $W, Z$ and $H$ and are  functions of the higgs mass $m_H$.
(The masses of the $W$ and $Z$ bosons enter  $H_i(h)$ via
the parameters $c, s$, defined by equation (\ref{22})).
The explicit form of the functions $H_i$ is given in Appendix I,
and their numerical values for various values of $m_H$
are given in Table 3.

\newpage
\begin{center}

{\bf Table 3}
\vspace{2mm}

\begin{tabular}{|r|r|r|r|r|}
\hline
$m_H$ & $h$ & $H_m$ & $H_A$ & $H_R$ \\
(GeV) & & & & \\ \hline
0.01 & 0.000 & 1.120 & -8.716 & 1.359 \\
0.10 & 0.000 & 1.119 & -5.654 & 1.354 \\
1.00 & 0.000 & 1.103 & -2.652 & 1.315 \\
10.00 & 0.012 & 0.980 & -0.133 & 1.016 \\
50.00 & 0.301 & 0.661 & 0.645 & 0.360 \\
100.00 & 1.203 & 0.433 & 0.653 & -0.022 \\
150.00 & 2.706 & 0.275 & 0.588 & -0.258 \\
200.00 & 4.810 & 0.151 & 0.518 & -0.430 \\
250.00 & 7.516 & 0.050 & 0.452 & -0.566 \\
300.00 & 10.823 & -0.037 & 0.392 & -0.679 \\
350.00 & 14.732 & -0.112 & 0.338 & -0.776 \\
400.00 & 19.241 & -0.178 & 0.289 & -0.860 \\
450.00 & 24.352 & -0.238 & 0.244 & -0.936 \\
500.00 & 30.065 & -0.292 & 0.202 & -1.004 \\
550.00 & 36.378 & -0.341 & 0.164 & -1.065 \\
600.00 & 43.293 & -0.387 & 0.128 & -1.122 \\
650.00 & 50.809 & -0.429 & 0.095 & -1.175 \\
700.00 & 58.927 & -0.469 & 0.064 & -1.223 \\
750.00 & 67.646 & -0.506 & 0.035 & -1.269 \\
800.00 & 76.966 & -0.540 & 0.007 & -1.311 \\
850.00 & 86.887 & -0.573 & -0.019 & -1.352 \\
900.00 & 97.410 & -0.604 & -0.044 & -1.390 \\
950.00 & 108.534 & -0.633 & -0.067 & -1.426 \\
1000.00 & 120.259 & -0.661 & -0.090 & -1.460 \\
\hline
\end{tabular}
\end{center}
\vspace{3mm}
The constants $C_i$ in (\ref{51}) include the contributions
of light fermions to the self-energy of the $W$ and $Z$ bosons,
and also to the diagrams of figure 3, describing the muon decay,
and of figures 6a--6c, describing the $Z$-boson decay. The
constants $C_i$ are relatively complicated functions of $s^2$
(see Appendix M).
We list
here their numerical values for $s^2 = 0.23110 - \delta s^2$:
\begin{equation}
C_m = -1.3497 + 4.13 \delta s^2\;\;,
\label{52}
\end{equation}
\begin{equation}
C_A = -2.2621 - 2.63 \delta s^2\;\;,
\label{53}
\end{equation}
\begin{equation}
C_R = -3.5045 - 5.72 \delta s^2\;\;,
\label{54}
\end{equation}
\begin{equation}
C_{\nu} = -1.1641 - 4.88 \delta s^2\;\;.
\label{55}
\end{equation}

\begin{center}
{\bf Corrections ${\bf{\delta V_i(t,h)}}$.}
\end{center}

\vspace{3mm}

Finally, the last term in equation (\ref{51}) includes the sum
of corrections of five different types. Their common feature is
that they are all quite small (except for $\delta^t_2 V_i$) and
that they represent two loops (with the exception of a one-loop
$\delta_1 V_i$ and a three-loop $\delta_3 V_i$).

\begin{enumerate}
\item{$\delta_1 V_i$ contains contributions of the $W$-boson and
the $t$-quark to the polarization of the electromagnetic vacuum
$\delta_W \alpha \equiv \Delta r^W_{em}$ and
$\delta_t \alpha \equiv \Delta r^t_{em}$ (see figures 7a-c and equation
\begin{figure}
\epsfxsize=400pt
\begin{center}
\parbox{\epsfxsize}{\epsfbox{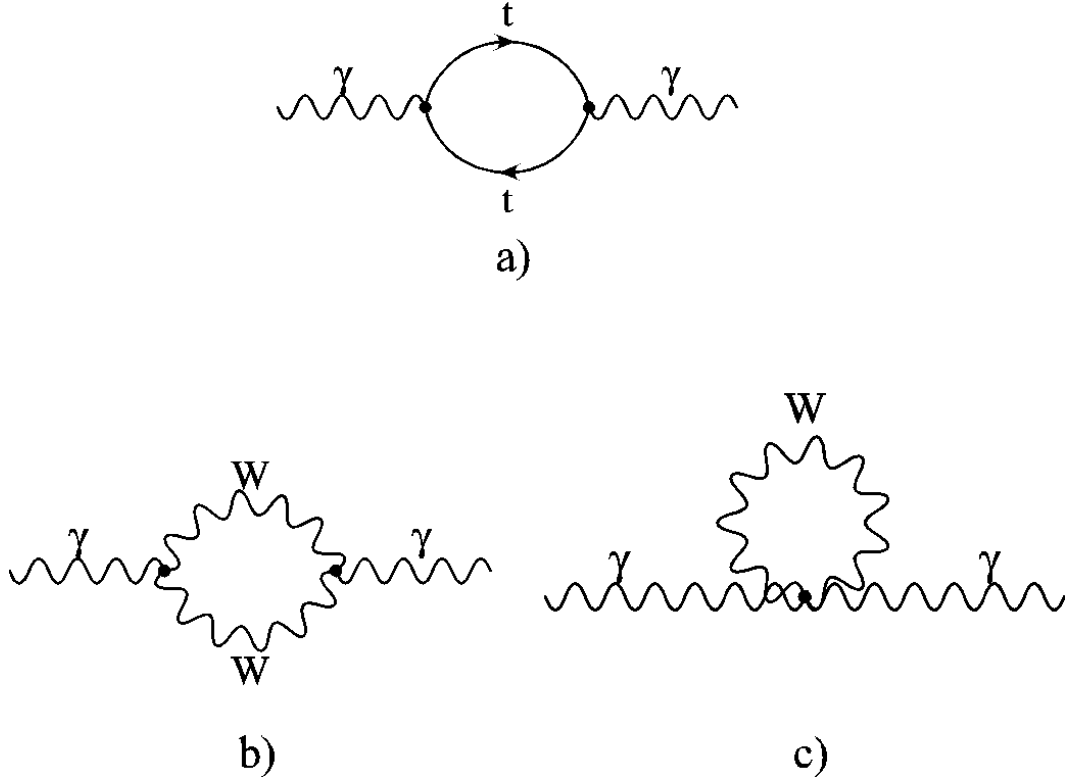}}
\end{center}
\caption{ Virtual $t$-quarks (a) and $W$ bosons (b), (c) in
the photon polarization of the vacuum. }
\end{figure}
(\ref{15})), which traditionally are not included into the
running of $\alpha(q^2)$, i.e. into $\bar{\alpha}$. It is
reasonable to treat them as electroweak corrections. This is
especially true for the $W$-loop that depends on the gauge
chosen for the description of the $W$ and $Z$ bosons. Only
after this loop is taken into account, the resultant electroweak
corrections become gauge-invariant, as it should indeed be for
physical observables. Here and hereafter in the calculations
the 't Hooft--Feynman gauge is used (see Appendix A),
\begin{equation}
\delta_1 V_m(t,h) =
-\frac{16}{3} \pi s^4 \frac{1}{\alpha}(\delta_W\alpha + \delta_t \alpha)
= -0.055 \;\; ,
\label{551}
\end{equation}
\begin{equation}
\delta_1 V_R(t,h) = -\frac{16}{3} \pi s^2 c^2
\frac{1}{\alpha}(\delta_W\alpha + \delta_t\alpha) = -0.181 \;\; ,
\label{552}
\end{equation}
\begin{equation}
\delta_1 V_A(t,h) = \delta_1 V_{\nu}(t,h) = 0\;,
\label{553}
\end{equation}
where
\begin{equation}
\frac{\delta_W\alpha}{\alpha} = \frac{1}{2\pi} [(3 + 4c^2)(1
-\sqrt{4c^2-1} \arcsin \frac{1}{2c}) - \frac{1}{3}] = 0.0686 \; ,
\label{554}
\end{equation}
\begin{equation}
\frac{\delta_t\alpha}{\alpha} = -\frac{4}{9\pi}[(1+2t)F_t(t) - \frac{1}{3}]
\simeq - \frac{4}{45\pi}\frac{1}{t} + ... \simeq
- 0.00768 \;\; .
\label{555}
\end{equation}
(Here and in what follows, unless specified otherwise,
we  use $m_t = 175$ GeV in numerical evaluations.)}

\item{The corrections $\delta_2V_i$ are caused by including virtual
gluons in electroweak loops in the order $\bar{\alpha}\hat{\alpha}_s$
(see figures 8a--c); 
\begin{figure}
\epsfxsize=350pt
\begin{center}
\parbox{\epsfxsize}{\epsfbox{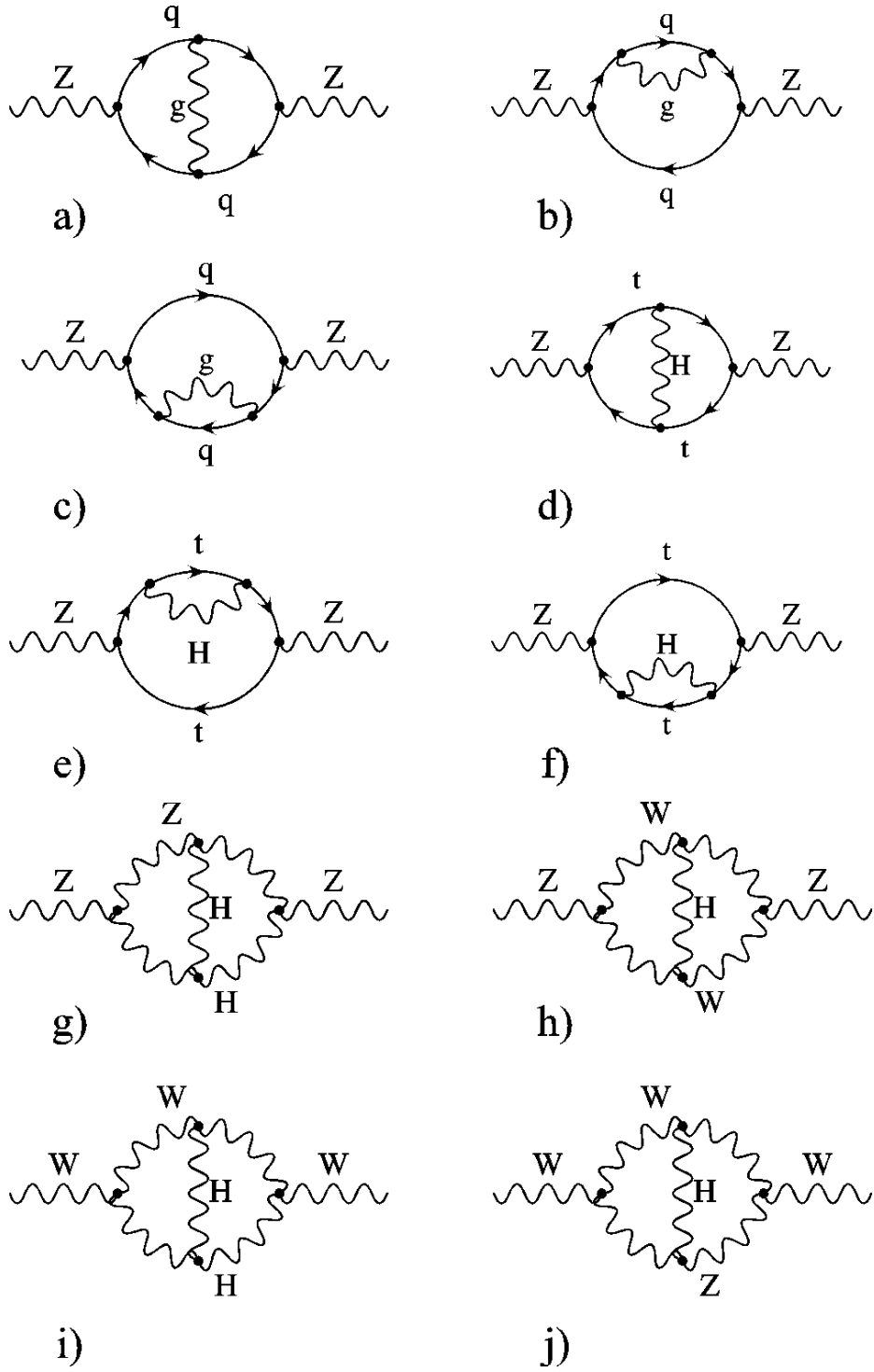}}
\end{center}
\caption{ Gluon corrections to the electroweak quark loop of
the proper energy of the $Z$ boson (a)-(c). 
Higgs corrections to the electroweak $t$ quark loop of the $Z$
boson (d)-(f). 
Two-loop higgs corrections (g)-(j). }
\end{figure}
similar diagrams can, of course, be drawn
for $W$-bosons. In addition to loops with light quarks
$q = u, d, s, c$, there exist similar loops with third-generation
quarks $t$ and $b$:
$$
\delta_2 V_i(t) = \delta^q_2 V_i + \delta^t_2 V_i(t)\;\;.
$$
The analytical expressions for corrections $\delta^q_2 V_i$ and
$\delta^t_2 V_i(t)$ are given in Appendix O. Here we only give
numerical estimates for them,
\begin{equation}
\delta^q_2 V_m =
-0.377\frac{\hat{\alpha}_s}{\pi}\;\;,
\label{56}
\end{equation}
\begin{equation}
\delta^q_2 V_A = 1.750 \frac{\hat{\alpha}_s}{\pi}\;\;,
\label{57}
\end{equation}
\begin{equation}
\delta^q_2 V_R = 0\;\;,
\label{58}
\end{equation}
\begin{equation}
\delta^t_2 V_m(t) = -11.67 \frac{\hat{\alpha}_s(m_t)}{\pi}
= -10.61 \frac{\hat{\alpha}_s}{\pi} \;\; ,
\label{59}
\end{equation}
\begin{equation}
\delta^t_2 V_A(t) = -10.10 \frac{\hat{\alpha}_s(m_t)}{\pi} = -9.18
\frac{\hat{\alpha}_s}{\pi} \;\; ,
\label{60}
\end{equation}
\begin{equation}
\delta^t_2 V_R(t) = -11.88 \frac{\hat{\alpha}_s(m_t)}{\pi} = -10.80
\frac{\hat{\alpha}_s}{\pi} \;\; ,
\label{61}
\end{equation}
since \cite{12}
\begin{equation}
\hat{\alpha}_s(m_t)
=\frac{\hat{\alpha}_s}{1+\frac{23}{12\pi}\hat{\alpha}_s\ln t}\;\;.
\label{62}
\end{equation}
(For numerical evaluation, we use $\hat{\alpha}_s\equiv
\hat{\alpha}_s(m_Z) = 0.125$.)}
We have mentioned already that the corrections $\delta^t_2 V_i(t)$,
whose numerical values were given in (\ref{59})--(\ref{61}),
are much larger than all other terms included in $\delta V_i$.
We emphasize that the term in $\delta^t_2 V_i$ that is leading
for high $t$ is universal: it is independent of $i$. As shown in
\cite{47}, this leading term is obtained by multiplying the
Veltman asymptotics $t$ by a factor
\begin{equation}
1 - \frac{2\pi^2 + 6}{9} \frac{\hat{\alpha}_s(m_t)}{\pi}\;\;,
\label{1162}
\end{equation}
or, numerically,
\begin{equation}
t \to t(1 - 2.86 \frac{\hat{\alpha}_s(m_t)}{\pi}) \;\; .
\label{1163}
\end{equation}
Qualitatively the factor (\ref{1162}) corresponds to the fact
that the running mass of the $t$-quark at momenta $p^2 \sim m^2_t$
that circulate in the $t$-quark loop is lower than
"on the mass-shell"
mass  of the $t$-quark. It is interesting to compare the
correction (\ref{1163}) with the quantity
\begin{equation}
\tilde{m}^2_t \equiv m^2_t(p^2_t =
- m^2_t) = m^2_t(1 - 2.78 \frac{\hat{\alpha}_s(m_t)}{\pi}) \;\; ,
\label{1164}
\end{equation}
calculated in the Landau gauge in \cite{48}, p~102. The
agreement is overwhelming. There is, therefore, a simple
mnemonic rule for evaluating the main gluon corrections for the
$t$-loop.
\item{Corrections $\delta_3 V_i$ of the order of
$\bar{\alpha}\hat{\alpha}^2_s$ were calculated in the literature
\cite{49} for the term leading in $t$ (i.e.
$\bar{\alpha}\hat{\alpha}^2_s t$).
They are independent of $i$ (in numerical estimates we use for
the number of quark flavors $N_f = 5$):
\begin{equation}
\delta_3 V_i(t)\simeq -(2.38 - 0.18 N_f) \hat{\alpha}_s^2(m_t)t
\simeq - 1.48 \hat{\alpha}_s^2(m_t)t = -0.07 \;\; .
\label{63}
\end{equation}

The corrections $\delta_1 V_i, \delta_2 V_i, \delta_3 V_i$,
are independent of $m_H$,  the corrections $\delta_4 V_i$
depend both on $m_t$, and on $m_H$, while the corrections $\delta_5
V_i$ are proportional to $m^2_H$. In contrast to all previous
corrections, they arise due to the electroweak
interaction in two loops, not one.}
\item{In the leading approximation in $t$ the correction
$\delta_4 V_i(t, h)$ produced by the diagrams of figures 8d-f
is independent of $i$ and takes the form
\begin{equation}
\delta_4 V_i(t, h) = -\frac{\bar{\alpha}}{16\pi s^2 c^2}
A(\frac{h}{t})t^2 \;\; ,
\label{64}
\end{equation}
where the function $A(h/t)$, calculated in \cite{50}, is given
in Table 4 for $m_H/m_t < 4$.

For $m_t = 175 GeV$ and $m_H = 300 GeV$
\begin{equation}
\delta_4 V_i = -0.11\;\; .
\label{65}
\end{equation}

The following expansion holds for $m_H/m_t > 4$:
\begin{eqnarray}
A(h/t) & = & -\frac{49}{4} - \pi^2 - \frac{27}{2} \ln r - \frac{3}{2}
\ln^2 r - \frac{1}{3}r (2 - 12\pi^2 + 12 \ln r - 27 \ln^2 r)
\nonumber \\
& - & \frac{r^2}{48} (1613 - 240 \pi^2 - 1500 \ln r - 720 \ln^2
r)\;\;,
\label{1181}
\end{eqnarray}
where $r = t/h$. $\delta_4 V_i(t,h)$ is the greatest of the
two-loop corrections in electroweak interaction; however, it is also
several times smaller than the main gluon corrections $\delta^t_2 V_i$.}

\newpage
\begin{center}
{\bf Table 4}
\vspace{2mm}

\begin{tabular}{|r|r|r||r|r|r|}
\hline
$m_H/m_t$ & $A(m_H/m_t)$ & $\tau^{(2)}(m_H/m_t)$ &
$m_H/m_t$ & $A(m_H/m_t)$ & $\tau^{(2)}(m_H/m_t)$ \\ \hline
0.00 & 0.739 & 5.710 & 2.10 & 9.655 & 1.373 \\
0.10 & 1.821 & 4.671 & 2.20 & 9.815 & 1.421 \\
0.20 & 2.704 & 3.901 & 2.30 & 9.964 & 1.475 \\
0.30 & 3.462 & 3.304 & 2.40 & 10.104 & 1.533 \\
0.40 & 4.127 & 2.834 & 2.50 & 10.235 & 1.595 \\
0.50 & 4.720 & 2.461 & 2.60 & 10.358 & 1.661 \\
0.60 & 5.254 & 2.163 & 2.70 & 10.473 & 1.730 \\
0.70 & 5.737 & 1.924 & 2.80 & 10.581 & 1.801 \\
0.80 & 6.179 & 1.735 & 2.90 & 10.683 & 1.875 \\
0.90 & 6.583 & 1.586 & 3.00 & 10.777 & 1.951 \\
1.00 & 6.956 & 1.470 & 3.10 & 10.866 & 2.029 \\
1.10 & 7.299 & 1.382 & 3.20 & 10.949 & 2.109 \\
1.20 & 7.617 & 1.317 & 3.30 & 11.026 & 2.190 \\
1.30 & 7.912 & 1.272 & 3.40 & 11.098 & 2.272 \\
1.40 & 8.186 & 1.245 & 3.50 & 11.165 & 2.356 \\
1.50 & 8.441 & 1.232 & 3.60 & 11.228 & 2.441 \\
1.60 & 8.679 & 1.232 & 3.70 & 11.286 & 2.526 \\
1.70 & 8.902 & 1.243 & 3.80 & 11.340 & 2.613 \\
1.80 & 9.109 & 1.264 & 3.90 & 11.390 & 2.700 \\
1.90 & 9.303 & 1.293 & 4.00 & 11.436 & 2.788 \\
2.00 & 9.485 & 1.330 & & & \\ \hline
\end{tabular}
\end{center}
\vspace{2mm}

\item{Corrections $\delta_5 V_i$ due to two-loop diagrams
of the type of figure 8g-8j. They are negligible, but for
the sake of completeness of the presentation, we list them
in Appendix P.}
\end{enumerate}

\begin{center}
{\bf Accidental (?) compensation and the mass of the
${\bf{t}}$-quark.}
\end{center}

\vspace{3mm}

Now that we have expressions for all terms in formula (\ref{51}),
it will be convenient to analyze their roles and the general
behavior of the functions $V_i(t,h)$.
As functions of $m_t$ at three fixed values of $m_H$, they
are shown in figures 9a, 10a, 11a.
\begin{figure}
\epsfxsize=190pt
\begin{center}
\parbox{\epsfxsize}{\epsfbox{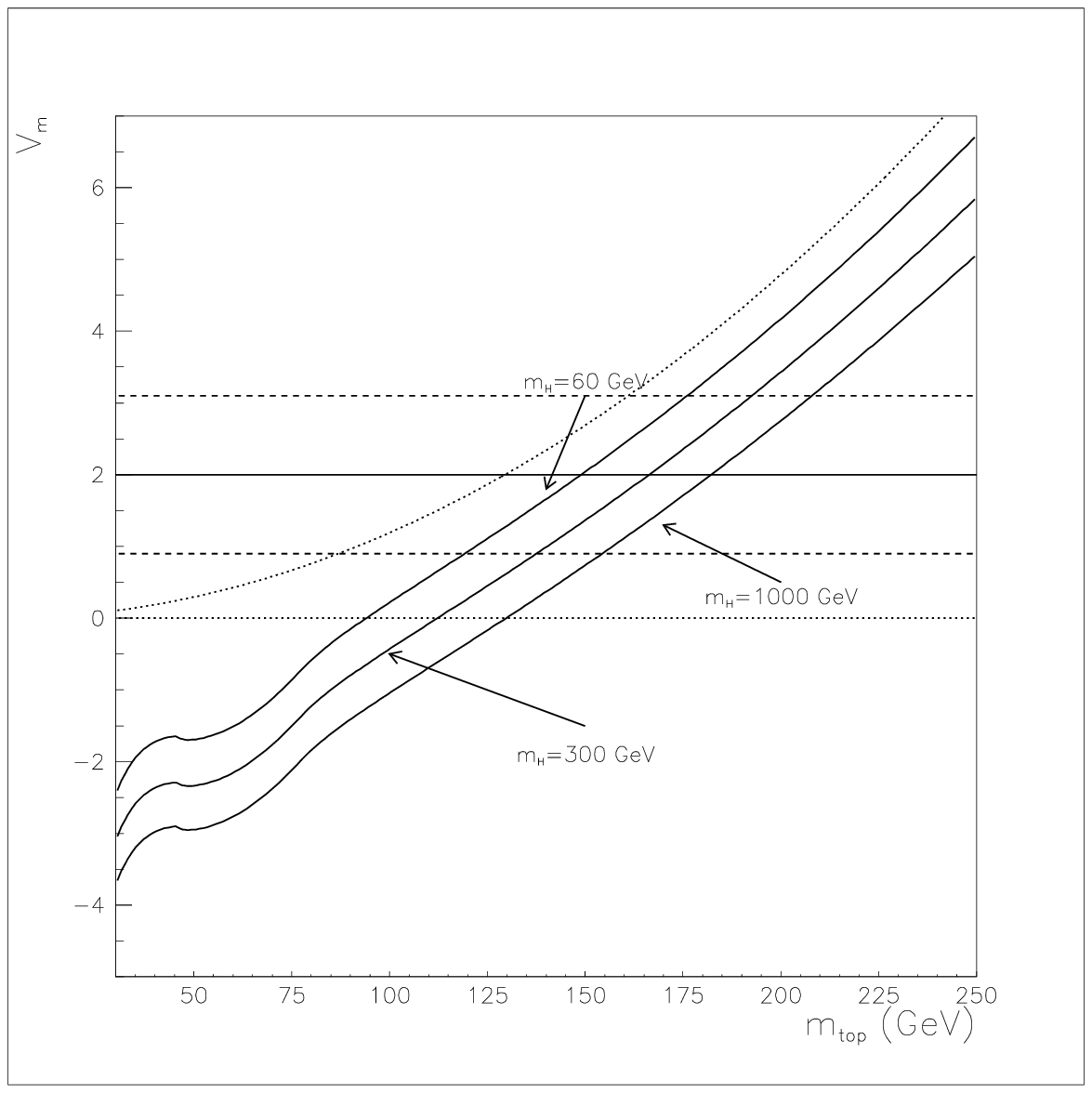}}

\epsfxsize=190pt
\parbox{\epsfxsize}{\epsfbox{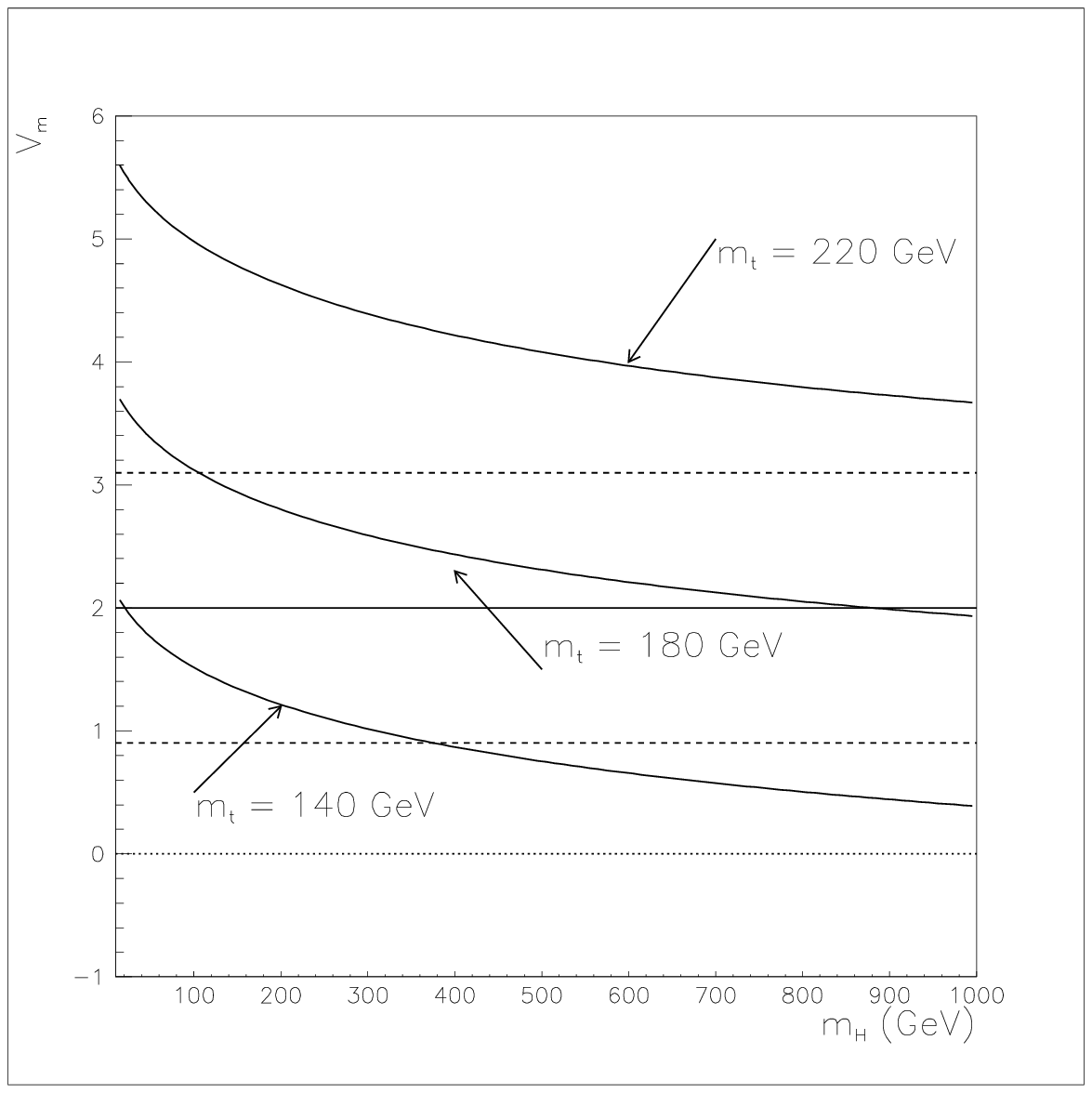}}
\end{center}
\caption{ $V_m$ as a function of $m_t$ for three values of $m_H$: 60, 300
and 1000 GeV, according to equation  (\ref{51}). The
dashed-curve parabola $t = (m_t/m_Z)^2$ corresponds to the
Veltman approximation. Solid horizontal line traces the experimental 
value of $V_m^{exp}$ while the long-dash horizontal lines give
its upper and lower limits of the $1\sigma$ level (a). 
   $V_m$ as a function of $m_H$ for three values of $m_t$: 150, 175
and 200 GeV, according to equation (\ref{51}). 
(Horizontal lines the same as in figure 9(a). )  }
\end{figure}
\begin{figure}
\epsfxsize=190pt
\begin{center}
\parbox{\epsfxsize}{\epsfbox{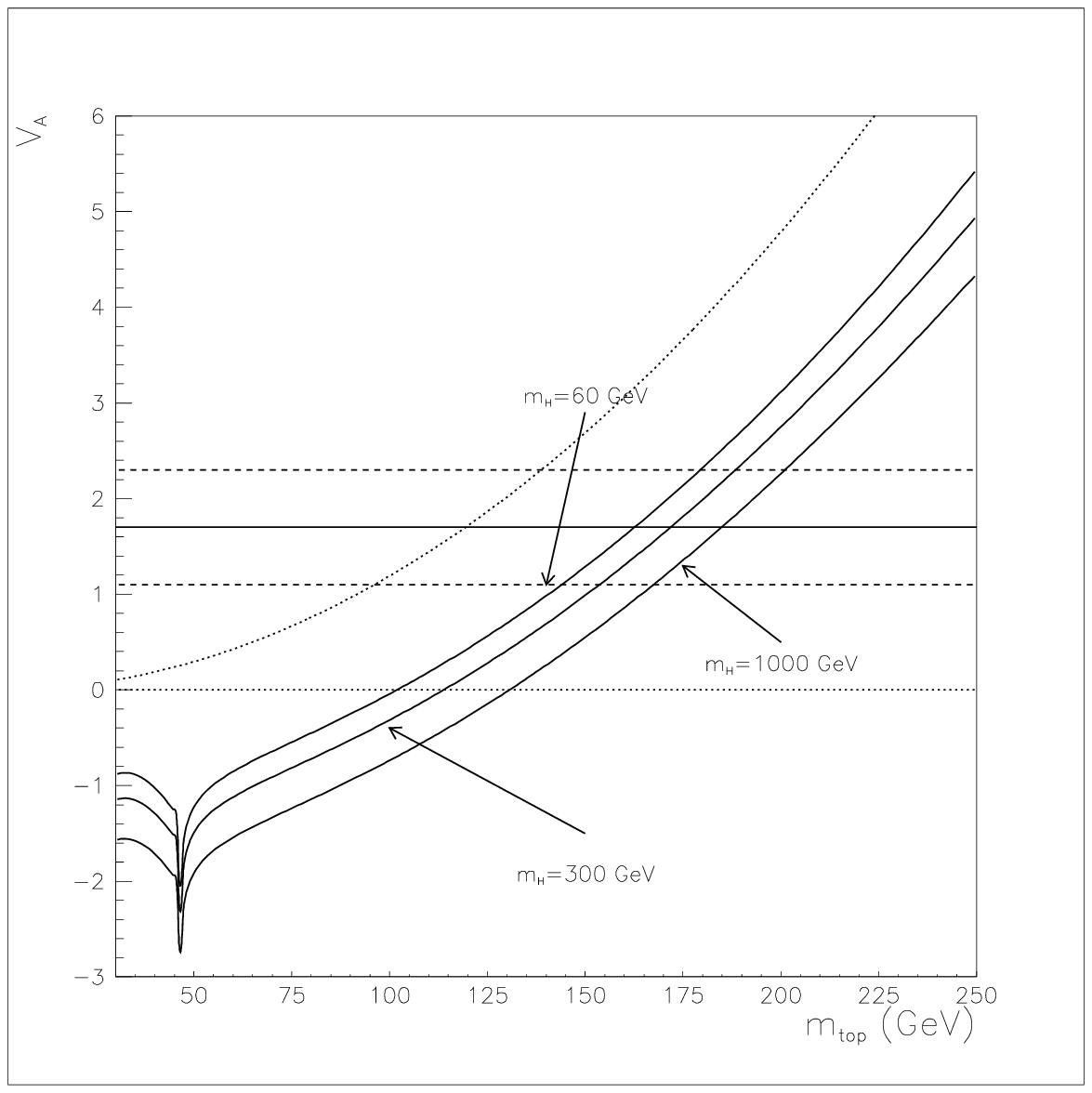}}

\epsfxsize=190pt
\parbox{\epsfxsize}{\epsfbox{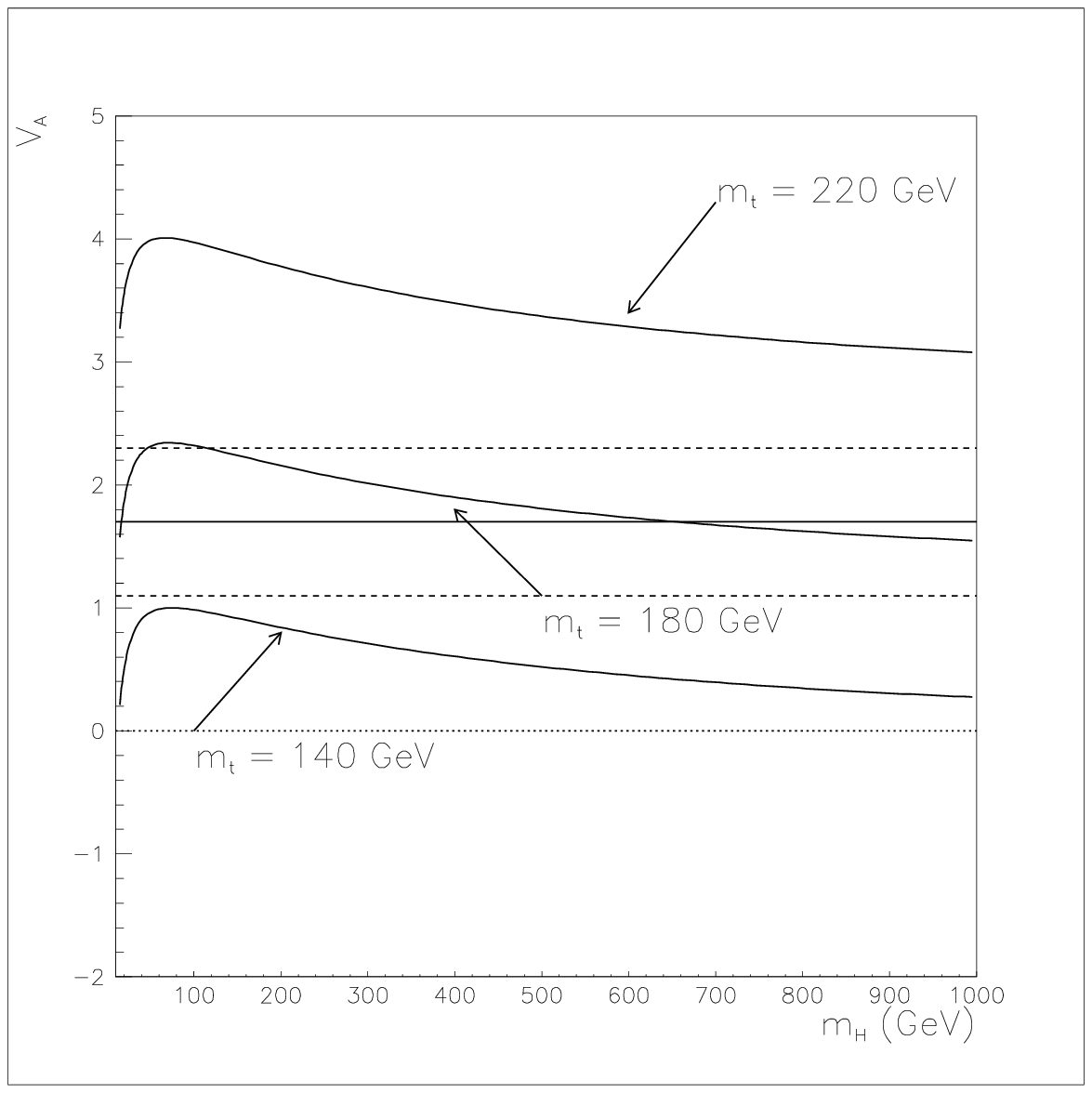}}
\end{center}
\caption{$V_A$ as a function of $m_t$. (The remaining
clarifications are similar to those given to figure 9) (a). 
  $V_A$ as a function $m_H$ for three values of $m_t$: 150, 175
and 200 GeV. (See caption to figure 9) (b) }
\end{figure}
\begin{figure}
\epsfxsize=190pt
\begin{center}
\parbox{\epsfxsize}{\epsfbox{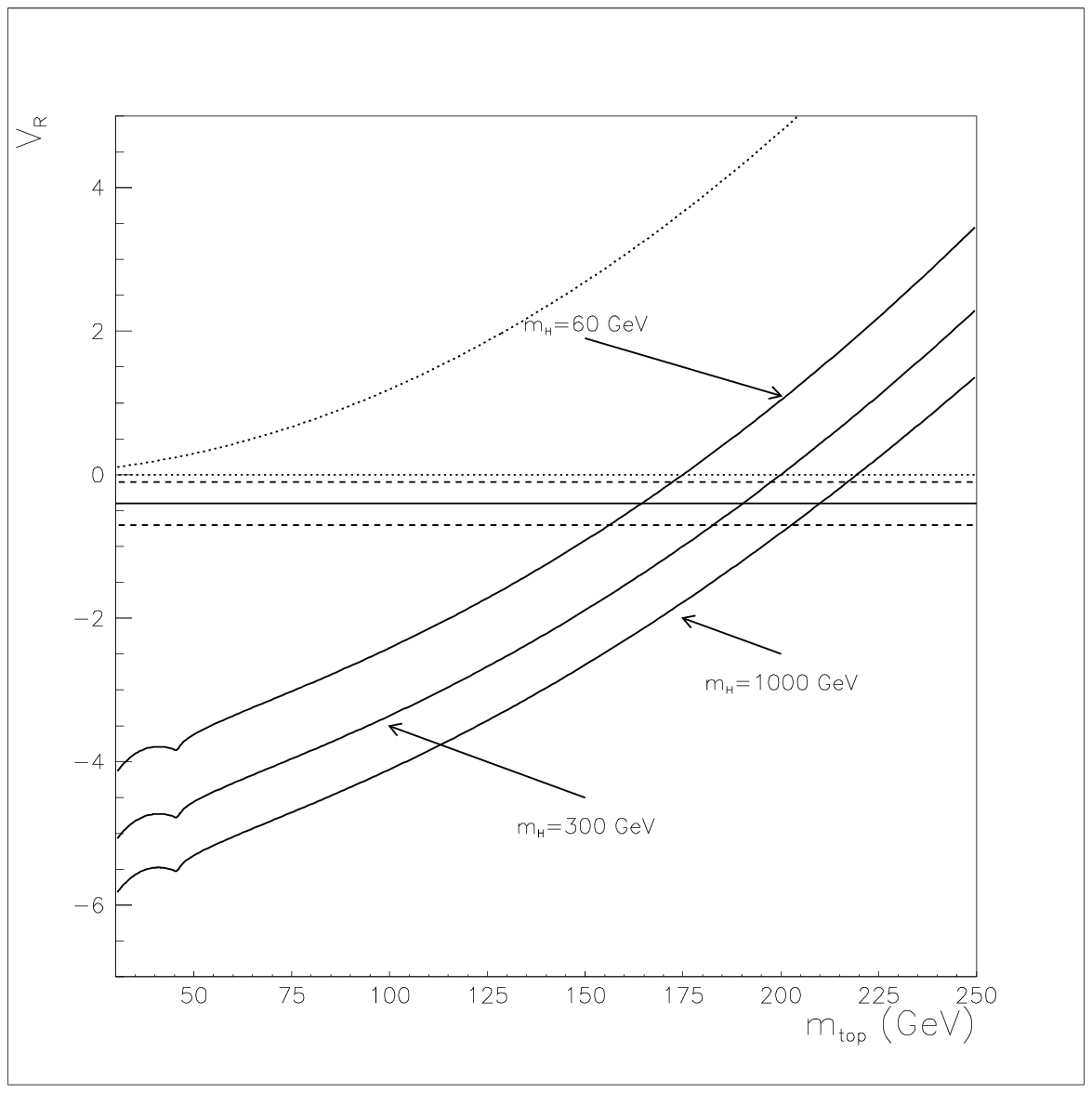}}

\epsfxsize=190pt
\parbox{\epsfxsize}{\epsfbox{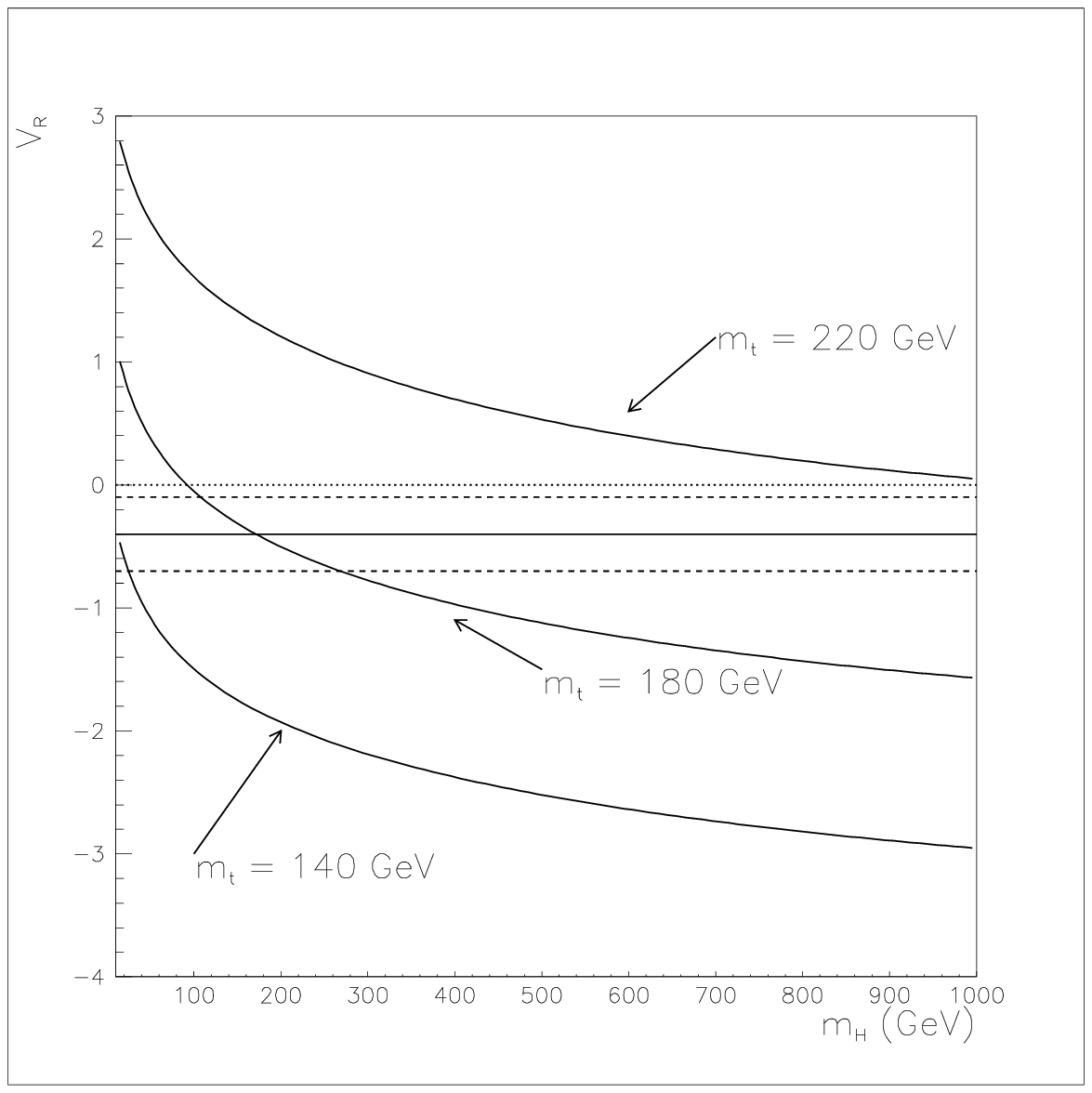}}
\end{center}
\caption{ $V_R$ as a function of $m_t$. (The remaining
clarifications are similar to those given to figure 9) (a). 
$V_R$ as a function of $m_H$ for three values of $m_t$: 150,
175 and 200 GeV. (See caption to figure 9) (b).   }
\end{figure}
 On all these figures, we see
a cusp at $m_t = m_Z/2$. This is a typical
threshold singularity that arises when the channel $Z \to t\bar{t}$
is opened. It is of no practical significance since
experiments give $m_t \gg m_Z/2$. What really impresses is that
the functions $V_i$ are nearly zero in the interval $m_t \sim 100-200$
GeV. This happens because of the compensation of the leading
term $t$ and the rest of the terms which produce a negative
aggregate contribution. This is especially well pronounced in
the function $V_R$ at $m_t \sim 180$ GeV. Here the main negative
contribution comes from the light fermions (the constant $C_R$).

If we neglect the small correction $\delta_4 V_i(t,h)$ which
depends  both on $t$ and on $h$, then each
function $V_i(t,h)$ is a sum of functions one of which is
$t$-dependent but independent of $h$, while the second is
$h$-dependent but independent of $t$ (plus, of course, a
constant which is independent of both $t$ and $h$). Therefore
the curves for $m_H =60$ and $1000$ GeV in figures 9a, 10a, 11a
are (mostly) produced by the parallel transfer of the curve
for $m_H= 300$ GeV.

We see in figures 9a, 10a and 11a that if the $t$-quark were light,
radiative corrections would be negative, and if it were very
heavy, they would be much larger. This looks like a
conspiracy of the
observable mass of the
$t$-quark and all other parameters of the electroweak theory,
as a result of which the electroweak correction $V_R$ becomes
anomalously small. Note that the corrections do not vanish
simultaneously because, if we fix the value of $m_H$, then
$V_m(t)$, $V_A(t)$ and $V_R(t)$ cross the horizontal lines
$V_i = 0$ at different values of $m_t$. What happens is an
approximate vanishing of the correction, which as if
corresponds to some broken symmetry. The nature of this symmetry
is not clear at all, and even its existence is very problematic.

One should specially note the dashed parabola in figures 9a, 10a
and 11a corresponding to the Veltman term $t$. We see that in
the interval $0 < m_t < 250$ GeV it lies much higher than
$V_A$ and $V_R$ and approaches $V_m$ only in the right-hand side
of figure 9a. Therefore, the so-called non-leading `small' corrections
that are typically replaced with ellipses in standard texts,
are found to be comparable with the leading term $t$.

A glance at figures 9a, 10a, 11a readily explains how the experimental
analysis of electroweak corrections allows, despite their
smallness, a prediction, within the framework of the minimal
standard model, of the $t$-quark mass. Even when the
experimental accuracy of LEP1 and SLC experiments was not
sufficient for detecting electroweak corrections, it was
sufficient for establishing the $t$-quark mass using the points
at which the curves $V_i(m_t)$ intersect the horizontal lines
corresponding to the experimental values of $V_i$ and the
parallel to them thin lines that show the band of one
standard deviation. The accuracy in determining $m_t$ is imposed
by the band width and the slope of $V_i(m_t)$ lines.

The dependence $V_i(m_H)$ for three fixed values of
$m_t = 150$, 175 and 200 GeV (figures 9b, 10b, 11b) can be
presented in a similar manner. As follows from the explicit
form of the terms $H_i(m_H)$, the dependence $V_i(m_H)$ is
considerably less steep (logarithmic). This is the reason why
the prediction of the higgs mass extracted from electroweak corrections
has such a high uncertainty. We will see later
(figures 13, 14, 15) that the accuracy of prediction of
$m_H$ will greatly depend on what the $t$-quark's mass is going
to be. If $m_t = 150 \pm 5$ GeV, then $m_H < 200$ GeV
at the $3 \sigma$ level. If $m_t = 200 \pm 5$ GeV, then
$m_H > 120$ GeV at the $3 \sigma$ level. If, however,
$m_t = 175 \pm 5$ GeV, we are hugely unlucky: there is
practically no constraint on $m_H$.

Before starting a discussion of hadronic decays of the $Z$ boson,
let us `go back to the roots' and recall how the equations for
$V_i(m_t, m_H)$ were derived.

\begin{center}
{\bf How to calculate ${\bf{V_i}}$? `Five steps'.}
\end{center}

\vspace{3mm}
An attentive reader should have already come up with the
question: what makes the amplitudes of the lepton decays of the $Z$ boson
in the one-loop approximation depend on the self-energy of the $W$
boson?  Indeed, the loops describing the self-energy of the $W$
boson appear in the decay diagrams of the $Z$ boson
only beginning with the two-loop approximation. The
answer to this question is this. We have already emphasized that
we find expressions for radiative corrections to
$Z$-boson decays in terms of $\bar{\alpha}$, $m_Z$ and $G_{\mu}$.
However, the expression for $G_{\mu}$ includes the self-energy
of the $W$ boson even in the one-loop approximation. The point
is thus in our expressing some observables (in this particular
case, $m_W/m_Z$, $g_A$, $g_V/g_A$) in terms of other, more
accurately measured observables ($\bar{\alpha}$, $m_Z$, $G_{\mu}$).

Let us trace how this is achieved, step by step.
There are altogether `five steps to happiness',
based on the one-loop approximation.

{\bf Step I.} We begin with the electroweak Lagrangian after it
had undergone the spontaneous violation of the
$SU(2) \times U(1)$-symmetry by the higgs vacuum condensate
(vacuum expectation value -- VEV)
$\eta$ and the $W$ and $Z$ bosons became massive. Let us
consider the bare coupling constants (the bare charges $e_0$ of the photon,
$g_0$ of the $W$-boson and $f_0$ of the $Z$-boson) and the bare
masses of the vector bosons:
\begin{equation}
m_{Z0} = \frac{1}{2} f_0 \eta \;\;,
\label{66}
\end{equation}
\begin{equation}
m_{W0} = \frac{1}{2} g_0 \eta \;\;,
\label{67}
\end{equation}
and also bare masses: $m_{t0}$ of the $t$-quark and $m_{H0}$ of
the higgs.

{\bf Step II.} We express $\bar{\alpha}$, $G_{\mu}$, $m_Z$ in
terms of $f_0$, $g_0$, $e_0$, $\eta$, $m_{t0}$, $m_{H0}$ and
$1/\varepsilon$ (see Appendix E). Here $1/\varepsilon$ appears
because we use the dimensional regularization, calculating
Feynman integrals in the space of $D$ dimensions (see Appendix A).
These integrals diverge at $D=4$ and are finite
in the vicinity of  $D = 4$.
By definition,
\begin{equation}
2\varepsilon = 4 - D \to 0 \;\;.
\label{68}
\end{equation}
Note that in the one-loop approximation $m_{t0} = m_t$, $m_{H0} = m_H$,
since we neglect the electroweak corrections to the masses of the
$t$-quark and the higgs. For the higgs this approximation is
quite legitimate, since the accuracy of extracting its mass
from radiative corrections is very poor. As for the $t$-quark,
this statement is also true at the current accuracy of
the experimental measurement of electroweak corrections; however,
this would become an unacceptably crude approximation if this
accuracy could be improved by an order of magnitude at
LEP and SLC. The situation here is analogous to that for
$G_{\mu}$ and the self-energy of the $W$-boson.

Step II is almost physics: we calculate Feynman diagrams
(we say `almost' to emphasize that observables are
expressed in terms of nonobservable, `bare', and generally
infinite quantities).

{\bf Step III.} Let us invert the expressions derived at step II
and write $f_0$, $g_0$, $\eta$ in terms of $\bar{\alpha}$,
$G_{\mu}$, $m_Z$, $m_t$, $m_H$ and $1/\varepsilon$. This step is
pure algebra.

{\bf Step IV.} Let us express $V_m$, $V_A$, $V_R$ (or the electroweak
one-loop correction to any other electroweak observable, all of
them being treated on an equal basis) in terms of $f_0$, $g_0$,
$\eta$, $m_t$, $m_H$ and $1/\varepsilon$. (Like step II, this
step is again almost physics.)

{\bf Step V.} Let us express $V_m$, $V_A$, $V_R$ (or any other
electroweak correction) in terms of $\bar{\alpha}$, $G_{\mu}$, $m_Z$,
$m_t$, $m_H$ using the results of steps III and IV. Formally
this is pure algebra, but in fact pure physics, since now we
have expressed certain physical observables in terms of other
observables. If no errors were made on the way, the terms
$1/\varepsilon$ cancel out. As a result, we arrive at formula
(\ref{51}) which gives $V_i$ as elementary functions of
$t$, $h$ and $s$.

The five steps outlined above are very simple and visually clear.
We obtain the main relations without using the `heavy artillery'
of quantum field theory with its counterterms in the Lagrangian
and the renormalization procedure. This simplicity and visual
clarity became possible owing to the one-loop electroweak
approximation (even though this approach to renormalizations is
possible in multiloop calculations, it becomes more cumbersome than
standard procedures). As for the QCD-corrections to quark electroweak
loops and the two-loop higgs contribution hidden in the terms
$\delta V_i$ in equation (\ref{51}), we take the relevant formulas
from the calculations of other authors.
\section{Hadronic decays of the ${\bf{Z}}$ boson. The leading quarks
and hadrons.}
As discussed above (see formulas (\ref{29})--(\ref{34})
and the subsequent section on `Asymmetry'), an analysis of hardronic
decays reduces to the calculation of decays to pairs of quarks:
$Z \to q\bar{q}$. The key role is played by the concept of leading
hadrons that carry away the predominant part of the energy.
For example, the $Z \to c\bar{c}$ decay
mostly produces two hadron jets flying in opposite directions,
in one of which the leading hadron is the one containing the
$\bar{c}$-quark, for example, $D^- = \bar{c}d$, and in the other
the hadron with the c-quark, for example, $D^0 = c\bar{u}$
or $\Lambda^+_c = udc$. Likewise, $Z \to b\bar{b}$ decays are
identified by the presence of high-energy $B$ or $\bar{B}$
mesons. If we select only particles with energy close to
$m_Z/2$, the identification of the initial quark channels is
unambiguous. The total number of such cases will, however, be
small. If we take into account as a signal less energetic
$B$-mesons, we face the problem of their origin. Indeed, a pair
$b\bar{b}$ can be created not only directly by a $Z$ boson but also
by a virtual gluon in, say, a $Z \to c\bar{c}$ decay (figure 12a)
\begin{figure}
\epsfxsize=400pt
\begin{center}
\parbox{\epsfxsize}{\epsfbox{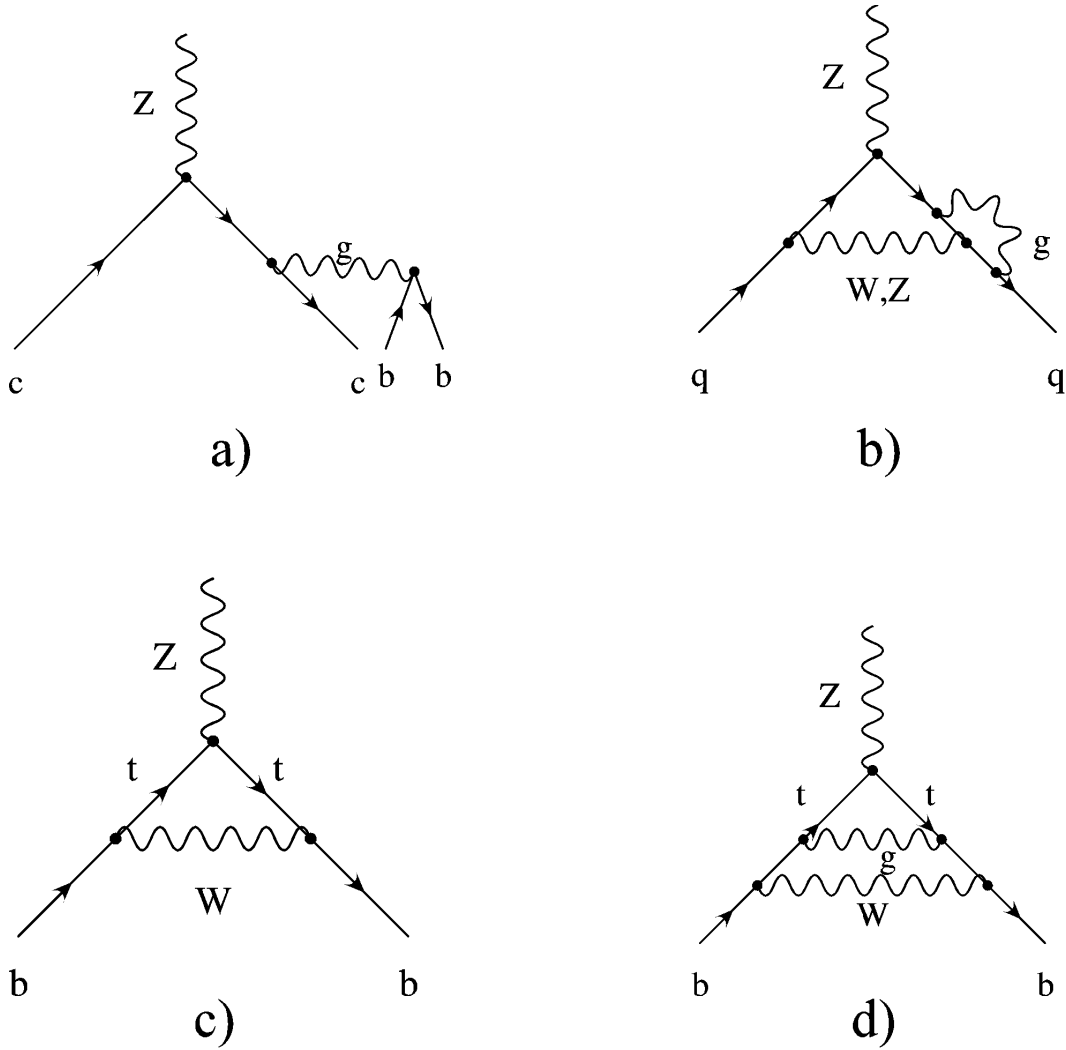}}
\end{center}
\caption{ The $Z\to c\bar{c}$ decay producing a secondary
pair $b\bar{b}$ created by a virtual gluon $g$ (a). 
  The $Z\to q\bar{q}$ decay with a virtual gluon $g$ which
connects a side of the quark triangle with the external quark
line. Diagrams of this type have not been calculated yet (b). 
  The verted electroweak diagram involving $t$-quarks and 
contributing to the $Z\to b\bar{b}$ decay (c). 
  One of the diagrams describing gluon corrections to the 
diagram of figure 12(c). }
\end{figure}
or $Z \to u\bar{u}$, or $s\bar{s}$. This example shows the
sort of difficulty encountered by experimentalists trying to
identify a specific quark--antiquark channel. Furthermore, owing
to secondary pairs, the total hadron width is not strictly equal to
the sum of partial quark widths.

We remind the reader that for the partial width of the $Z \to q\bar{q}$
decay we had (\ref{29})
\begin{equation}
\Gamma_q \equiv \Gamma(Z \to q\bar{q}) =
12 \Gamma_0 [g^2_{Aq} R_{Aq} + g^2_{Vq} R^2_{Vq}] \;\;,
\label{69}
\end{equation}
where the standard width $\Gamma_0$ is, according to (\ref{25}),
\begin{equation}
\Gamma_0 = \frac{G_{\mu} m^3_Z}{24 \sqrt{2}\pi} =
82.944(6) {\rm MeV} \;\;,
\label{70}
\end{equation}
and the radiators $R_{Aq}$ and $R_{Vq}$ are given in Appendix F.
As for the electroweak corrections, they are included in the
coefficients $g_{Aq}$ and $g_{Vq}$. The sum of the Born and one-loop
terms has the form
\begin{equation}
g_{Aq} = T_{3q}[1 + \frac{3\bar{\alpha}}{32 \pi s^2c^2}
V_{Aq}(t,h)] \;\;,
\label{71}
\end{equation}
\begin{equation}
R_q \equiv g_{Vq}/g_{Aq} = 1 - 4|Q_q| s^2 + \frac{3|Q_q|}{4\pi(c^2
- s^2)} \bar{\alpha} V_{Rq}(t,h) \;\;.
\label{72}
\end{equation}

\vspace{3mm}

\begin{center}
{\bf Decays to pairs of light quarks.}
\end{center}

\vspace{3mm}

Here, as in the case of hadronless observables, the quantities $V$
that characterize corrections are normalized in the standard
way: $V \to t$ as $t \gg 1$. Naturally, those terms in $V$ that
are due to the self-energies of vector bosons are identical for both
leptons and quarks. The deviation of the differences $V_{Aq}- V_{A}$
and $V_{Rq} - V_{R}$ from zero are caused by the differences in
radiative corrections to vertices $Z \to q\bar{q}$ and $Z \to
l\bar{l}$.  For four light quarks we have \begin{equation}
V_{Au}(t,h) = V_{Ac}(t,h) = V_{A}(t,h)
 + \frac{128\pi s^3 c^3}{3 \bar{\alpha}} (F_{Al}+F_{Au})\;\;,
\label{73}
\end{equation}
\begin{equation}
V_{Ad}(t,h) = V_{As}(t,h) = V_{A}(t,h) + \frac{128\pi s^3 c^3}{3
\bar{\alpha}} (F_{Al}-F_{Ad}) \;\;,
\label{74}
\end{equation}
\begin{eqnarray}
V_{Ru}(t,h) = V_{Rc}(t,h) = V_{R}(t,h)+ \frac{16\pi sc(c^2 - s^2)}{3
\bar{\alpha}} \times \\ \nonumber
\times [F_{Vl}-(1-4s^2)F_{Al} + \frac{3}{2}(-(1-\frac{8}{3}s^2)
F_{Au}+F_{Vu})] \;\;,
\label{75}
\end{eqnarray}
\begin{eqnarray}
V_{Rd}(t,h) = V_{Rs}(t,h) = V_{R}(t,h)+ \frac{16\pi sc(c^2 - s^2)}{3
\bar{\alpha}} \times \\ \nonumber
\times [F_{Vl}-(1-4s^2)F_{Al} + 3((1-\frac{4}{3}s^2)F_{Ad}-F_{Vd})]
\;\;,
\label{76}
\end{eqnarray}
where (see Appendix M):
\begin{equation} F_{Al} =
\frac{\bar{\alpha}}{4\pi}(3.0099 + 16.4 \delta s^2)\;,
\label{77}
\end{equation}
\begin{equation}
F_{Vl} = \frac{\bar{\alpha}}{4\pi}(3.1878 + 14.9 \delta s^2)\;,
\label{78}
\end{equation}
\begin{equation}
F_{Au} = - \frac{\bar{\alpha}}{4\pi}(2.6802 + 14.7 \delta s^2)\;,
\label{79}
\end{equation}
\begin{equation}
F_{Vu} = - \frac{\bar{\alpha}}{4\pi}(2.7329 + 14.2 \delta s^2)\;,
\label{80}
\end{equation}
\begin{equation}
F_{Ad} = \frac{\bar{\alpha}}{4\pi}(2.2221 + 13.5 \delta s^2)\;,
\label{81}
\end{equation}
\begin{equation}
F_{Vd} = \frac{\bar{\alpha}}{4\pi}(2.2287 + 13.5 \delta s^2)\;.
\label{82}
\end{equation}
The values of $F$ are given here for $s^2 = 0.23110 - \delta s^2$.
The accuracy to five decimal places is purely arithmetic. The
physical uncertainties introduced by neglecting higher-order
loops manifest themselves already in the third decimal place. It
is necessary to point out that corrections of the type of that
shown in figure 12b have not yet been calculated.
\begin{center}
{\bf Decays to ${\bf b\bar{b}}$ pair. }
\end{center}

\vspace{3mm}
In the $Z \to b\bar{b}$ decay it is necessary to take into account
additional $t$-dependent vertex corrections:
\begin{equation}
V_{Ab}(t,h) =
V_{Ad}(t,h)-\frac{8s^2 c^2}{3(3-2s^2)}(\phi (t) + \delta \phi (t)),
\label{83}
\end{equation}
\begin{equation}
V_{Rb}(t,h) = V_{Rd}(t,h)-\frac{4s^2 (c^2 -s^2)}{3(3-2s^2)}(\phi (t) + \delta
\phi (t))  .
\label{84}
\end{equation}
Here the term $\phi(t)$ calculated in \cite{51} corresponds to
figure 12c and the term $\delta \phi(t)$ calculated in \cite{52},
\cite{50} corresponds to the leading gluon and higgs corrections
to the term $\phi(t)$ (see figure 12d). Expressions for $\phi(t)$ and
$\delta\phi(t)$ are given in Appendix N. For $m_t = 175$ GeV,
$\hat{\alpha}_s(m_Z) = 0.125$, $m_H = 300$ GeV,
\begin{equation}
\phi(t) = 29.9\;\; ,
\label{85}
\end{equation}
\begin{equation}
\delta \phi(t) = -3.0 \;\; ,
\label{86}
\end{equation}
and correction terms in equations (\ref{83}) and (\ref{84}) are
very large: they equal $-5.0$ and $-1.8$, respectively.
\section{Comparison of theoretical results and experimental
LEP1 and SLC data.}

\begin{center}
{\bf LEPTOP code.}
\end{center}

\vspace{3mm}

A number of computer programs (codes) were written for
comparing high-precision data of LEP1 and SLC. The best known of
these programs in Europe is ZFITTER \cite{ZF}, which takes into
account not only electroweak radiative corrections but also all
purely electromagnetic ones, including, among others, the
emission of photons by colliding electrons and positrons. Some of
the first publications in which the $t$ quark mass was
predicted on the basis of precision measurements \cite{EL},
were based on the code ZFITTER. Other European codes, BHM \cite{BH},
WOH, TOPAZO \cite{TOP}, somewhat differ from ZFITTER. The best
known in the USA are the results generated by the code used by
Langacker \cite{29}. %

The original idea of the authors of this review in 1991--1993
was to derive simple analytical formulas for electroweak
radiative corrections, which would make it possible to predict
the $t$-quark mass using no computer codes, just by analyzing
experimental data on a sheet of paper. Alas, the diversity of
hadron decays of $Z$ bosons, depending on the constants of strong
gluon interaction $\hat{\alpha}_s$, was such that it was
necessary to convert analytical formulas into a computer program
which we jokingly dubbed LEPTOP \cite{53}. The LEPTOP
calculates the electroweak observables in the framework of the
Minimal Standard Model and fits experimental data so as to
determine the quantities $m_t$, $M_H$ and $\hat{\alpha}_s(M_Z)$.
The logical structure of LEPTOP is clear from the preceding
sections of this review and is shown in the Flowchart. The code of
LEPTOP can be downloaded from the Internet home page:
http://cppm.in2p3.fr./leptop/intro$\_$leptop.html

A comparison of the codes ZFITTER, BHM, WOH, TOPAZO and LEPTOP
carried out in 1994--95 \cite{16} has demonstrated that their
predictions for all electroweak observables coincide
with accuracy that is much better than the accuracy of
the experiment. The results of processing the experimental data
using LEPTOP are shown below.

\newpage
\begin{center}
{\bf Flowchart.}
\end{center}
\vspace{2mm}

\setlength{\unitlength}{1mm}
\begin{picture}(160,210)
\put(0,195){\framebox(160,15)}
\put(80,205){\makebox(0,0){\large
Choose three observables measured with the highest accuracy:}}
\put(80,199){\makebox(0,0){\large $G_{\mu}\;,
M_Z\;, \alpha(M_Z) \equiv \bar{\alpha}$ }}
\put(80,190){\line(0,1){5}}

\put(0,175){\framebox(160,15)}
\put(80,185){\makebox(0,0){\large
Determine angle $\theta$ ( $s \equiv \sin \theta, c \equiv \cos
\theta$)}}
\put(80,180){\makebox(0,0){\large in terms of $G_{\mu}\;, M_Z\;,
\bar{\alpha}$:
$G_{\mu} = (\pi / \sqrt{2}) \bar{\alpha} / s^2 c^2 M^2_Z$}}
\put(80,170){\line(0,1){5}}

\put(0,150){\framebox(160,20)}
\put(80,165){\makebox(0,0){\large
Introduce bare coupling constants in the framework of MSM}}
\put(80,160){\makebox(0,0){\large ($\alpha_0\;, \alpha_{Z0}\;,
\alpha_{W0}$), bare masses ($M_{Z0}, M_{W0}, M_{H0}, m_{t0},
m_{q0}$) }}
\put(80,155){\makebox(0,0){\large
and the vacuum expectation value (VEV) of the higgs field $\eta$.}}
\put(80,145){\line(0,1){5}}

\put(0,125){\framebox(160,20)}
\put(80,140){\makebox(0,0){\large
Express $\alpha_0\;, \alpha_{Z0}\;, M_{Z0}$ in terms of
$G_{\mu}\;, M_Z\;, \bar{\alpha}$}}
\put(80,135){\makebox(0,0){\large
in one-loop approximation, using dimensional}}
\put(80,130){\makebox(0,0){\large
regularization $(1/\varepsilon\;, \mu$).}}
\put(80,120){\line(0,1){5}}

\put(0,95){\framebox(160,25)}
\put(80,115){\makebox(0,0){\large
Express one-loop electroweak corrections to all}}
\put(80,110){\makebox(0,0){\large
electroweak observables in terms of
$\alpha_0\;, \alpha_{Z0}\;, M_{Z0}\;, m_t\;, M_H$,}}
\put(80,105){\makebox(0,0){\large
and hence, in terms of $G_{\mu}\;, M_Z\;, \bar{\alpha}\;,
m_t\;, M_H$. Check}}
\put(80,100){\makebox(0,0){\large
cancellation of the terms $(1/\varepsilon\;, \mu)$.}}
\put(80,90){\line(0,1){5}}

\put(0,65){\framebox(160,25)}
\put(80,85){\makebox(0,0){\large
Introduce gluon corrections to quark loops }}
\put(80,80){\makebox(0,0){\large
and QED (and QCD) final state interactions }}
\put(80,75){\makebox(0,0){\large
(in hadron decays), given in terms of}}
\put(80,70){\makebox(0,0){\large
$\bar{\alpha}\;,
\hat{\alpha}_s(M_Z)\;, m_b(M_Z)\;, m_t$.}}
\put(80,60){\line(0,1){5}}

\put(0,40){\framebox(160,20)}
\put(80,55){\makebox(0,0){\large
Compare the predictions of the Born approximation }}
\put(80,50){\makebox(0,0){\large
and the approximation Born + one loop with the experimental }}
\put(80,45){\makebox(0,0){\large
data on $Z$-decays.}}
\put(80,35){\line(0,1){5}}

\put(0,20){\framebox(160,15)}
\put(80,30){\makebox(0,0){\large
Perform global fit for the three parameters}}
\put(80,25){\makebox(0,0){\large $m_t\;,
\hat{\alpha}_s(M_Z)\;, M_H$ or for the first two, with the third fixed.}}
\put(80,15){\line(0,1){5}}

\put(0,-5){\framebox(160,20)}
\put(80,10){\makebox(0,0){\large
Derive theoretical predictions of the central values }}
\put(80,5){\makebox(0,0){\large
for all electroweak observables and of }}
\put(80,0){\makebox(0,0){\large
the corresponding uncertainties (`errors').}}
\end{picture}

\newpage
\begin{center}
{\bf General fit.}
\end{center}

\vspace{3mm}
Table 5 shows experimental values of the electroweak
observables, obtained by averaging the results of four LEP
detectors (part a), and also SLC data (part b) and the data on
$W$ boson mass (part c). (The data on the W boson mass
from  the $p\bar{p}$-colliders
are also shown, for the reader's convenience, in the form
of $s^2_W$, while the data on $s_W^2$ from $\nu N$-experiments
are also shown in the form of $m_W$. These two numbers are
given in italics, emphasizing that they are not independent
experimental data.) We take experimental data from the paper
\cite{4400}.

Table 5 sums up the experimental data used for determining
(fitting) the parameters of the standard model $m_t$ and
$\hat{\alpha}_s(m_Z)$ (see Table 7). The central values in the third
column of Table 5
were calculated for $m_H = 300$ GeV. Shown in brackets
is the uncertainty of the last significant decimal places due to the
uncertainty of the fitted values $m_t$ and $\hat{\alpha}_s$.
Above and below we give the shifts in the last significant
decimal places corresponding to $m_H = 1000$ GeV and $m_H = 60$ GeV,
respectively. The last column shows the value of the `pull'. By
definition, the pull is the difference between the experimental and
the theoretical values divided by experimental uncertainty. The
pull values show that the maximum discrepancy between the experimental
data and the Minimal Standard Model is found for $R_b$
($3.8 \sigma$). Deviations at the level of $2.5 \sigma$ are also
found for $R_c$ and $\sin^2 \theta^{lept}_{eff} \equiv s^2_l$ in $A_{LR}$
(SLC). For most observables the discrepancy is less than $1\sigma$.
At the same time, Table 6 shows that the value $s^2_l =
0.23186(34)$ extracted from all asymmetries at LEP agrees quite
well with the fitted MSM value $s^2_l = 0.2321(4)$ from the LEP
data, and for all sets of data.

\newpage
\begin{center}
{\bf Table 5}
\vspace{2mm}

\begin{tabular}{|l|c|c|c|}
\hline
& Experimental & Fit & \\
Observable & data & standard & Pull \\
& & model & \\ \hline
a) \underline{LEP} & & & \\
& & & \\
shape of $Z$-peak & & & \\
and lepton asymmetries: & & & \\
$m_Z$ [GeV] & 91.1884(22) & & \\
$\Gamma_Z$ [GeV] & 2.4963(32) & 2.4976(26)$^{+6}_{-16}$
& -0.4 \\
$\sigma_h [nb]$ & 41.488(78) & 41.450(20)$^{+3}_{-7}$
& 0.5 \\
$R_l$ & 20.788(32) & 20.770(24)$^{-5}_{+11}$
& 0.6 \\
$A^l_{FB}$ & 0.0172(12) & 0.0158(6)$^{-2}_{+3}$
& 1.2 \\
$\tau$-polarization: & & & \\
$A_{\tau}$ & 0.1418(75) & 0.1450(26)$^{-7}_{+13}$ & -0.4 \\
$A_e$ & 0.1390(89) & 0.1450(26)$^{-7}_{+13}$ & -0.7 \\
Results for $b$ and $c$ & & & \\
quarks: & & & \\
$R_b$ & 0.2219(17) & 0.2155(3)$^{-7}_{+7}$
& 3.8 \\
$R_c$ & 0.1543(74) & 0.1724(1)$^{+2}_{-2}$ & -2.5 \\
$A^b_{FB}$ & 0.0999(31) & 0.1017(19)$^{-6}_{+10}$
& -0.6 \\
$A^c_{FB}$ & 0.0725(58) & 0.0726(14)$^{-4}_{+7}$ & 0.0 \\
Charge asymmetry for pairs & & & \\
of light quarks $q\bar{q}$: & & & \\
$s^2_l(<Q_{FB}>)$ & 0.2325(13) & 0.2318(3)$^{+1}_{-2}$ & 0.6
\\
\hline
b) \underline{SLC} & & & \\
$A_{LR}$ & 0.1551(40) & 0.1450(26)$^{-7}_{+13}$ & 2.5 \\
$s^2_l(A_{LR})$ & {\it 0.2305(5)} & {\it 0.2318(3)$^{+1}_{-2}$}
& {\it -2.5} \\
$R_b$ & 0.2171(54) & 0.2155(3)$^{-7}_{+7}$ & 0.3 \\
$A_b$ & 0.8410(530) & 0.9345(2)$^{-3}_{+4}$ & -1.8 \\
$A_c$ & 0.6060(900) & 0.6670(12)$^{-3}_{+6}$ & -0.7 \\
\hline
c) \underline{$p\bar{p}$ and $\nu N$} & & & \\
& & & \\
$m_W$ [GeV]~~ $(p\bar{p})$ & 80.26(16) &80.35(5)$^{+1}_{-3}$
& -0.5 \\
&{\it 0.2253(31)} & & \\
$1-m^2_W/m^2_Z$~~ $(\nu N)$ & 0.2257(47) & 0.2237(9)$^{-2}_{+5}$
& 0.4 \\
&{\it 80.24(24)} & & \\
\hline
\end{tabular}
\end{center}

\newpage
\begin{center}
{\bf Table 6}
\vspace{3mm}

\begin{tabular}{|l||c|c|cc|}
\hline
Observable & $s^2_l$ & Average over & Cumulative & $\chi^2/d.o.f.$\\
& & groups of observations & average &
\\ \hline \hline
$A^l_{FB}$ & 0.23096(68) & & &\\
$A_{\tau}$ & 0.23218(95) & & &\\
$A_e$ & 0.2325(11) & 0.23160(49) & 0.23160(49)
&1.9/2 \\ \hline
$A^b_{FB}$ & 0.23209(55) & & & \\
$A^c_{FB}$ & 0.2318(13) & 0.23205(51) & 0.23182(35)
& 2.4/4 \\ \hline
$<Q_{FB}>$ & 0.2325(13) & 0.2325(13) & 0.23186(34) & 2.6/5 \\
\hline
$A_{LR}$ (SLD) & 0.23049(50) & 0.23049(50) & 0.23143(28)
& 7.8/6 \\ \hline
\end{tabular}
\end{center}

Table 6 gives experimental values of $s^2_l$. The third column was
obtained by averaging of the second column, and the fourth by
cumulative averaging of the third; it also lists the values of
$\chi^2$ over the number of degrees of freedom.
\vspace{3mm}

\begin{center}
{\bf Table 7}
\vspace{3mm}

\begin{tabular}{|l||c|c|c|c|}
\hline
Physical & LEP & LEP + SLC & LEP + $M_W$ & LEP + SLC \\
quantities & & & & + $M_W$ \\
\hline \hline
$m_t$ (GeV) & $171(9)^{18}_{-21}$ & $182(7)^{+18}_{-22}$ &
$170(8)^{+17}_{-21}$ & $180(7)^{+18}_{-21}$ \\ \hline
$\hat{\alpha}_s$ & $0.125(4)^{+2}_{-2}$ & $0.123(4)^{+2}_{-2}$ &
$0.125(4)^{+2}_{-2}$ & $0.124(4)^{+2}_{-2}$ \\ \hline
$\chi^2/d.o.f.$ & 18/9 & 29/13 & 18/11 & 30/15 \\ \hline
\hline
$s^2_l$ & $0.2321(4)^{+1}_{-2}$ & $ 0.2317(3)^{+1}_{-2}$ &
$0.2321(4)^{+1}_{-2}$ & $ 0.2318(3)^{+1}_{-2}$ \\ \hline
$g_{Vl}/g_{Al}$ & $0.07116(16)^{-4}_{+8}$
& $0.0732(12)^{-4}_{+8}$ & $0.0716(16)^{-4}_{+8}$ &
$0.0728(12)^{-4}_{+8}$ \\ \hline
$s^2_W$ & $0.2247(9)^{-2}_{+4}$ & $0.2234(9)^{-2}_{+5}$ &
$ 0.2237(9)^{-2}_{+4}$ & $0.2237(9)^{-4}_{+5}$ \\ \hline
$m_W/m_Z$ & $0.8805(5)^{+1}_{-2}$ &
$ 0.8813(5)^{+1}_{-3}$ & $0.8804(5)^{+1}_{-2}$ &
$0.8811(5)^{+2}_{-3}$ \\ \hline
$m_W$ (GeV) & $ 80.29(5)^{+1}_{-2}$ & $80.36(5)^{+1}_{-3}$
& $ 80.28(5)^{+1}_{-2}$ & $80.35(5)^{+1}_{-3}$ \\ \hline
\end{tabular}
\end{center}

Table 7 gives the fitted values $m_t$ and $\hat{\alpha}_s
\equiv \hat{\alpha}_s(m^2_Z)$, and also the values of $\chi^2$
times the number of degrees of freedom for various sets of data, where
$M_W$ stands for both the data of $M_W$ measurements in $p\bar{p}$
collisions and the values of $s^2_W$, extracted from experiments
with $\nu N$. The lower part of the table gives the values of
$s^2_l \equiv \sin^2 \theta^{lept}_{eff} \equiv \frac{1}{4}(1
- g_{Vl}/g_{Al})$ and $s^2_W\equiv 1 - m^2_W/m^2_Z$, calculated
in one-loop electroweak approximation in the framework of the
Minimal Standard Model using fitted values of $m_t$ and $\hat{\alpha}_s$.
Errors given in parentheses are due to uncertainties in
$m_t$, $\hat{\alpha}_s$ and $\bar{\alpha}$, and were calculated
by summation of squares, ignoring correlations.
Note that the errors in the values of $s^2_W$ calculated using the
fits are substantially lower than in the experimental values of
this quantity (see Table 5); at the same time, the errors for $s^2_l$
are practically identical for the experimental (Table 6) and the
theoretical (Table 7) values. Note that the first and second
rows of the lower part of Table 7 carry identical information;
the same is true for the third, fourth and fifth rows.

\newpage
\section{Conclusions.}
\begin{center}
{\bf Achievements.}
\end{center}

\vspace{2mm}
What are the main results obtained with four detectors of the
LEP1 collider and one detector of the SLC?

Judging by the criteria of accelerator and experimental
techniques, the highest imaginable level has been achieved in the
team creativity. The impossible became the reality owing to a
never before dreamt-of sophistication of equipment of gigantic
high-precision detectors. Twenty million decays of $Z$ bosons were
measured with better accuracy than that admired in gemstone cutting.

From the physics standpoint, the main result is the experimental
proof that there exist only
three generations of quarks and leptons with light
neutrinos. The number of light neutrinos, found from the
sum width of the invisible $Z$ boson decays is
\begin{equation}
N_{\nu} = 2.990(16)
\label{Z1}
\end{equation}
The lower limit on the masses of the heavy neutrinos in the
additional generations, provided they exist, is close to $m_Z/2$
and equals 44 GeV.

No new particles were found in $Z$ boson decays. In particular,
the higgs was not found. From LEP1 data, the lower limit
on the higgs mass is
\begin{equation}
m_H > 60\ {\rm GeV}\;\;.
\label{Z2}
\end{equation}

The high-precision measurement of the $Z$ boson mass, its total
and partial decay widths and also of the $P$- and $C$-violating
asymmetries made it possible to experimentally determine
the electroweak radiative corrections. A comparison of these
experimental values with the results of theoretical calculations
based on the Minimal Standard Model led to prediction
of the $t$-quark mass $m_t$ and the constant of strong
interaction for gluons $\hat{\alpha}_s$:
\begin{equation}
m_t = 180(7)^{+18}_{-21} {\rm GeV}\;\;,
\label{Z3}
\end{equation}

\begin{equation}
\hat{\alpha}_s = 0.124(4)^{+2}_{-2}\;\;.
\label{Z4}
\end{equation}
Shown in parentheses here is the uncertainty (one standard
deviation) due to the uncertainty of the experimental data.
The central value corresponds to the assumption that $m_H = 300$ GeV,
and the upper and lower `shifts' correspond to $m_H = 1000$ GeV and $60$
GeV, respectively. Radiative corrections depend only weakly on
$m_H$, so one cannot extract the value of $m_H$ from them.
The $t$-quark mass predicted on the basis of the radiative corrections
within the currently known experimental uncertainties is in
excellent agreement with the results of direct measurements of
$m_t$ in CDF and D0 detectors at the Tevatron \footnote{
At the spring 1996 conferences more accurate data have been
presented:
$$
m_t = 175.6 \pm 5.7\pm 7.4 \; {\rm GeV} \;\; {\rm CDF \cite{54}}
$$
$$
m_t = 170 \pm 15 \pm 10 \; {\rm GeV} \;\; {\rm D0 \cite{55}}
$$}:
\begin{equation}
m_t = 176(13) \; {\rm GeV} \;\;\; {\rm \cite{9}}\;\; {\rm CDF}\;\;.
\label{Z5}
\end{equation}
\begin{equation}
m_t = 199(30) \; {\rm GeV}\;\;\; {\rm \cite{10}}\;\; {\rm D0}\;\;.
\label{Z6}
\end{equation}

After the $t$-quark mass uncertainty is further
reduced, radiative corrections can be used
for determining the region in which the higgs mass can lie.
Figures 13--15 show that the results will greatly depend on
luck. If $m_t = 150(5)$ GeV, figure 13 demonstrates that $m_H < 150$ GeV
\begin{figure}
\epsfxsize=400pt
\begin{center}
\parbox{\epsfxsize}{\epsfbox{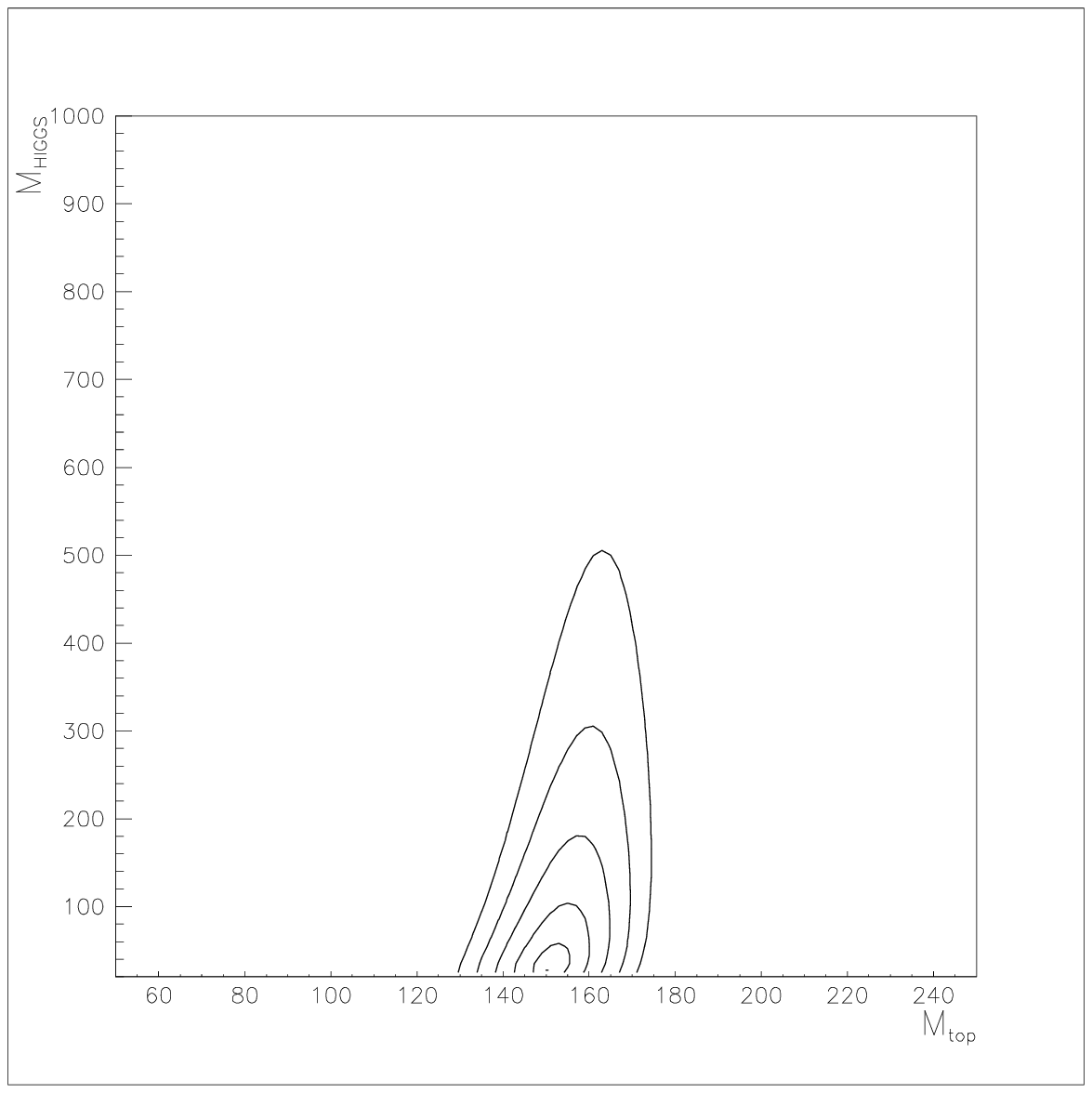}}
\end{center}
\caption{Isolines of $\chi^2$ in the $m_t, m_h$ plane,
obtained by fitting the electroweak corrections under the
assumption that direct measurements of the $t$ quark mass
will give $m_t = 150\pm 5$ GeV . }
\end{figure}
\begin{figure}
\epsfxsize=400pt
\begin{center}
\parbox{\epsfxsize}{\epsfbox{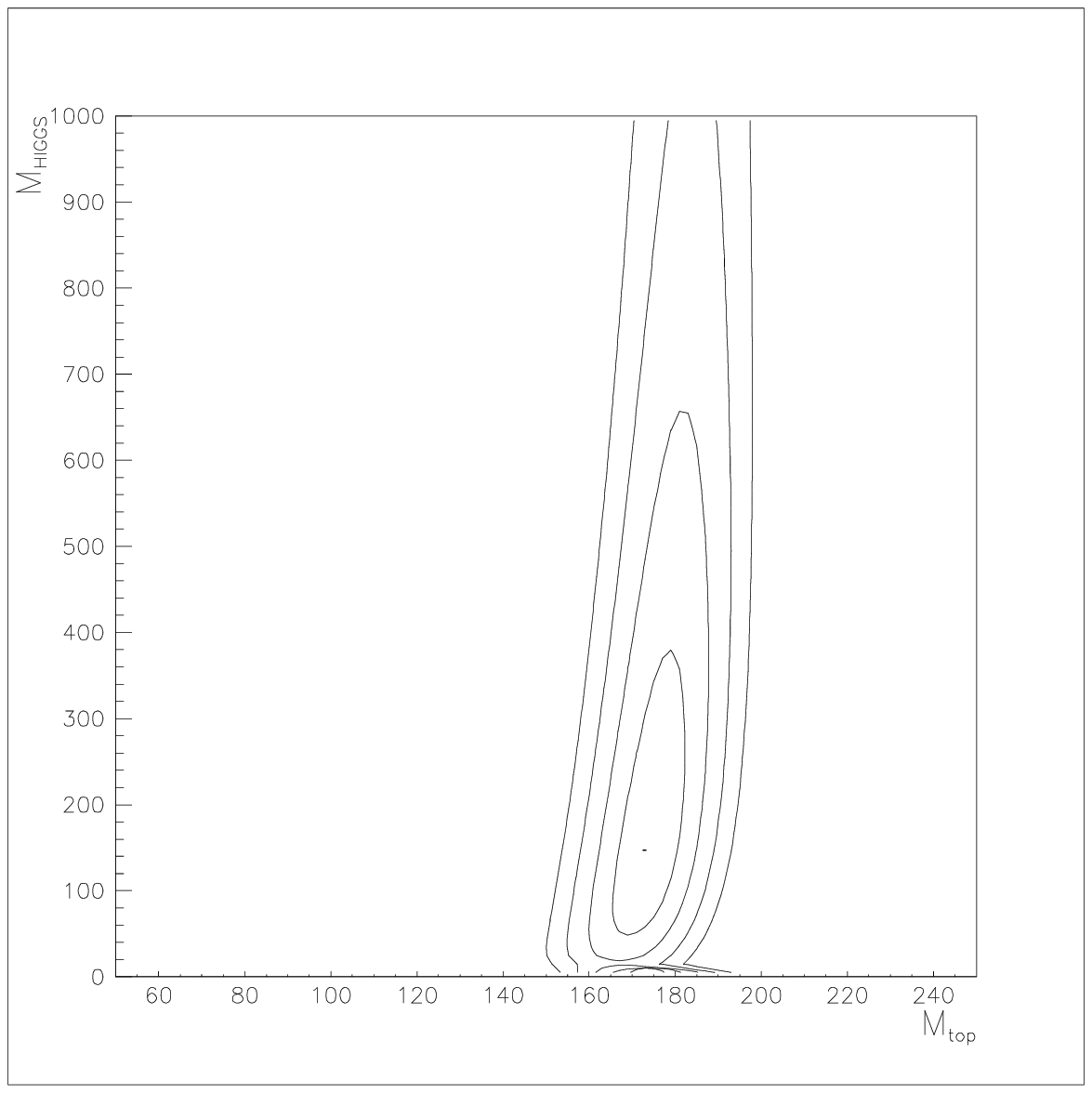}}
\end{center}
\caption{ Same as in figure 13, for $m_t = 175\pm 5$ GeV. }
\end{figure}
\begin{figure}
\epsfxsize=400pt
\begin{center}
\parbox{\epsfxsize}{\epsfbox{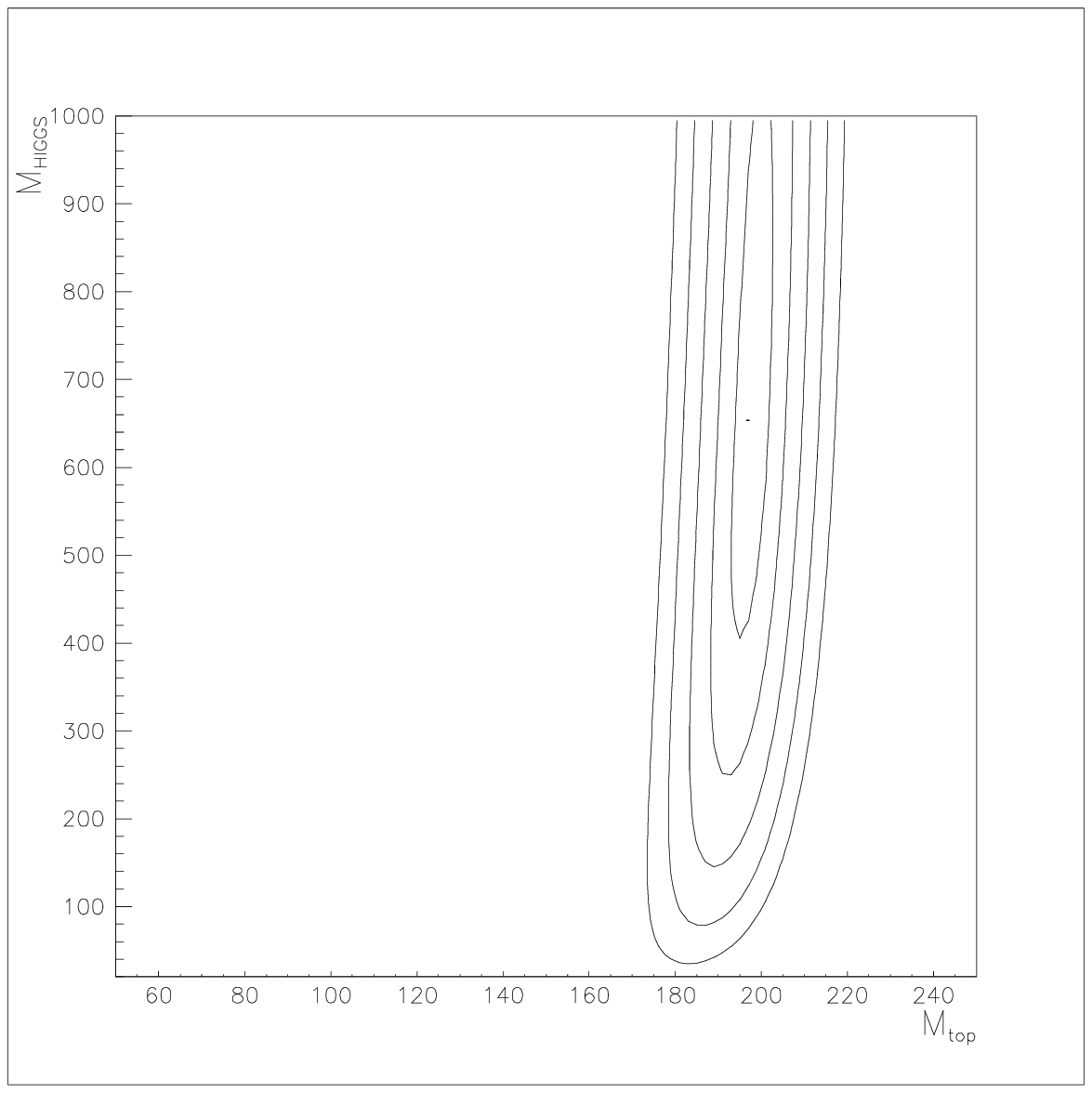}}
\end{center}
\caption{ Same as in figure 13, for $m_t = 200\pm 5$ GeV.  }
\end{figure}
at the $3\sigma$ level. If $m_t = 200(5)$ GeV, then $m_H > 120$ GeV
at the $3\sigma$ level (figure 15). If, however, $m_t = 175(5)$,
then the higgs can have any mass within $3\sigma$.

\begin{center}
{\bf Problems.}
\end{center}

A cursory glance at Table 5 is sufficient for identifying the
main problem of the $Z$-boson physics: the discrepancy between the
measured width of decay into a pair $b\bar{b}$ and its theoretical
prediction.

The supersymmetrization of the standard model may help solving
this problem \cite{131313} (see figure 16a, this diagram
with superpartners increases $R_b$ ).
\begin{figure}
\epsfxsize=400pt
\begin{center}
\parbox{\epsfxsize}{\epsfbox{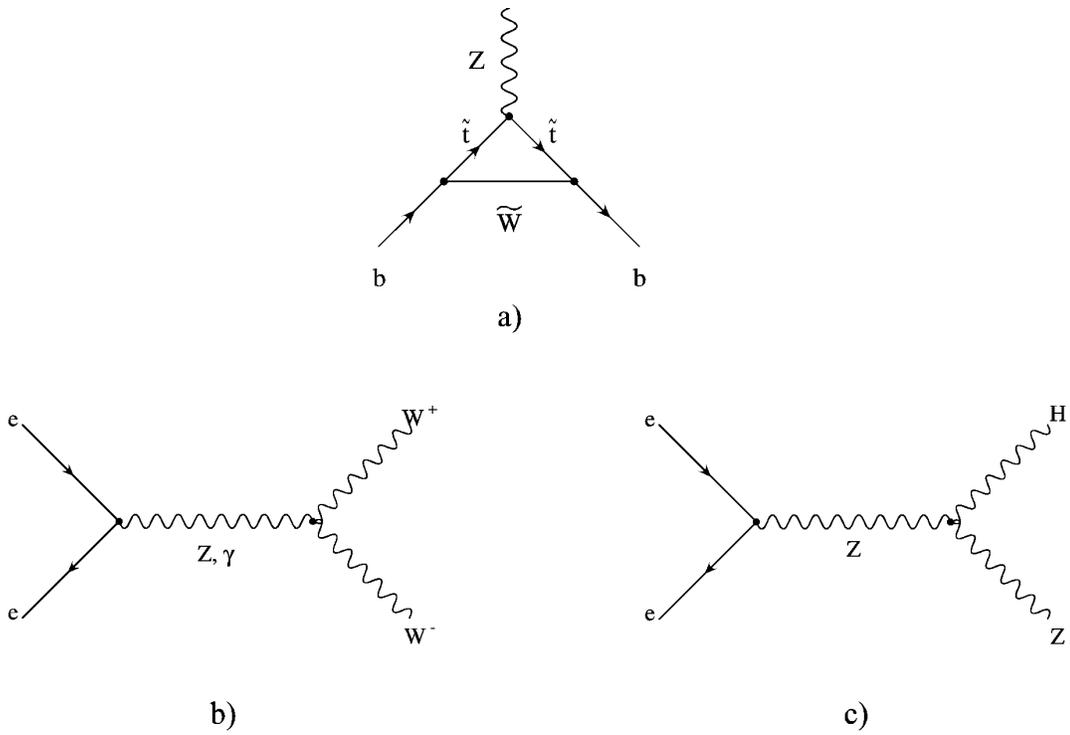}}
\end{center}
\caption{A vertex with virtual $\tilde{t}$ squarks and
winos, $\tilde{W}$ (a). 
 The reaction $e^+e^- \to W^+ W^-$ with a virtual photon or $Z$ boson (b). 
 The reaction $e^+e^- \to HZ$ (c).  }
\end{figure}
Theories with extra neutral vector boson$Z^{'}$ also may be relevant.
Let us turn now to $\hat{\alpha}_s(m_Z)$. A number of papers
\cite{56}, \cite{57} pointed out that the value $0.124(4)^{+2}_{-2}$,
shown in Table 7, is in contradiction with the measurements of
$\hat{\alpha}_s(q^2)$ for $q^2 \la (10 {\rm GeV})^2$ in
deep inelastic scattering \cite{58}, in hadron decays of the
$\Upsilon$-meson \cite{59} and especially in the spectrum of
upsilonium levels \cite{60}. If the low-energy values
$\hat{\alpha}_s(q^2\la(10{\rm GeV})^2)$ are extrapolated in the
framework of the standard QCD to $q^2 = m_Z^2$, we find
$\hat{\alpha}_s(m_Z^2) = 0.110 - 0.118$. As for the uncertainty
in this range, the authors of \cite{57}-\cite{60} do not come to
a common opinion. The most cautious evaluate it as $\pm 0.005$
\cite{58}-\cite{59}. The bravest one insists on $\pm 0.001$ \cite{57},
\cite{60}. In the last case, there is an obvious contradiction
with the value derived by analyzing the $Z$ boson decay data.
This contradiction served as a basis for hypotheses \cite{56}
that MSM predictions for hadron widths are modified by the contribution
of some new unknown particles to electroweak radiative
corrections, for example, squarks and gluino, that is, the
supersymmetric partners of quarks and gluons. For loops with
these particles to result in sufficiently strong deviations from MSM,
it is necessary that squarks and gluino were sufficiently light,
with masses of the order of 100 GeV.
A deviation of the observable value of $R_b$ by more than $3\sigma$
from the values predicted in MSM on the basis of the global fit (see Table 5)
is also considered as an independent argument in favor of the
view that the new physics `lurks round the corner'. Two other
deviations from MSM in Table 5 are less serious: $R_c$ and $A_{LR}$
are off by 2.5 standard deviations. Note that in the latter case
we witness a discrepancy not only with MSM but also with the
measurements of the lepton asymmetries at LEP1, since, according to equation
(\ref{40}), $A_{LR} = A_l$.

Various parametrizations of the manifestations of the new physics
can be found in the literature. The better known ones are the
parameters $S, T, U$ \cite{131315} and $\varepsilon_1$, $\varepsilon_2$,
$\varepsilon_3$, $\varepsilon_b$ \cite{131316}.

\begin{center}
{\bf Prospects.}
\end{center}

The LEP collider completed its work in the LEP1 mode in autumn
1995 and began working in the LEP2 mode. It is expected that the
total energy of the electron and positron collision
will be raised to 192 GeV.
What are the main goals of LEP2 \cite{131314}?

A careful measurement of the cross section of the reaction
$e^+e^- \to W^+W^-$ needed to measure the $W$-boson mass with
accuracy of the order of 50 MeV and to test whether the
interaction of $W$ bosons with photons and with the $Z$ boson
agrees with the standard model (see figure 16b).

A search for a higgs with a mass up to 92 GeV in the reaction
$e^+e^- \to HZ$ (figure 16c).

A search for light superparticles (sleptons, squarks).

A search for the unanticipated.

Further prospects for testing the Standard Model and for
searching for `new physics' beyond its limits are rooted in the
Large Hadron Collider (LHC) (a decision to build it has already
been made at CERN) and in the so-called `Next Linear Collider' for
electrons and positrons, which is so far at the stage of
discussion of competing projects.

We are grateful to A. V. Novikov for help in preparing this
review; A. R. is grateful to CNRS-IN 2 P3/CPPM for
support; M.V., V.N. and L.O.
are grateful to RFBR for the grant 93-02-14431;
M.V. and V.N.  are grateful to INTAS for the grants 93-3316 and
94-2352, which helped to carry out the work presented here.

\newpage

\setcounter{equation}{0}
\renewcommand{\theequation}{A.\arabic{equation}}
\begin{center}
{\large \bf Appendix A.}\\

\vspace{3mm}

{\Large\bf Feynman rules in electroweak theory.}
\end{center}
\vspace{3mm}

A consistent derivation of Feynman rules for theories with
spontaneous violation of gauge symmetry can be found in
textbooks (see, for example, Itzykson, Zuber; Ramond; Slavnov,
Faddeev \cite{15}). In this appendix we only give a summary of
results, accompanied by brief comments.

\begin{center}
{\bf Gauges and propagators.}
\end{center}

\vspace{3mm}

Quantization of gauge fields (in MSM this is $W_{\mu}^{\pm}$,
$Z_{\mu}$ and $A_{\mu}$) requires fixing gauge.
The most popular is the so-called $R_{\xi}$ gauge which
corresponds to adding gauge fixing
new terms $\delta{\cal L}_{GF}$ to the
classical Lagrangian:
\begin{eqnarray}
\delta{\cal L}_{GF}= -\frac{1}{2\xi_A}(\partial_{\mu}A_{\mu})^2
-\frac{1}{2\xi_Z}(\partial_{\mu}Z_{\mu}-m_Z \xi_Z G^0)^2 - \nonumber
\\ \\
-\frac{1}{\xi_W}(\partial_{\mu}W_{\mu}^+ -im_W \xi_W
G^+)(\partial_{\mu}W_{\mu}^- +im_W\xi_W G^-) \;\;, \nonumber
\label{A.1}
\end{eqnarray}
where $G^{\pm}, G^0$ and $H$ are the components of the higgs
doublet $\Phi$ in the parametrization
\begin{equation}
\Phi = \left( \begin{array}{c}
G^+(x) \\
\frac{1}{\sqrt{2}}(\eta +H(x) +i G^0(x))
\end{array}
\right)
\label{A.2}
\end{equation}

In what follows we use the particular case of $R_{\xi}$
gauge, namely
$$
\xi_A = \xi_W = \xi_Z = \xi
$$
With gauge fixed, it is possible to determine the
propagators
$D_{\mu\nu}^W(p)$, $D_{\mu\nu}^Z(p)$ and $D_{\mu\nu}^A(p)$
of the fields
$W_{\mu}^{\pm}$, $Z_{\mu}$ and $A_{\mu}$:
\begin{eqnarray}
\nonumber
D_{\mu\nu}^W(p) = -\frac{i}{p^2 -m_W^2 +i\varepsilon}\left\{
g_{\mu\nu}-(1-\xi)\frac{p_{\mu}p_{\nu}}{p^2 -m_W^2\xi
+i\varepsilon}\right\} \equiv \\
\\
\equiv -\frac{i}{p^2 -m_W^2 +i\varepsilon}(g_{\mu\nu}-\frac{p_{\mu}
p_{\nu}}{m_W^2}) -i\frac{p_{\mu} p_{\nu}}{m_W^2} \frac{1}{p^2
-m_W^2\xi +i\varepsilon} \;\;, \nonumber
\label{A.3}
\end{eqnarray}

\begin{eqnarray}
\nonumber
D_{\mu\nu}^Z(p) = -\frac{i}{p^2 -m_Z^2 +i\varepsilon}\left\{
g_{\mu\nu}-(1-\xi)\frac{p_{\mu}p_{\nu}}{p^2 -m_Z^2\xi
+i\varepsilon}\right\} = \\
\\
= -\frac{i}{p^2 -m_Z^2 +i\varepsilon}(g_{\mu\nu}-\frac{p_{\mu}
p_{\nu}}{m_Z^2}) -i\frac{p_{\mu} p_{\nu}}{m_Z^2} \frac{1}{p^2
-m_Z^2\xi +i\varepsilon} \;\;, \nonumber
\label{A.4}
\end{eqnarray}
\begin{equation}
D_{\mu\nu}^A(p) = -\frac{i}{p^2 +i\varepsilon}\{
g_{\mu\nu}-(1-\xi)\frac{p_{\mu}p_{\nu}}{p^2 +i\varepsilon}\} \;\;,
\label{A.5}
\end{equation}

The case $\xi =1$ corresponds to the Feynman--'tHooft gauge,
$\xi =0$ to the Landau gauge, $\xi \to\infty$ to the
Proca gauge.

As follows from (A.3) and (A.4), the propagators of massive
vector fields can be written as sums of a propagator in the Proca
gauge that describes the propagation of physical degrees
of freedom of a vector particle, and a scalar propagator with a
gauge-dependent pole which corresponds to the propagation
of non-physical degrees of freedom. As a result,
diagrams with virtual $W^{\pm}$, $Z$ bosons contain non-physical
threshold singularities whose positions
depend on the gauge parameter $\xi$. These non-physical
singularities partially cancel out after the appropriate
diagrams with virtual Goldstone bosons $G^{\pm}$, $G^0$
(arising from the
higgs doublet $\Phi$ (\ref{A.2})) are added. The
Goldstone boson propagators have the form:
\begin{equation}
D_{G^+}(p) =\frac{i}{p^2 -m_W^2 \xi +i\varepsilon}
\label{A.6}
\end{equation}
\begin{equation}
D_{G^0}(p) =\frac{i}{p^2 -m_Z^2 \xi +i\varepsilon}  .
\label{A.7}
\end{equation}

A complete restoration of unitarity (cancellation of
non-physical singularities) and of gauge invariance
(validity of the
Ward identities) are achieved if one takes into account the
diagrams with the Faddeev--Popov ghosts $\eta^{\pm}$, $\eta^Z$ and
$\eta^A$, which interact only with gauge fields and the
Goldstone fields and which do not correspond to any physical degrees
of freedom.

Ghost propagators take the form
\begin{equation}
D_{\eta^+}(p) =\frac{i}{p^2 -m_W^2 \xi +i\varepsilon} \;\;,
\label{A.8}
\end{equation}
\begin{equation}
D_{\eta^Z}(p) =\frac{i}{p^2 -m_Z^2 \xi +i\varepsilon} \;\;,
\label{A.9}
\end{equation}
\begin{equation}
D_{\eta^A}(p) =\frac{i}{p^2 +i\varepsilon}
\label{A.10}
\end{equation}
Ghosts obey the Fermi statistics, so an additional sign $(-1)$
must be ascribed to ghost loops, as one does for fermion loops.

The propagators of other fields are written as
\begin{equation}
{\rm for~~the~~higgs~~field:}~~ D_H(p) =\frac{i}{p^2-m_H^2
+i\varepsilon}
\label{A.11}
\end{equation}
\begin{equation}
{\rm for~~fermion~~ fields:}~~ \hat{D}_f(p) =\frac{i}{\hat{p}-m_f
+i\varepsilon}
\label{A.12}
\end{equation}
In order to describe numerous three-particle vertices, it is
convenient to unify the notations. Let us fix the momenta once
and for all, as shown in figure 17, 
\begin{figure}
\epsfxsize=200pt
\begin{center}
\parbox{\epsfxsize}{\epsfbox{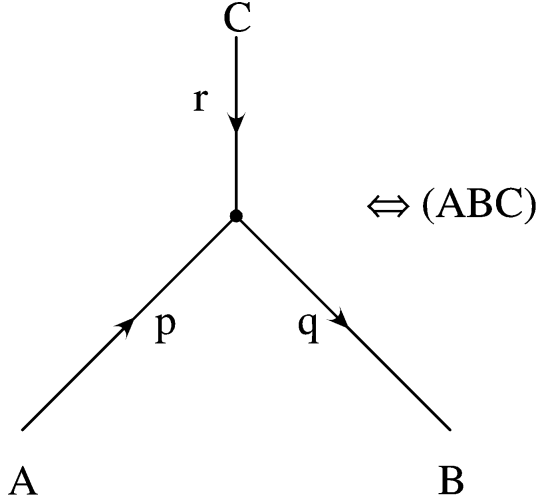}}
\end{center}
\caption{Three-particle vertex.  }
\end{figure}
and let us denote a vertex by
a set of fields in the following order: (ACB). The Feynman
rules for three-particle vertices are then written as  follows.

\vspace{3mm}
\begin{center}
{\bf Interaction between gauge fields and fermions. }
\end{center}
\vspace{2mm}
\begin{eqnarray}
\begin{array}{ll}
(f A_{\mu}f) & -ie Q_f \gamma_{\mu} \;\;, \\
& \\
(f Z_{\mu}f) & -i\frac{f}{4}[g_V \gamma_{\mu}+g_A\gamma_{\mu}
\gamma_5] \;\;, \\
& \\
(\nu_e W_{\mu}^- l) \;\;\; & -i
\frac{g}{2\sqrt{2}}\gamma_{\mu}(1+\gamma_5) \;\;, \\
& \\
(U W_{\mu}^- D) \;\;\; & -i \frac{g}{2\sqrt{2}}V_{DU}
\gamma_{\mu}(1+\gamma_5) \;\;;
\end{array}
\label{A.13}
\end{eqnarray}
where $Q_f$ is the charge of the
fermion  $f$, $T_3^f$ is the third projection
of
the isotopic  spinor, describing the  left-handed component
of the
fermion $f$, $g_A =2T_3^f$, $g_V =2T_3^f - 4Q_f
\sin^2\theta$; $U$ and $D$ denote any of the quarks with $T_3^f
=\frac{1}{2}$ and $T_3^f =-\frac{1}{2}$, respectively, and $V_{DU}$
is an element of the Kobayashi--Maskawa matrix.

\vspace{3mm}
\begin{center}
{\bf Interaction of scalar fields with fermions. }
\end{center}
\vspace{2mm}
\begin{eqnarray}
\begin{array}{ll}
(f H f) & -\frac{ig}{2m_W} m_f \;\;,\\
& \\
(f G^0 f) & -\frac{g}{4}\frac{m_f}{m_W} T_3^f \;\;,\\
& \\
(U G^- D) & -\frac{i}{2\sqrt{2}}\frac{g}{m_W}[(m_D -m_U)+\gamma_5(m_D
+m_U)] \;\;, \\
& \\
(D G^+ U) & -\frac{i}{2\sqrt{2}}\frac{g}{m_W}[(m_D -m_U)-\gamma_5(m_D
+m_U)]\;\;.
\end{array}
\label{A.14}
\end{eqnarray}

\vspace{3mm}
\begin{center}
{\bf Three-boson interactions.}
\end{center}
\vspace{2mm}

Three gauge bosons:
\begin{eqnarray}
\begin{array}{ll}
(W_{\lambda}^+ A_{\nu} W_{\mu}^+) & ie[r+q)_{\lambda} g_{\mu\nu}
-(q+p)_{\nu} g_{\lambda\mu} +(p-r)_{\mu} g_{\nu\lambda}] \;\;,\\
& \\
(W_{\lambda}^+ Z_{\nu} W_{\mu}^+) & ig\cos\theta[r+q)_{\lambda}
g_{\mu\nu} -(q+p)_{\nu} g_{\lambda\mu} +(p-r)_{\mu} g_{\nu\lambda}]\\
\end{array}
\label{A.15}
\end{eqnarray}

Two gauge bosons and one scalar boson ($G$ or $H$):
\begin{eqnarray}
\begin{array}{ll}
(W_{\mu}^+ G^- A_{\nu}) & ie m_W g_{\mu\nu} \;\;, \\
& \\
(W_{\mu}^+ G^- Z_{\nu}) & -ig m_Z\sin^2\theta g_{\mu\nu} \;\;, \\
& \\
(W_{\mu}^+ H W_{\nu}^+) & ig m_W g_{\mu\nu} \;\;, \\
& \\
(Z_{\mu} H Z_{\nu}) & ig \frac{m_Z^2}{m_W} g_{\mu\nu} \;\;; \\
\end{array}
\label{A.16}
\end{eqnarray}

One gauge boson and two scalar bosons ($GG$ or $GH$):
\begin{eqnarray}
\begin{array}{ll}
(G^+ A_{\mu} G^+) & -ie(p+q)_{\mu} \;\;, \\
& \\
(G^+ Z_{\mu} G^+) & -ig\frac{\cos
2\theta}{2\cos\theta}(p+q)_{\mu}\;\;, \\
& \\
(G^0 W_{\mu}^+ G^+) & \frac{1}{2}g(p+q)_{\mu}\;\;, \\
& \\
(H W_{\mu}^+ G^+) & -\frac{1}{2}ig(p+q)_{\mu}\;\;, \\
& \\
(G^+ W_{\mu}^- G^0) & -\frac{1}{2}g(p+q)_{\mu}\;\;, \\
& \\
(G^+ W_{\mu}^- H) & -\frac{1}{2}ig(p+q)_{\mu}\;\;, \\
& \\
(H Z_{\mu} G^0) & \frac{g}{\cos\theta}(p+q)_{\mu}\;\;;
\end{array}
\label{A.17}
\end{eqnarray}

Interaction of higgses with goldstones and among themselves:
\begin{eqnarray}
\begin{array}{ll}
(G^+ H G^+) & \frac{i}{2}g\frac{m_H^2}{m_W} \;\;, \\
& \\
(G^- H G^-) & -\frac{i}{2}g\frac{m_H^2}{m_W} \;\;, \\
& \\
(G^0 H G^0) & -\frac{i}{2}g\frac{m_H^2}{m_W} \;\;, \\
& \\
(H H H) & -\frac{3i}{2}g\frac{m_H^2}{m_W}\;\;.
\end{array}
\label{A.18}
\end{eqnarray}
\vspace{3mm}

Interaction between ghosts and gauge fields:
\begin{eqnarray}
\begin{array}{ll}
(\eta^+ A_{\mu} \eta^+) & -ieq_{\mu} \;\;, \\
& \\
(\eta^- A_{\mu} \eta^-) & ieq_{\mu} \;\;, \\
& \\
(\eta^- W_{\mu}^+ \eta^{\gamma}) & -ieq_{\mu} \;\;, \\
& \\
(\eta^+ W_{\mu}^- \eta^{\gamma}) & ieq_{\mu} \;\;, \\
& \\
(\eta^{\gamma} W_{\mu}^+ \eta^+) & ieq_{\mu} \;\;, \\
& \\
(\eta^{\gamma} W_{\mu}^- \eta^-) & -ieq_{\mu} \;\;, \\
& \\
(\eta^+ Z_{\mu} \eta^+) & -ig\cos\theta\; q_{\mu} \;\;, \\
& \\
(\eta^- Z_{\mu} \eta^-) & ig\cos \theta\; q_{\mu} \;\;,
\end{array}
\label{A.19}
\end{eqnarray}

\newpage
$$
\begin{array}{ll}
(\eta^- W_{\mu}^+ \eta^Z) & -ig\cos\theta \;q_{\mu} \;\;, \\
& \\
(\eta^+ W_{\mu}^- \eta^Z) & ig\cos\theta \;q_{\mu} \;\;, \\
& \\
(\eta^Z W_{\mu}^+ \eta^+) & ig\cos\theta \;q_{\mu} \;\;, \\
& \\
(\eta^Z W_{\mu}^- \eta^-) & -ig\cos\theta \;q_{\mu} \;\;;
\end{array}
$$

Interaction of ghosts with a higgs or a goldstone:
\begin{eqnarray}
\begin{array}{ll}
(\eta^- H \eta^-) & -\frac{i}{2}g\xi m_W \;\;, \\
& \\
(\eta^- G^0 \eta^-) & -\frac{1}{2}g\xi m_W \;\;, \\
& \\
(\eta^+ G^0 \eta^+) & \frac{1}{2}g\xi m_W \;\;, \\
& \\
(\eta^{\gamma} G^+ \eta^+) & -ie\xi m_W \;\;, \\
& \\
(\eta^Z G^+ \eta^+) & -\frac{i}{2}g \frac{\cos
2\theta}{\cos\theta}\xi m_W \;\;,\\
& \\
(\eta^- G^+ \eta^Z) & \frac{i}{2}g \xi m_Z\;\;.
\end{array}
\label{A.20}
\end{eqnarray}

\vspace{3mm}
\begin{center}
{\bf Four-boson interactions.}
\end{center}
\vspace{2mm}

To describe four-boson vertices, we introduce the notation (ABCD),
see figure 18.
\begin{figure}
\epsfxsize=200pt
\begin{center}
\parbox{\epsfxsize}{\epsfbox{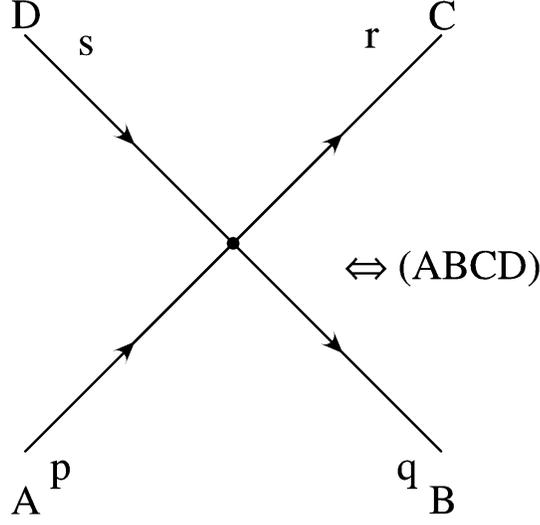}}
\end{center}
\caption{ Four-particle vertex.  }
\end{figure}
In this notation, the interactions of four vector bosons take
the form
\begin{eqnarray}
\begin{array}{ll}
(W_{\lambda}^+ W_{\mu}^-
W_{\nu}^+ W_{\rho}^-) & ig^2 [2g_{\lambda\nu} g_{\mu\rho} -
g_{\mu\nu}g_{\lambda\rho} - g_{\mu\lambda}g_{\nu\rho}] \;\;, \\
& \\
(W_{\lambda}^+ W_{\mu}^- A_{\nu} A_{\rho}) & -ie^2
[2g_{\nu\rho} g_{\mu\lambda} - g_{\mu\rho}g_{\nu\lambda} -
g_{\mu\nu}g_{\lambda\rho}] \;\;, \\
& \\
(W_{\lambda}^+ W_{\mu}^- Z_{\nu} Z_{\rho}) & -ig^2\cos^2\theta
[2g_{\nu\rho} g_{\mu\lambda} - g_{\mu\rho}g_{\nu\lambda} -
g_{\mu\nu}g_{\lambda\rho}] \;\;, \\
& \\
(W_{\lambda}^+ W_{\mu}^- A_{\nu} Z_{\rho}) & -ieg\cos\theta
[2g_{\nu\rho} g_{\mu\lambda} - g_{\mu\rho}g_{\nu\lambda} -
g_{\mu\nu}g_{\lambda\rho}]\;\;.
\end{array}
\label{A.21}
\end{eqnarray}

\begin{center}
Interaction between two vector bosons and $HH$, $GG$ or $HG$.
\end{center}
\begin{eqnarray}
(W_{\mu}^+ W_{\nu}^- H H) &~~~& \frac{i}{2} g^2 g_{\mu\nu} \;\;,
\nonumber \\
\nonumber \\
(W_{\mu}^+ W_{\nu}^- G^0 G^0) &~~~& \frac{i}{2} g^2
g_{\mu\nu}\;\;, \nonumber \\
\nonumber \\
(W_{\mu}^+ W_{\nu}^- G^- G^+) &~~~& \frac{i}{2}
g^2 g_{\mu\nu} \;\;, \nonumber \\
\nonumber \\
(A_{\mu} A_{\nu} G^- G^+) &~~~& 2ie^2 g_{\mu\nu}\;\;. \nonumber
\nonumber \\
\nonumber \\
Z_{\mu} Z_{\nu} H H &~~~& i\frac{1}{2} g^2 \sec^2 \theta
g_{\mu\nu}\;\;, \nonumber \\
Z_{\mu} Z_{\nu} G^0 G^0 &~~~& i\frac{1}{2} g^2 \sec^2 \theta
g_{\mu\nu}\;\;, \nonumber \\
Z_{\mu} Z_{\nu} G^- G^+ &~~~& i \frac{1}{2} g^2 \sec^2 \theta \cos^2
2\theta g_{\mu\nu}\;\;, \\
A_{\mu} W^+_{\nu} G^+ H &~~~& i\frac{1}{2} egg_{\mu\nu} \;\;,
\nonumber \\
A_{\mu} W^+_{\nu} G^- H &~~~&
i\frac{1}{2}egg_{\mu\nu}\;\;,\nonumber\\
\nonumber \\
\nonumber \\
A_{\mu} W^+_{\nu} G^+ G^0 &~~~& -\frac{1}{2} egg_{\mu\nu}\;\;,
\nonumber \\
A_{\mu} W_{\nu}^- G^- G^0 &~~~& \frac{1}{2} egg_{\mu\nu}\;\;,
\nonumber \\
A_{\mu} Z_{\nu} G^- G^+ &~~~& i eg \sec \theta \cos 2\theta
g_{\mu\nu}\;\;, \nonumber\\
Z_{\mu} W^+_{\nu} G^+ H &~~~& i\frac{1}{2} g^2 \sec \theta
(\frac{1}{2} \cos 2\theta - 1) g_{\mu\nu}\;\;,
\nonumber \\
Z_{\mu} W^-_{\nu}
G^- H &~~~& i\frac{1}{2} g^2 \sec \theta(\frac{1}{2} \cos 2\theta -
1)g_{\mu\nu}\;\;,
\nonumber \\
Z_{\mu} W^+_{\nu} G^+ G^0 &~~~& -\frac{1}{2} g^2
\sec \theta(\frac{1}{2} \cos 2\theta - 1)g_{\mu\nu}\;\;,
\nonumber \\
Z_{\mu} W^-_{\nu} G^- G^0 &~~~& \frac{1}{2} g^2 \sec \theta
(\frac{1}{2} \cos 2\theta - 1)g_{\mu\nu}\;\;. \nonumber
\label{A.22}
\end{eqnarray}

\begin{center}
Interactions of $GGGG$, $HHHH$ or $GGHH$.
\end{center}
\begin{eqnarray}
\begin{array}{ll}
(G^+ G^- G^- G^+) & -\frac{i}{2} g^2 m_H^2/m_W^2 \;\;, \\
& \\
(G^+ G^- G^0 G^0) & -\frac{i}{4} g^2 m_H^2/m_W^2 \;\;, \\
& \\
(G^+ G^- H H) & -\frac{i}{4} g^2 m_H^2/m_W^2 \;\;, \\
& \\
(H H H H) & -i \frac{3}{4} g^2 m_H^2/m_W^2 \;\;, \\
& \\
(G^0 G^0 G^0 G^0) & -i \frac{3}{4} g^2 m_H^2/m_W^2 \;\;, \\
& \\
(G^0 G^0 H H) & -\frac{i}{4} g^2 m_H^2/m_W^2 \;\;.
\end{array}
\label{A.23}
\end{eqnarray}

\newpage

{\bf Regularization of Feynman integrals.}
\vspace{2mm}

Integrals corresponding to diagrams with loops formally diverge and
thus need regularization. Note that there does not exist yet a consistent
regularization of electroweak theory in all loops. A dimensional
regularization can be used in the first several loops; this corresponds
to a transition to a $D$-dimensional spacetime in which the
following finite expression is assigned to the diverging
integrals:
\begin{eqnarray}
\int\frac{d^D p}{\mu^{D-4}} \frac{(p^2)^s}{(p^2 +m^2)^{\alpha}}
& = &
\frac{\pi^{\frac{D}{2}}}{\Gamma(\frac{D}{2})}
\frac{\Gamma(\frac{D}{2} +s)\Gamma(\alpha - \frac{D}{2}
-s)}{\Gamma(\alpha)} \times \nonumber \\
& \times & \frac{(m^2)^{\frac{D}{2} -\alpha +s}}{\mu^{D-4}} \;\;,
\label{A.24}
\end{eqnarray}
where $\mu$ is a parameter with mass dimension, introduced to
conserve the dimension of the original integral.

This formula holds in the range of convergence of the integral.
In the range of divergence, a formal expression (\ref{A.24})
is interpreted as the analytical continuation. Obviously, the
integral allows a shift in integration variable in the
convergence range as well. Therefore, a shift $p\to p+q$ for
arbitrary $D$ can also be done in (\ref{A.24}). This factor is
decisive in proving the gauge invariance of dimensional
regularization.

At $D=4$ the integrals in (\ref{A.24}) contain a pole term
\begin{equation}
\Delta =\frac{2}{4-D} +\ln 4\pi -\gamma -\ln\frac{m^2}{\mu^2} \;\;,
\label{A.25}
\end{equation}
where $\gamma =0.577...$ is the Euler constant.
Choice of  constant
terms in (\ref{A.25}) is a matter of convention.

The algebra of $\gamma$-matrices in the $D$-dimensional space is
defined by the relations
\begin{equation}
\gamma_{\mu}\gamma_{\nu} +\gamma_{\nu}\gamma_{\mu} =2g_{\mu\nu}
\times I \;\;,
\label{A.26}
\end{equation}
\begin{equation}
g_{\mu\mu} =D \;\;,
\label{A.27}
\end{equation}
\begin{equation}
\gamma_{\mu}\gamma_{\nu}\gamma_{\mu} =(2-D)\gamma_{\nu} \;\;,
\label{A.28}
\end{equation}
where $I$ is the identity matrix.

As for the dimensionality of spinors, different approaches can
be chosen in the continuation to the $D$-dimensional space.
One possibility is to assume that the $\gamma$ matrices are $4\times 4$
matrices, so that
\begin{equation}
Sp I =4 \;\;.
\label{A.29}
\end{equation}

The $D$-dimensional regularization creates difficulties when one
has to define the absolutely antisymmetric tensor and (or)
$\gamma_5$ matrix. For calculations in several first loops,
a formal definition of $\gamma_5$,
\begin{equation}
\gamma_5 \gamma_{\mu} +\gamma_{\mu}\gamma_5 = 0 \;\;,
\label{A.30}
\end{equation}
\begin{equation}
\gamma_5^2 =I
\label{A.31}
\end{equation}
does not lead to contradictions.

Thus, the amplitudes of physical processes, once they are
expressed in terms of bare charges and bare masses, contain
pole terms $\sim 1/(D-4)$.

If we eliminate bare quantities and express some physical
observables in terms of other physical observables, then all
pole terms cancel out. The general property of renormalizability
guarantees this cancellation. (We have verified this cancellation
directly in \cite{35}.) This renormalization procedure is employed
in this review.

In order to avoid divergences in intermediate expressions, one
can agree to subtract from each Feynman integral the pole
terms $\sim 1/(4-D)$, since they will cancel out anyway in the
final expressions. Depending on which constant terms (in
addition to pole terms) are subtracted from the diagrams,
different subtraction schemes arise: the $\overline{MS}$ scheme
corresponds to subtracting the universal combination
$$
\frac{2}{4-D} -\gamma + \ln 4\pi \;\;.
$$

\newpage

\setcounter{equation}{0}
\renewcommand{\theequation}{B.\arabic{equation}}
\begin{center}
{\large \bf Appendix B.}\\

\vspace{3mm}

{\Large\bf Relation between {\boldmath $\bar{\alpha}$} and
{\boldmath $\alpha(0)$}.}
\end{center}

\vspace{2mm}

We begin with the following famous relation of quantum
electrodynamics \cite{76D}:
\begin{equation}
\alpha(q^2) = \frac{\alpha(0)}{1 + \Sigma_{\gamma}(q^2)/q^2 -
\Sigma'_{\gamma}(0)}\;\;.
\label{B.1}
\end{equation}

Here the fine structure constant $\alpha \equiv \alpha(0)$ is a
physical quantity. It can be measured as a residue of the
Coulomb pole $1/q^2$ in the scattering amplitude of charged
particles. As for the running coupling constant $\alpha(q^2)$,
it can be measured from the scattering of particles
with large masses $m$
at low momentum transfer: $m \gg \sqrt{|q^2|}$. In the standard model
we have the $Z$-boson, and the contribution of the photon cannot be
identified unambiguously if $q^2 \neq 0$. Therefore, the
definition of the running constant $\alpha(q^2)$ becomes
dependent on convention and on details of calculations.

At $q^2 = m^2_Z$, the contribution of $W$-bosons to $\bar{\alpha}
\equiv \alpha(m^2_Z)$ is not large, so it is convenient to make use
of the definition accepted in QED:  \begin{equation} \bar{\alpha} =
\frac{\alpha}{1 - \delta\alpha}\;\;, \label{B.2} \end{equation} where
\begin{eqnarray}
\nonumber
\delta\alpha = -\Pi_{\gamma}(m^2_Z) + \Sigma'_{\gamma}(0)\;\;, \\
\Pi_{\gamma}(m^2_Z) = \frac{1}{m^2_Z} \Sigma_{\gamma}(m^2_Z)\;\;.
\label{B.3}
\end{eqnarray}

One-loop expression for the self-energy of the photon can be
rewritten as \cite{61}:
\begin{eqnarray}
\nonumber
\Sigma_{\gamma}(s) &=& (\alpha/3\pi) \sum_f N^f_c Q^2_f [s\Delta_f +
(s + 2m^2_f) F (s, m_f, m_f) - 3/2] - \\
& - & (\alpha/4\pi) [3s\Delta_W + (3s + 4m^2_W) F (s, m_W, m_W)]\;\;,
\label{B.4}
\end{eqnarray}
where $s \equiv q^2$, the subscript $f$ denotes fermions, the sum
$\Sigma_f$ runs through lepton and quark flavors, and $N^f_c$ is the
number of colors. The contribution of fermions to $\Sigma_{\gamma}(q^2)$
is independent of gauge. The last term in (\ref{B.4})
refers to the gauge-dependent contribution of $W$-bosons; the
 't~Hooft--Feynman gauge was used in equation
(\ref{B.4}).

The singular term $\Delta_i$ is:
\begin{equation}
\Delta_i = \frac{1}{\epsilon} - \gamma + \ln 4\pi - \ln
\frac{m^2_i}{\mu^2}\;\;,
\label{B.5}
\end{equation}
where $2\epsilon = 4 - D$ ($D$ is the variable dimension of
spacetime, $\epsilon \to 0$), $\gamma = -\Gamma'(1) =
0.577...$ is the Euler constant and $\mu$ is an arbitrary parameter.
Both $1/\epsilon$ and $\mu$ vanish in relations between
observables.

The function $F(s, m_1, m_2)$ is defined by the contribution
to self-energy of a scalar particle at $q^2 = s$, owing to a loop with
two scalar particles (with masses $m_1$ and $m_2$) and with the
coupling constant equal to unity:
\begin{eqnarray}
\nonumber
F(s,m_1, m_2) &=& -1 + \frac{m^2_1 - m^2_2}{m^2_1 - m^2_2} \log
\frac{m_1}{m_2} - \\
&-& \int\limits^1_0 dx \log \frac{x^2s - x(s + m^2_1 - m^2_2) + m^2_1
-i\epsilon}{m_1 m_2}\;\;.
\label{B.6}
\end{eqnarray}
The function $F$ is normalized in such a way that it vanishes at $q^2 = 0$,
which corresponds to subtracting the self-energy at $q^2 = 0$:
\begin{equation}
F(0, m_1, m_2) = 0\;\;,
\label{B.7}
\end{equation}

The following formula holds for $m_1 = m_2 = m$:
\begin{eqnarray}
F(s,m,m)\equiv F(\tau) = \left\{ \begin{array}{ll}
2\left[1-\sqrt{4\tau -1}\arcsin\frac{1}{\sqrt{4\tau}}\right] \;\;,
& 4\tau > 1 \;\;, \nonumber \\
2\left[ 1-\sqrt{1-4\tau}\ln
\frac{1+\sqrt{1-4\tau}}{\sqrt{4\tau}}\right] \;\;,
& 4\tau < 1 \;\;,
\end{array}
\right.
\label{B.77}
\end{eqnarray}
where $\tau = m^2/s$.

To calculate the contributions of light fermions, the $t$-quark and
the $W$-boson to $\delta\alpha$, we need the asymptotics $F(\tau)$
for small and large $\tau$:
\begin{equation}
F(\tau) \simeq \ln \tau + 2 +...,\;\; |\tau | \ll 1\;\;,
\label{B.8}
\end{equation}
\begin{equation}
F(\tau) \simeq \frac{1}{6\tau} + \frac{1}{60 \tau^2} +..., \;\;
|\tau | \gg 1\;\;,
\label{B.9}
\end{equation}
\begin{equation}
F'(s, m, m) = \frac{d}{ds} F (s, m, m) \stackrel{s \to 0}{\simeq}
\frac{1}{m^2} \left[ \frac{1}{6} + \frac{1}{30 \tau} \right]\;\;.
\label{B.10}
\end{equation}

As a result we obtain
\begin{eqnarray}
\Pi_{\gamma}(m^2_Z) &\equiv & \frac{\Sigma_{\gamma}(m^2_Z)}{m^2_Z} =
\frac{\alpha}{3\pi} \sum_8 N^f_c Q^2_f (\Delta_Z + \frac{5}{3}) +
\nonumber \\
&+& \frac{\alpha}{\pi} Q^2_f \left[ \Delta_t + (1 + 2t) F(t) -
\frac{1}{3} \right] - \nonumber \\
&-& \frac{\alpha}{4\pi} [3\Delta_W + (3 + 4c^2) F(c^2)]\;\;,
\label{B.11}
\end{eqnarray}
where $t = m^2_t/m^2_Z$, and
\begin{equation}
\Sigma'_{\gamma}(0) = \frac{\alpha}{3\pi} \sum_9 N^f_c Q^2_f \Delta_f
- \frac{\alpha}{4\pi} (3\Delta_W + \frac{2}{3})\;\;,
\label{B.12}
\end{equation}
\begin{eqnarray}
\nonumber
\delta\alpha &=& \frac{\alpha}{\pi} \left\{
\sum_8 \frac{N^f_c Q^2_f}{3} \left( \ln \frac{m^2_Z}{m^2_f} -
\frac{5}{3} \right) - \right. \\
&-& \left. Q^2_t \left[ (1 + 2t) F(t) - \frac{1}{3} \right] +
\left[ \left( \frac{3}{4} + c^2 \right) F(c^2) - \frac{1}{6} \right]
\right\}\;\;.
\label{B.13}
\end{eqnarray}
Therefore, $\delta \alpha$ is found as a sum of four terms,
\begin{equation}
\delta\alpha = \delta\alpha_l + \delta\alpha_h + \delta\alpha_t +
\delta\alpha_W\;\;,
\label{B.14}
\end{equation}

\begin{equation}
\delta\alpha_l = \frac{\alpha}{3\pi} \sum_3 \left[ \ln
\frac{m^2_Z}{m^2_l} - 5/3 \right] = 0.03141\;\;.
\label{B.15}
\end{equation}

\begin{equation}
\delta\alpha_t \simeq -\frac{\alpha}{\pi} \frac{4}{45} \left(
\frac{m_Z}{M_t} \right)^2 = -0.00005(1)\;\;,
\label{B.16}
\end{equation}
where we have used that $m_t = 175 \pm 10$ GeV. Note that
$\delta\alpha_t$ is negligible  and has the antiscreening sign
(the screening of the $t$-quark loops in QED begins at $q^2 \gg m^2_t$,
while in our case $q^2 = m^2_Z < m^2_t)$.

Finally, the $W$ loop gives
\begin{equation}
\delta\alpha_W = \frac{\alpha}{2\pi} \left[ (3 + 4c^2) \left( 1 -
\sqrt{4c^2 - 1} \arcsin \frac{1}{2c} \right) - 1/3 \right] =
0.00050\;\;.
\label{B.17}
\end{equation}
The value of $\delta\alpha_W$ depends on gauge \cite{62};
here we give the result of calculations in the 't~Hooft--Feynman
gauge. Traditionally, the definition of $\bar{\alpha}$ takes
into account the contributions
of leptons and
five light quarks; the terms $\delta\alpha_t$
and $\delta\alpha_W$ are taken into account in the electroweak
radiative corrections. In our approach, these terms give
the corrections $\delta_1 V_i$.

\newpage

\setcounter{equation}{0}
\renewcommand{\theequation}{C.\arabic{equation}}
\begin{center}
{\large \bf Appendix C}

\vspace{3mm}
{\Large\bf Summary of the  results for
{\boldmath $\bar{\alpha}$}.}
\end{center}
\vspace{3mm}

Among the three input parameters $\bar{\alpha}$, $G_{\mu}$ and
$M_Z$, the first one has the maximum uncertainty; this
uncertainty leads to the uncertainty $\pm 5$ GeV in the value
of the $t$-quark mass extracted from the measurements of $m_W$
and the decay parameters of the $Z$-boson.
According to the formulas of Appendix B,
$$
\bar{\alpha}=\frac{\alpha}{1-\delta\alpha} \;\;,
\delta\alpha =\delta\alpha_l +\delta\alpha_h \;\;,
$$
$$
\delta\alpha_l =0.0314 \;\;.
$$

In (\cite{20}, Burkh) $\delta\alpha_h$ was calculated by
substituting experimental data on
$\sigma_{e^+e^- \to{\rm hadrons}}$ into the dispersion
integral at $\sqrt{s}< 40$ GeV and  the parton model result at
$\sqrt{s}>40$ GeV:
\begin{equation}
\delta\alpha_h^{[{\rm Burkh}]} = 0.0282(9) \;\;, \;\;
\bar{\alpha}^{[{\rm Burkh}]} =[128.87(12)]^{-1} \;\;.
\label{C.1}
\end{equation}

In (\cite{20}, Vys) it was pointed out that the simplest model
(the lightest vector meson + the QCD-improved parton continuum
in each flavor channel) produces a surprisingly close result:
\begin{equation}
\delta\alpha_h^{[{\rm Vys}]} =0.0280(4) \;\;, \;\;
\bar{\alpha}^{[{\rm Vys}]} = [128.90(6)]^{-1} \;\;.
\label{C.2}
\end{equation}

The model with infinite number of poles (\cite{20}, Gesh) yields
the determination of $\delta\alpha_h$ with very high accuracy:
\begin{equation}
\delta\alpha_h^{[{\rm Gesh}]} =0.0275(2) \;\;, \;\;
\bar{\alpha}^{[{\rm Gesh}]} =[128.96(3)]^{-1} \;\;.
\label{C.3}
\end{equation}

A recent analysis of experimental data (\cite{20}, Sw, \cite{20},
Ma) yielded considerably lower values of $\delta\alpha_h$:
\begin{equation}
\delta\alpha_h^{[{\rm Sw}]} = 0.0265(8) \;\;, \;\;
\bar{\alpha}^{[{\rm Sw}]} = [129.10(12)]^{-1} \;\;,
\label{C.4}
\end{equation}
\begin{equation}
\delta\alpha_h^{[{\rm Ma}]} = 0.0273(4) \;\;, \;\;
\bar{\alpha}^{[{\rm Ma}]} = [128.99(6)]^{-1} \;\;.
\label{C.5}
\end{equation}

In this review we make use of the results of a recent analysis
(\cite{20}, Eid),
\begin{equation}
\delta\alpha_h^{[{\rm Eid}]} = 0.0280(7) \;\;, \;\;
\bar{\alpha}^{[{\rm Eid}]} = [128.896(90)] \;\;.
\label{C.6}
\end{equation}

\newpage

\setcounter{equation}{0}
\renewcommand{\theequation}{D.\arabic{equation}}
\begin{center}
{\large\bf Appendix D.}

\vspace{3mm}

{\Large\bf How {\boldmath$\alpha_W(q^2)$} and
{\boldmath$\alpha_Z(q^2)$} `crawl'.}
\end{center}
\vspace{2mm}

The effect of `running' of electromagnetic coupling constants $\alpha(q^2)$
(logarithmic dependence of the effective charge on
momentum transfer $q^2$) is known for more than four decades) \cite{76D}.
In contrast to $\alpha(q^2)$, the effective constants of $W$ and
$Z$ bosons $\alpha_W(q^2)$ and $\alpha_Z(q^2)$ in the region $0<q^2 \la
m_Z^2$ `crawl' rather than run \cite{77D}.

If we define the effective gauge coupling constant
$g^2(q^2)$ in terms of the bare charge $g_0^2$ and the bare mass $m_0$,
and sum up the geometric series with the self-energy $\Sigma(q^2)$
inserted in the gauge boson propagator, this gives the expression
\begin{equation}
g^2(q^2) = \frac{g_0^2}{1+g_0^2 \frac{\Sigma(q^2)-\Sigma(m^2)}
{q^2 -m^2}} \;\;.
\label{D.1}
\end{equation}
Here $m$ is the physical mass, and $\Sigma(q^2)$ contains the
contribution of fermions only, since loops with $W$, $Z$ and $H$ bosons
do not contain large logarithms in the region $|q^2|\leq m_Z^2$.

The bare coupling constant in the difference $1/g^2(q^2) - 1/g^2(0)$
is eliminated, which gives a finite expression.
The result is
\begin{equation}
1/\alpha_Z(q^2) - 1/\alpha_Z(0) =b_Z F(x) \;\;, \;\;
{\rm where} \;\; x=q^2/m_Z^2 \;\;,
\label{D.2}
\end{equation}
\begin{equation}
1/\alpha_W(q^2) - 1/\alpha_W(0) =b_W F(y) \;\;, \;\;
{\rm where} \;\; y=q^2/m_Z^2 \;\;,
\label{D.3}
\end{equation}
\begin{equation}
F(x) =\frac{x}{1-x} \ln |x|
\label{D.4}
\end{equation}

If $x\gg 1$, equations (\ref{D.2}) and (\ref{D.3}) define the
logarithmic running of charges owing to leptons and quarks, and
$b_Z$ and $b_W$ represent the contribution of fermions to the
first coefficient of the Gell-Mann--Low function:
\begin{eqnarray}
b_Z & = & \frac{1}{48\pi}\{N_u 3[1+(1-\frac{8}{3}s^2)^2] +N_d 3[1+(-1
+ \frac{4}{3}s^2)^2] + N_l[2+(1+(1-4s^2)^2)]\} \;\;, \nonumber \\
b_W & = & \frac{1}{16\pi}\{6N_q +2N_l] \;\;, 
\label{D.5}
\end{eqnarray}
where $N_{u,d,q,l}$ are the numbers of quarks and leptons with
masses that are considerably lower than $\sqrt{q^2}$.

For $q^2 \la m_Z^2$, the numerical values of the coefficients $b_{Z,W}$ are
\cite{77D}
$$
b_Z \simeq 0.195
$$
$$
b_W \simeq 0.239
$$
The massive propagator $\frac{1}{q^2 -m^2}$ in (\ref{D.1})
greatly suppresses the running of $\alpha_W(q^2)$ and $\alpha_Z(q^2)$.
Thus, according to (\ref{D.2}) and (\ref{D.3}), the constant
$\alpha_Z(q^2)$ grows by 0.85\% from $q^2 =0$ to $q^2 = m_Z^2$,
$$
1/\alpha_Z(m_Z^2) = 22.905
$$
\begin{equation}
1/\alpha_Z(m_Z^2) -1/\alpha_Z(0) = -0.195 \;\;,
\label{D.6}
\end{equation}
and the constant $\alpha_W(q^2)$ grows by 0.95\%,
$$
1/\alpha_W(m_Z^2) = 28.74
$$
\begin{equation}
1/\alpha_W(m_Z^2) -1/\alpha_W(0) = -0.272 \;\;,
\label{D.7}
\end{equation}
while the electromagnetic constant $\alpha(q^2)$ increases by
6.34\%:
\begin{equation}
1/\alpha(m_Z^2) -1/\alpha_W(0) = 128.90 - 137.04 = -8.14
\label{D.8}
\end{equation}

With the accuracy indicated above, we can thus assume
\begin{eqnarray}
\alpha_Z(m_Z^2) \simeq \alpha_Z(0) \nonumber \\
\alpha_W(m_Z^2) \simeq\alpha_W(0) .
\label{D.9}
\end{eqnarray}

At the same time, $\alpha(m_Z^2)$ differs greatly from $\alpha(0)$;
therefore the latter has no connection to the electroweak physics
but only to the purely electromagnetic physics.

\newpage

\setcounter{equation}{0}
\renewcommand{\theequation}{E.\arabic{equation}}
\begin{center}
{\large\bf Appendix E}

\vspace{3mm}
{\Large\bf Relation between
{\boldmath$\bar{\alpha}$},
{\boldmath$G_{\mu}$}, {\boldmath$m_Z$}
 and the bare quantities.}
\end{center}

\vspace{3mm}

The bare quantities are marked by the subscript `0'. In the electroweak
theory, three bare charges $e_0$, $f_0$ and $g_0$ that describe
the interactions of $\gamma$, $Z$ and $W$ are related by a single
constraint:
\begin{equation}
(e_0/g_0)^2 +(g_0/f_0)^2 =1 \;\;.
\label{E.1}
\end{equation}

The bare masses of the vector bosons are defined by the bare
vacuum expectation value of the higgs field $\eta_0$:
\begin{equation}
m_{Z0}=\frac{1}{2}f_0 \eta_0 \;\;, \;\;
m_{W0} =\frac{1}{2}g_0\eta_0 \;\;.
\label{E.2}
\end{equation}

The fine structure constant $\alpha = e^2/4\pi$ is related to
he bare charge $e_0$ by the formula
\begin{equation}
\alpha\equiv\alpha(q^2 =0) =\frac{e_0^2}{4\pi}
\left(1-\Sigma'_{\gamma}(0)-2\frac{s}{c} \frac{\Sigma_{\gamma Z}(0)}
{m_Z^2} \right) \;\;,
\label{E.3}
\end{equation}
where $\Sigma'(0) =\lim_{q^2\to 0} \Sigma(q^2)/q^2$. In the Feynman
gauge $\Sigma_{\gamma Z}(0)\approx -(\alpha/2\pi)
(m^2_W/cs)(1/\epsilon)$, where the dimension of spacetime
is $D=4-2\varepsilon$. In the unitary gauge $\Sigma_{\gamma Z}(0)=0$.

The simplest way  to verify the presence the term
$2(s/c)\Sigma_{\gamma Z}(0)/m_Z^2$ is by considering the
interaction of a photon with the right-handed electron $e_R$.
Note that in this case
there are no weak vertex corrections due to the $W$-boson
exchange. (Note also that the left-handed neutrino remains neutral
even when loop corrections are taken into account, since the diagram
with the $\gamma -Z - \nu_L \bar{\nu}_L$ interaction is compensated
for by the vertex diagram with the $W$ exchange).

Our first basic equation is the renormalization group improved
relation between $\bar{\alpha}=\alpha(q^2 =m_Z^2)$ and
$\alpha_0$:
\begin{equation}
\bar{\alpha}=\alpha_0 \left[1-\Pi_{\gamma}(m_Z^2)-2\frac{s}{c}
\Pi_{\gamma Z}(0) \right] \;\;,
\label{E.4}
\end{equation}
where $\Pi_{\gamma}(q^2)=\Sigma_{\gamma}(q^2)/m_Z^2$,
$\Pi_{\gamma Z}(q^2)=\Sigma_{\gamma Z}(q^2)/m_Z^2$.

The  second basic equation is:
\begin{equation}
m_Z^2 =m_{Z0}^2 [1-\Pi_Z(m_Z^2)] =m_{W0}^2 /c_0^2
[1-\Pi_Z(m_Z^2)]\;\;.
\label{E.5}
\end{equation}
A similar equation holds for $m_W^2$:
\begin{equation}
m_W^2 =m_{W0}^2 [1-\Pi_W(m_W^2)] \;\;,
\label{E.6}
\end{equation}
where $\Pi_i(q^2) = \Sigma_i(q^2)/m_i^2$, $i=W, Z$.

Finally, the third basic equation is
\begin{equation}
G_{\mu} =\frac{g_0^2}{4\sqrt{2}m_{W0}^2} [1+\Pi_W(0)+D] \;\;,
\label{E.7}
\end{equation}
where $\Pi_W(0) = \Sigma_W(0)/m_W^2$ comes from the propagator of
$W$,
while $D$ is the contribution of the box and the vertex diagrams
(minus the electromagnetic corrections to the four-fermion
interaction) to the muon decay amplitude. According to Sirlin
\cite{17}, \begin{equation} D=\frac{\bar{\alpha}}{4\pi
s^2}(6+\frac{7-4s^2}{2s^2} \ln c^2 +4\Delta_W) \;\;, \label{E.8}
\end{equation}
where
\begin{equation}
\Delta_W \equiv \Delta(m_W) =\frac{2}{4-D} +\ln 4\pi -\gamma
-\ln(m_W^2 /\mu^2)
\label{E.9}
\end{equation}
\newpage

\setcounter{equation}{0}
\renewcommand{\theequation}{F.\arabic{equation}}
\begin{center}
{\large\bf Appendix F}\\
\vspace{3mm}
{\Large\bf The radiators ${\bf R_{Aq}}$ and
${\bf R_{Vq}}$.}
\end{center}
\vspace{2mm}

For decays to light quarks $q = u,d,s$, we neglect the
quark masses and take into account the gluon exchanges in the final
state up to terms $\sim\alpha^3_s$ \cite{A} - \cite{D}, and also
one-photon exchange in the final state and the interference of
the photon and the gluon exchanges \cite{4300}. These corrections
are slightly different for the vector and the axial channels.

For decays to quarks we have
\begin{equation}
\Gamma_q = \Gamma(Z \to q\bar{q}) = 12[g^2_{Aq} R_{Aq} + g^2_{Vq}
R_{Vq}]\Gamma_0
\label{F.1}
\end{equation}
where the factors $R_{A,V}$ are responsible for the interaction in
the final state (in our previous papers, we used letter $G$ instead of
$R$). The results presented in \cite{A} - \cite{4300} are
\begin{eqnarray}
R_{Vq} &=& 1 +
\frac{\hat{\alpha}_s}{\pi} + \frac{3}{4} Q^2_q
\frac{\bar{\alpha}}{\pi} - \frac{1}{4} Q^2_q \frac{\bar{\alpha}}{\pi}
\frac{\hat{\alpha}_s}{\pi} + \nonumber \\
&+& [1.409 + (0.065 + 0.015 \ln t)\frac{1}{t}](\frac{\hat{\alpha}_s}{\pi})^2
-12.77(\frac{\hat{\alpha}_s}{\pi})^3
+12 \frac{\hat{m}^2_q}{m^2_Z} \frac{\hat{\alpha}_s}{\pi} \delta_{vm}
\label{F.2}
\end{eqnarray}
$$
R_{Aq} = R_{Vq} - (2T_{3q})[I_2(t)(\frac{\hat{\alpha}_s}{\pi})^2 +
I_3(t)(\frac{\hat{\alpha}_s}{\pi})^3]
-
$$
\begin{eqnarray}
-12 \frac{\hat{m}^{2}_{q}}{m^{2}_{Z}} \frac{\hat{\alpha}_s}{\pi} \delta_{vm} -
6\frac{\hat{m}^2_q}{m^2_Z} \delta^1_{am}
- 10 \frac{\hat{m}^2_q}{m^2_t}(\frac{\hat{\alpha}_s}{\pi})^2 \delta^2_{am}
\label{F.3}
\end{eqnarray}
where $\hat{m}_q$ is the running quark mass (see below),
\begin{eqnarray}
\delta_{vm} = 1+8.7
(\frac{\hat{\alpha}_s}{\pi})+45.15(\frac{\hat{\alpha}_s}{\pi})^2,
\label{F.4}
\end{eqnarray}
\begin{eqnarray}
\delta^1_{am} = 1 + 3.67(\frac{\hat{\alpha}_s}{\pi}) +(11.29 - \ln t)
(\frac{\hat{\alpha}_s}{\pi})^2,
\label{F.5}
\end{eqnarray}
\begin{eqnarray}
\delta^2_{am} = \frac{8}{81}+\frac{\ln t}{54},
\label{F.6}
\end{eqnarray}
\begin{equation}
I_2(t) = -3.083 - \ln t + \frac{0.086}{t} + \frac{0.013}{t^2}\;\;,
\label{F.7}
\end{equation}
\begin{eqnarray}
I_3(t) &=& -15.988 - 3.722 \ln t + 1.917 \ln^2 t\;\;, \\ \nonumber
t &=& m^2_t/m^2_Z\;\;.
\label{F.8}
\end{eqnarray}
Terms of the order of $(\frac{\hat{\alpha}_s}{\pi})^3$ caused
by the diagrams with three gluons in intermediate state were
calculated in \cite{63}. For $R_{Vq}$ they are numerically very
small, $\sim 10^{-5}$; for this reason, we dropped them from
formula (\ref{F.2}).

For the $Z \to b\bar{b}$ decay, the $b$-quark mass is not
negligible; it reduces $\Gamma_b$ by about 1 MeV ($\sim 0.5\%$).
The gluon corrections result in a replacement of the pole mass
$m_b \simeq 4.7 \mbox{\rm GeV}$ by the running mass, the
virtuality 
being $m_Z\;:\;\; m_b \to \hat{m}_b(m_Z)$. We express $\hat{m}_b(m_Z)$
in terms of $m_b$, $\hat{\alpha}_s(m_Z)$ and $\hat{\alpha}_s(m_b)$
using standard two-loop equations in the $\overline{MS}$ scheme (see
\cite{E}).

For the $Z \to c\bar{c}$ decay, the running mass $\hat{m}_c(m_Z)$
is of the order of $0.5$ GeV and the corresponding contribution
to $\Gamma_c$ is of the order of $0.05$ MeV. We have included
this infinitesimal term in the LEPTOP code, since it is taken
into account in other codes (see, for example, \cite{16}).

We need to remark in connection with $\Gamma_c$ that the term
$I_2(t)$, given by equation (\ref{F.7}), contains
interference terms $\sim(\hat{\alpha}_s/\pi)^2$. These terms
 are related to three types of final states:
 one quark pair, a quark pair and
a gluon, two quark pairs. This last contribution comes to about 5\%
of $I_2$ and is infinitesimally small at the currently
achievable experimental accuracy. Nevertheless, in principle
these terms require special consideration, especially if these
quark pairs are of different flavors, for example, $b\bar{b}c\bar{c}$.
Such mixed quark pairs must be discussed separately.

Note that $\hat{\alpha}_s$ stands for the strong interaction constant
in the $\overline{MS}$ subtraction scheme, with $\mu^2 =m^2_Z$.

\newpage

\setcounter{equation}{0}
\renewcommand{\theequation}{G.\arabic{equation}}
\begin{center}
{\large\bf Appendix G.}\\

\vspace{3mm}

{\Large\bf Derivation of formulas for the asymmetries.}
\end{center}

\vspace{2mm}

Asymmetry in the processes $e^+e^- \to Z \to f\bar{f}$ is
calculated with the masses of $e$ and $f$ neglected in comparison
with the $Z$ boson mass.
(Mass corrections for $f = b$ are of the order of $2\cdot 10^{-3}$
and will be taken into account below).
The amplitude (20) of the interactions of the $Z$ boson with
massless fermions $f\bar{f}$ can be conveniently rewritten in
the form
\begin{equation}
M(Z \to f\bar{f}) = \frac{1}{2} \bar{f} Z_{\alpha} [g^f_L
j^L_{\alpha} + g^f_R j^R_{\alpha}]\;\;,
\label{G.1}
\end{equation}
where
$$
j^{L,R}_{\alpha} = \bar{\psi}_{L,R} \gamma_{\alpha} \psi_{L,R}\;\;,
$$
$$
\psi_{L,R} = \frac{1}{2}(1\pm\gamma_5)\psi \;\;,
$$
$$
g^f_{L,R} = g_{Vf} \pm g_{Af}\;\;.
$$
The chirality is a conserved quantum number for massless fermions
(anomalies do not yet manifest themselves in the approximations
we deal with here) and coincides with a fermion's helicity up to
the sign.

Therefore, the pairs $e^-_L e^+_L$ and $e^-_R e^+_R$ do not
transform into the $Z$ boson at all, and the pairs $e^-_L e^+_R$
and $e^-_R e^+_L$ create a $Z$ boson with the polarization
$\pm 1$, respectively (along the positron beam). The scattering
amplitudes thus have the form
\begin{eqnarray}
T(e^-_{L,R} e^+ \to f_{L,R} \bar{f}) = g^e_{L,R}
g^f_{L,R} T_0 (1 + \cos \theta) \nonumber \\
T(e^-_{L,R} e^+ \to f_{R,L} \bar{f}) = g^e_{L,R}
g^f_{R,L} T_0 (1 - \cos \theta) \;\;,
\label{G.2}
\end{eqnarray}
where coefficient
$T_0$ is nonimportant at the moment (it can be
reconstructed from (\ref{G.1})), and $\theta$ is the angle
between the momenta of $e^-$ and $f$. The sign in front of $\cos
\theta$ is chosen for the helicity to be conserved in forward and
backward scattering.

Once the form of the amplitude (\ref{G.2}) is known, all
asymmetries are immediately found.

(a) Left--right asymmetry $A_{LR}$ is defined as the ratio
$$
A_{LR} = \frac{\sigma_L - \sigma_R}{\sigma_L + \sigma_R}\;\;,
$$
where $\sigma_{L,R} = \sigma(e_{L,R} e^+ \to f\bar{f})$. Hence
\begin{equation}
A_{LR} = \frac{(g^e_L)^2 - (g^e_R)^2}{(g^e_L)^2 + (g^e_R)^2}
\equiv A^e
\label{G.3}
\end{equation}

(b) Longitudinal polarization $P_{\tau}(\cos \theta)$ is defined
as the ratio of the difference to the sum of differential cross
sections,
$(\frac{d\sigma}{d\theta})_{R,L} =
\frac{d\sigma}{d\theta} (e\bar{e} \to
\tau_{R,L} \bar{\tau})$:
\begin{equation}
P_{\tau}(\cos \theta) = \frac{(\frac{d\sigma}{d\theta})_R -
(\frac{d\sigma}{d\theta})_L}
{(\frac{d\sigma}{d\theta})_R + (\frac{d\sigma}{d\theta})_L}\;\;,
\label{G.4}
\end{equation}
where
\begin{eqnarray}
(\frac{d\sigma}{d\theta})_R = \frac{1}{2m^2_Z} |T_0|^2 (g^{\tau}_R)^2
[(g^e_R)^2 (1+\cos \theta)^2 + (g^e_L)^2 (1 - \cos \theta)^2]
\nonumber \\
(\frac{d\sigma}{d\theta})_L = \frac{1}{2m^2_Z} |T_0|^2
(g^{\tau}_L)^2[(g^e_L)^2 (1+\cos \theta)^2 + (g^e_R)^2 (1 - \cos
\theta)^2]
\label{G.5}
\end{eqnarray}
Substituting (\ref{G.5}) into the definition (\ref{G.4}), we
obtain
\begin{equation}
P_{\tau}(\cos\theta)
= -\frac{A_{\tau} (1 + \cos^2 \theta) + 2A_e \cos \theta}
{(1+ \cos^2 \theta) + 2A_e A_{\tau} \cos \theta} \;\;,
\label{G.6}
\end{equation}
where $A_e$ and $A_{\tau}$ are defined according to (\ref{G.3}).
The longitudinal polarization $P^{\tau}$, averaged over
directions of $\tau$-leptons, is defined as the following
ratio of the total cross sections $\sigma_{L,R} = \int^1_{-1} d \cos \theta
(\frac{d\sigma}{d\theta})_{L,R}$
\begin{eqnarray}
P_{\tau} &=& \frac{\sigma_R^{\tau} - \sigma_L^{\tau}}
{\sigma_R^{\tau} +
\sigma_L^{\tau}} = - \frac{\int^1_{-1} d\cos\theta [A_{\tau}(1 +
\cos^2\theta) + 2A_e \cos\theta)]}{\int^1_{-1} d\cos\theta [1 +
\cos^2\theta + 2A_e A_{\tau} \cos\theta]} \nonumber \\ &=& - A_{\tau}
\label{G.7}
\end{eqnarray}

(c) Forward--backward asymmetry $A^f_{FB}$ is calculated more
simply in terms of $g_{A,V}$. The squared matrix element of the
process $e\bar{e} \to f\bar{f}$ is proportional to
\begin{eqnarray}
|M|^2 \sim \left\{ [ g^2_{Ae} + g^2_{Ve}]\;
\left[ [g^2_{Af} + g^2_{Vf}](1+
v^2 \cos^2\theta) \right. \right. \nonumber \\
+\left. \left. [g^2_{Af} - g^2_{Af}] (1 - v^2)\right] +
\frac{1}{2}(g_{Ve} g_{Ae} g_{Vf} g_{Af}) v\cos\theta \right\}\;\;,
\label{G.8}
\end{eqnarray}
where $\theta$ is the scattering angle and $v = 1 - \frac{4m^2_f}{m^2_Z}$
is the  velocity of fermion $f$. This immediately implies that
\begin{equation}
A^f_{FB} = \frac{3}{4} A_e \left[ \frac{2g_{Af} g_{Vf} v}
{g^2_{Af} v^2 + g^2_{Vf} \frac{3 - v^2}{2}} \right]
\label{G.9}
\end{equation}

The mass $m_f$ is negligible in all channels with the exception
of $f = b$, where the nonzero mass produces effects of the order of
$2 \cdot 10^{-3}$. Gluon corrections in the
final state (see Appendix F) replace the pole mass
$m_b \simeq 4.7$ GeV in equations (\ref{G.8})--(\ref{G.9})
by the  running mass at the $m_Z$ scale: $m_b \to
\hat{m}_b(m_Z)$.

Note that
starting from gluon corrections of the order of
$(\frac{\alpha_s}{\pi})^2$, it is impossible to unambiguously
separate different quark channels, since additional pairs of
`alien' quarks are created in this order. We do not consider
corrections $(\frac{\hat{\alpha}_s}{\pi})^2$ in asymmetries. In
our approximation the ratio $g_{Vf}/g_{Af}$ is not renormalized
by the  gluonic interaction in the final state.
Therefore, the expected accuracy of formula (\ref{G.9}) is
$(\frac{\alpha_s}{\pi})^2 \sim 2\cdot 10^{-3}$, which is by an
order of magnitude better than the experimental accuracy.

\newpage

\setcounter{equation}{0}
\renewcommand{\theequation}{H.\arabic{equation}}
\begin{center}
{\large\bf Appendix H. }

\vspace{2mm}

{\Large\bf Corrections proportional to ${\bf{m_t^2}}$.}
\end{center}
\vspace{3mm}

This appendix gives a simple mnemonic recipe for the derivation
of corrections proportional to $m^2_t$. A rigorous derivation
requires careful regularization of Feynman integrals.

The terms proportional to $m_t^2$ contribute to radiative corrections
to  bare masses (squared) of the $W$- and $Z$-bosons, but not to
the corrections to the bare coupling constants. This follows
from dimensional arguments. Indeed, the dimension of
self-energy  $\Sigma$  for the
boson, equals $m^2$; therefore, the terms $\sim m_t^2$ remain in
$\Sigma(q^2)$ in the limit $q^2 \to 0$. On the other hand, the corrections
to coupling constants are proportional to $d\Sigma/ d q^2$ and
do not contain terms $\sim m_t^2$. Therefore, it is easy to
evaluate the contribution of the $t$-quark to the parameter $\rho =
(\alpha_Z/\alpha_W)(m_W^2/m_Z^2)$ in the approximation $\sim \alpha
m_t^2$ (the Veltman approximation \cite{21}), neglecting the
terms $\sim \alpha$:
\begin{eqnarray}
\nonumber
\rho &\simeq & \frac{\alpha_{Z0}}{\alpha_{W0}} \frac{m^2_{W0}-
\Sigma_W(m^2_W)} {m^2_{Z0}-\Sigma_Z(m^2_Z)} \simeq 1 +
\frac{\Sigma_Z(0)}{m^2_Z} - \frac{\Sigma_W(0)}{m^2_W} \equiv \\
&\equiv & 1 + \Pi_Z(0) - \Pi_W(0)\;\;.
\label{H.1}
\end{eqnarray}
The evaluation of the difference $\Pi_Z(0) - \Pi_W(0)$ is
elementary:
\begin{eqnarray}
\nonumber
\Pi_W(0)& = & \frac{\Sigma_W(0)}{m^2_W} = \\
& = & \frac{3\alpha_W}{8\pi m^2_W} \left(\int^{\infty}_0 \frac{p^2 d
p^2}{p^2 + m^2_t} = \int^{\infty}_0 d p^2 - m^2_t \int^{\infty}_0
\frac{d p^2}{p^2 + m^2_t} \right)_{.}
\label{H.2}
\end{eqnarray}
(As we have neglected the mass of the $b$-quark, the
propagator of the $b$-quark compensates  the factor $p^2$ in the
numerator.)
\begin{eqnarray}
\nonumber
\Pi_Z(0) = \frac{\Sigma_Z(0)}{m^2_Z}&=& \\
\nonumber
&=& \frac{3\alpha_Z}{8\pi m^2_Z}
\left(\frac{1}{2} \int^{\infty}_0 d p^2 + \frac{1}{2}
\int^{\infty}_0 \frac{p^4 d p^2}{(p^2 + m^2_t)^2}= \right. \\
&=& \left. \frac{1}{2} \int^{\infty}_0 d p^2 + \frac{1}{2}
\int^{\infty}_0 - m^2_t \int^{\infty}_0 \frac{d p^2}{p^2 + m^2_t} +
\frac{1}{2} m^2_t \right)\;\;.
\label{H.3}
\end{eqnarray}

Taking into account that in one-loop approximation
we can set in (\ref{H.2}) and (\ref{H.3})
\begin{equation}
\frac{\alpha_W}{m^2_W} = \frac{\alpha_Z}{m^2_Z}\;\;,
\label{H.4}
\end{equation}
we see that quadratic and logarithmic divergences cancel out and
that finally
\begin{equation}
\rho \simeq 1 + \Delta\rho_t = 1 +
\frac{3\alpha_Z}{16\pi} \frac{m^2_t}{m^2_Z} = 1 +
\frac{3\alpha_Z}{16\pi} t\;\;,
\label{H.5}
\end{equation}

\begin{equation}
\Delta \rho_t = \frac{3\bar{\alpha}}{16\pi s^2c^2} t \;\;,
\label{H.6}
\end{equation}
where $t = m^2_t/m^2_Z$ and we assume that $t \gg 1$.

Let us express  the leading in
$t$  corrections to the main quantities $m_W/m_Z$, $g_{A_l}$ and $g_{V_l}$
in terms of $\Delta \rho_t$.

We define
\begin{equation}
c^2_{\alpha} =
\alpha_W/\alpha_Z\;, s^2_{\alpha} = 1 - c^2_{\alpha}\;\;.
\label{H.7}
\end{equation}
Then
\begin{equation}
G_{\mu} = \frac{\pi\alpha_W}{\sqrt{2} m^2_W} =
\frac{\pi}{\sqrt{2}\rho} \frac{\bar{\alpha}}{c^2_{\alpha}
s^2_{\alpha} m^2_Z}\;\;,
\label{H.8}
\end{equation}
and hence
\begin{equation}
s^2_{\alpha} c^2_{\alpha} \simeq s^2 c^2/(1+ \Delta\rho_t)\;\;.
\label{H.9}
\end{equation}
Solving the last equation, we obtain
\begin{equation}
c^2_{\alpha} \simeq c^2 \left( 1 + \frac{s^2}{c^2 - s^2} \Delta\rho_t
\right)\;\;,
\label{H.10}
\end{equation}
\begin{equation}
s^2_{\alpha} \simeq s^2 \left( 1 - \frac{c^2}{c^2 - s^2}
\Delta\rho_t\right)\;\;,
\label{H.11}
\end{equation}
and therefore,
\begin{equation}
\frac{m^2_W}{m^2_Z} \simeq c^2_{\alpha} (1 + \Delta\rho_t) \simeq
c^2 \left(1 + \frac{c^2}{c^2 - s^2} \Delta\rho_t \right)\;\;,
\label{H.12}
\end{equation}
and
\begin{equation}
\frac{m_W}{m_Z} \simeq c+\frac{3\bar{\alpha}}{32\pi} \frac{c}{(c^2 -
s^2) s^2} t\;\;.
\label{H.13}
\end{equation}
Likewise,
\begin{eqnarray}
\nonumber
g^2_{\nu} & \simeq& g^2_{A_l} \simeq \frac{1}{4} \left(
\frac{\bar{\alpha}}{c^2_{\alpha} s^2_{\alpha}}/
\frac{\bar{\alpha}}{c^2 s^2} \right) \simeq \frac{1}{4}( 1 +
\Delta\rho_t) \simeq \\
& \simeq & \frac{1}{4} \left( 1 + \frac{3\bar{\alpha}}{16\pi s^2 c^2}
t \right)\;\;.
\label{H.14}
\end{eqnarray}

\begin{eqnarray}
\nonumber
\frac{g_{V_l}}{g_{A_l}} &\simeq & (1 - 4 s^2_{\alpha}) \simeq (1 -
4s^2) + \frac{4 c^2 s^2}{c^2 - s^2} \Delta\rho_t = \\
& = & (1 - 4 s^2) + \frac{3\bar{\alpha}}{4\pi(c^2 - s^2)} t\;\;.
\label{H.15}
\end{eqnarray}

The corrections proportional to $m_t^2$ were first pointed out
by Veltman \cite{21}, who emphasized the appearance of such
corrections for a large difference $m^2_t - m^2_b$ which violates
the isotopic symmetry. In this review, the coefficients
in front of  the
factors $t$ in equations (\ref{H.13})--(\ref{H.15}) are used as
 coefficients for normalized radiative corrections $V_i$.

\newpage

\setcounter{equation}{0}
\renewcommand{\theequation}{I.\arabic{equation}}
\begin{center}
{\large\bf Appendix I}\\

\vspace{3mm}

{\Large\bf Explicit form of the functions
${\bf T_i(t)}$ and ${\bf H_i(h)}$.}

\end{center}
\vspace{2mm}

The equations for $T_i(t)$ and $H_i(h)$ are
\cite{35}, \cite{53}:

$$
\underline{i = m}
$$
$$
T_m(t) = (\frac{2}{3} - \frac{8}{9}s^2)\ln t - \frac{4}{3} + \frac{32}{9}
s^2 +
$$
$$
+ \frac{2}{3}(c^2 - s^2)(\frac{t^3}{c^6} - \frac{3t}{c^2} + 2) \ln \mid 1 -
\frac{c^2}{t} \mid +
$$
\begin{equation}
+ \frac{2}{3} \frac{c^2-s^2}{c^4} t^2 + \frac{1}{3} \frac{c^2 - s^2}{c^2} t
+ [\frac{2}{3} - \frac{16}{9} s^2 -
\frac{2}{3} t - \frac{32}{9} s^2t] F_t(t)\;;
\label{I.1}
\end{equation}
$$
H_m(h) = -\frac{h}{h-1} \ln h + \frac{c^2 h}{h-c^2}\ln \frac{h}{c^2} -
\frac{s^2}{18c^2} h - \frac{8}{3}s^2 +
$$
$$
+ (\frac{h^2}{9} -\frac{4h}{9} + \frac{4}{3})F_h(h) -
$$
$$
-(c^2 - s^2)(\frac{h^2}{9c^4} - \frac{4}{9} \frac{h}{c^2} + \frac{4}{3})
F_h(\frac{h}{c^2}) +
$$
$$
+ (1.1203 - 2.59\delta s^2)\;;
$$
where $\delta s^2 = 0.23110 - s^2$ (note the sign!).

$$ \underline{i = A} $$
$$
T_A(t) = \frac{2}{3} -
\frac{8}{9} s^2 + \frac{16}{27} s^4 - \frac{1 - 2tF_t(t)}{4t-1} +
$$

\begin{equation}
+(\frac{32}{9} s^4 - \frac{8}{3} s^2 - \frac{1}{2})[ \frac{4}{3}tF_t(t) -
\frac{2}{3}(1+2t)\frac{1-2tF_t(t)}{4t-1}] \;\;;
\label{I.2}
\end{equation}
$$
H_A(h) = \frac{c^2}{1-c^2/h}\ln \frac{h}{c^2} - \frac{8h}{9(h-1)}\ln h +
$$
$$
+(\frac{4}{3} - \frac{2}{3}h+ \frac{2}{9}h^2)F_h(h) - (\frac{4}{3} -
\frac{4}{9}h + \frac{1}{9} h^2) F'_h(h)-
$$
$$
-\frac{1}{18}h + [0.7752 + 1.07\delta s^2]
\;.$$

$$
\underline{i = R}
$$
$$
T_R(t) = \frac{2}{9}\ln t + \frac{4}{9} -
\frac{2}{9}(1+ 11t)F_t(t)\;;
$$
\begin{equation}
H_R(h) = -\frac{4}{3} - \frac{h}{18} + \frac{c^2}{1 - c^2/h}\ln
\frac{h}{c^2} +
\label{I.3}
\end{equation}
$$
+ (\frac{4}{3} - \frac{4}{9}h + \frac{1}{9}h^2)F_h(h) + \frac{h}{1-h}\ln h +
$$
$$
+ (1.3590 + 0.51\delta s^2)
$$
$$
\underline{i=\nu}
$$
$$
T_{\nu}(t) = T_A(t)
$$
\begin{equation}
H_{\nu}(h) = H_A(h)
\label{I.4}
\end{equation}

The functions $F_t$ and $F_h$ are the limiting cases of the function $F(s,
m_1, m_2)$, described in Appendix B. The explicit formulas for $F_t(t)$ and
$F_h(h)$ are given in Appendix J (equation (J.3))
and Appendix L (equation (L.11)), respectively.

\newpage

\setcounter{equation}{0}
\renewcommand{\theequation}{J.\arabic{equation}}
\begin{center}
{\large\bf Appendix J.}\\
\vspace{3mm}

{\Large\bf The contribution of heavy fermions to the self-energy
  of the vector bosons.}
\end{center}

\vspace{2mm}

Let us give the expressions for the contribution of
third-generation quarks $(t,b)$ to the polarization operators
(self-energies)
of the vector bosons. We use the following notations:
$t = m^2_t/m^2_Z$, $b = m^2_b/m^2_Z$, $(b \ll 1)$, $h = m^2_H/m^2_Z$,
$\Pi_{\gamma}(q^2) = \Sigma_{\gamma \gamma}(q^2)/m^2_Z$,
$\Pi_{\gamma Z}(q^2) = \Sigma_{\gamma Z}(q^2)/m^2_Z$, $\Pi_W(q^2)
= \Sigma_W(q^2)/m^2_W$.

The dimensional regularization yields the terms
\begin{equation}
\Delta_i = \frac{2}{4 - D} - \gamma + \ln 4\pi - \ln
\frac{m^2_i}{\mu^2}\;\;,
\label{J.1}
\end{equation}
where $i = t, b, W, Z...$, $D$ is the variable dimension of
spacetime, $(4- D = 2\varepsilon$, $\varepsilon \to 0)$,
$\gamma = -\Gamma'(1) = 0.577\dots$ (we follow \cite{64}, p. 53).

We begin with an auxiliary function $F_t(t)$, obtained
as a limiting case of
the function $F(s, m_1, m_2)$
(see Appendix B and \cite{64}, p~54;
\cite{82}, p~88),
\begin{equation}
F_t(t) \equiv F(s = m^2_Z, m_t, m_t) = F(1, t, t)\;\;,
\label{J.2}
\end{equation}
and get, using \cite{64},
\begin{equation}
F_t(t) =
\left\{
\begin{array}{ll}
2[1 - \sqrt{4t - 1} \arcsin \frac{1}{\sqrt{4t}}]\;\;, & \;\; 4t >
1\;\;, \\
2[1 - \sqrt{1 - 4t} \ln \frac{1 + \sqrt{1 - 4t}}{\sqrt{4t}}\;\;, &
\;\; 4t < 1\;\;.
\end{array}
\right. \label{J.3}
\end{equation}

The asymptotics of $F_t$ are
\begin{eqnarray}
\nonumber
F_t \simeq \ln t + 2 \;\; & \;\; t \to 0\;\;, \\
F_t \simeq 1/6t + 1/60 t^2 \;\; & \;\; t \to \infty\;\;.
\label{J.4}
\end{eqnarray}

Differentiation gives
\begin{equation}
F'_t \equiv m^2_Z \frac{dF}{d m^2_Z} = -t \frac{d}{dt} F_t = \frac{1
- 2 t F_t}{4t -1}\;\;.
\label{J.5}
\end{equation}

In this Appendix, $\Pi_i$ stands for the contribution of the doublet
$(t,b)$ to the corresponding polarization operator:
\begin{equation}
\Pi_{\gamma}(0) = 0\;\;.
\label{J.6}
\end{equation}

\begin{equation}
\Pi_{\gamma}(m^2_Z) = \frac{\bar{\alpha}}{\pi} \left[ Q^2_t
\left( \Delta_t + (1 + 2t) F_t(t) - \frac{1}{3} \right) +
Q^2_b \left( \Delta_b + \frac{5}{3} + \ln b \right) \right]\;\;.
\label{J.7}
\end{equation}

\begin{equation}
\Pi_{\gamma Z}(0) = 0\;\;,
\label{J.8}
\end{equation}

\begin{eqnarray}
\Pi_{\gamma Z}(m^2_Z) & = &
\frac{\bar{\alpha}}{c s \pi}
\left\{ \left( \frac{Q_t}{4} - s^2 Q^2_t \right)
\left( \Delta_t + (1 + 2t) F_t(t) - \frac{1}{3} \right)
- \right.  \nonumber\\
& - & \left. \left( \frac{Q_b}{4} + s^2 Q^2_b \right)
\left( \Delta_b +\frac{5}{3} + \ln b \right) \right\} \;\;.
\label{J.9}
\end{eqnarray}

\begin{equation}
\Pi_W(0) = -\frac{\bar{\alpha}}{4\pi s^2 c^2} \left[ \frac{3}{2}
t\Delta_t + \frac{3}{4} t \right]\;\;.
\label{J.10}
\end{equation}

\begin{eqnarray}
\Pi_W(m^2_W) & = & \frac{\bar{\alpha}}{4\pi s^2} \left\{ \left( 1 -
\frac{3t}{2c^2} \right) \Delta_t + \frac{5}{3} - \frac{t}{c^2}
- \frac{t^2}{2c^4} - \right.  \nonumber \\
& - & \left. \left( 1 - \frac{3t}{2c^2} + \frac{t^3}{2c^6} \right) \ln | 1 -
\frac{c^2}{t} | \right\}\;\;.
\label{J.11}
\end{eqnarray}

\begin{eqnarray}
\Pi_Z(m^2_Z) & = & \frac{\bar{\alpha}s^2}{\pi c^2} \left[ Q_t^2 \left(
\Delta_t + (1 + 2t) F_t(t) - \frac{1}{3} \right) + Q_b^2 \left(
\Delta_b + \frac{5}{3} + \ln b \right) \right] - \nonumber \\
& - & \frac{\bar{\alpha}}{2\pi c^2} \left[ Q_t \left( \Delta_t + (1 +
2t) F_t(t) - \frac{1}{3} \right) - Q_b \left( \Delta_b +
\frac{5}{3} + \ln b \right) \right] + \nonumber \\
& + & \frac{\bar{\alpha}}{8\pi s^2 c^2} \left[ (2 - 3t)
\Delta_t + (1 - t) F_t(t) + \frac{4}{3} + \ln t \right]\;\;.
\label{J.12}
\end{eqnarray}

\begin{equation}
\Pi_Z(0) = - \frac{\bar{\alpha}}{4\pi s^2 c^2} \left(
\frac{3}{2} t\Delta_t \right)\;\;.
\label{J.13}
\end{equation}

\begin{eqnarray}
\nonumber
\Sigma'(m^2_Z) &=& \frac{\bar{\alpha}s^2}{\pi c^2} \left[ Q^2_t
\left( \Delta_t+ F_t - \frac{1}{3} + (1 + 2t) F'_t \right) + Q^2_b
\left( \Delta_b + \frac{5}{3} + \ln b \right) \right] - \\
\nonumber
& - & \frac{\bar{\alpha}}{2\pi c^2} \left[ Q_t (\Delta_t + F_t -
\frac{1}{3} + (1 + 2t) F'_t) - Q_b \left( \Delta_b + \frac{5}{3} +
\ln b \right) \right] + \\
\nonumber
& + & \frac{\bar{\alpha}}{8 \pi s^2 c^2}\left[ 2\Delta_t + F_t
+ \frac{4}{3} + \ln t + (1-t) F'_t - 1 \right] = \\
\nonumber
& = & \frac{\bar{\alpha} s^2}{\pi s^2} \left[ Q^2_t \left( \Delta_t +
\frac{4 + 2t}{3(4t - 1)} + \frac{1 - 2t + 4t^2}{1 - 4t}
F_t \right) + Q^2_b \left( \Delta_b + \frac{5}{3} + \ln b
\right) \right] - \\
\nonumber
& - & \frac{\bar{\alpha}}{2 \pi c^2} \left[ Q_t \left(
\Delta_t + \frac{4 + 2t}{3(4t - 1)} + \frac{1 - 2t +
4t^2}{ 1 - 4t} F_t \right) - Q_b \left( \Delta_b +
\frac{5}{3} + \ln b \right) \right] + \\
&+& \frac{\bar{\alpha}}{ 8\pi s^2 c^2} \left[ 2\Delta_t + \ln t +
\frac{2 + t}{3(4t - 1)} + \frac{2t^2 + 2t - 1}{4t - 1} F_t
\right]\;\;.
\label{J.14}
\end{eqnarray}

\begin{eqnarray}
\nonumber
\Pi_Z(m^2_Z) - \Sigma'_Z(m^2_Z) = [2t F_t(t) - (1 + 2t) F'_t(t)]
[\frac{\bar{\alpha} s^2}{\pi c^2} Q^2_t - \frac{\bar{\alpha}}{2\pi
c^2} Q_t - \\
- \frac{\bar{\alpha}}{16 \pi s^2 c^2}] + \frac{3\bar{\alpha}}{8\pi
s^2 c^2} t\Delta_t + \frac{\bar{\alpha}}{16 \pi s^2 c^2}(2 -
3F'_t(t)) + \frac{\bar{\alpha}}{2 \pi c^2} Q_b + \frac{\bar{\alpha}
s^2}{\pi c^2} Q^2_b\;\;.
\label{J.15}
\end{eqnarray}

Substituting the expressions for the polarization operators into
the formulas for physical observables, we verify the cancelation
of the terms $\sim\Delta_i$, and also the terms proportional to
$\ln b$, since the limit $m_b \to 0$ does not produce
divergences. It is convenient to get rid of the terms
$\sim \Delta_b$ and $\sim \ln b$ already in the expression for
the polarization operators using
equation
\begin{equation}
\Delta_b + \ln(b) = \Delta_t + \ln(t)
\label{J.16}
\end{equation}
and then making sure that the terms $\sim \Delta_t$ are indeed
eliminated.

Our definition of the wavefunction of the $Z$-boson differs in sign
from that assumed in \cite{64}; hence the quantity $\Pi_{\gamma Z}$
we use also differs in sign from the expression given in
\cite{64}. With our definition, the interaction of $Z$ bosons
with Weyl fermions is $-i e \bar{f}\gamma_{\mu} f(T_3 - Q s^2)
Z_{\mu}$, and that of photons is $-ie Q\bar{f}\gamma_{\mu} f
A_{\mu}$.  The latter vertex coincides with the one given in
equations (8)--(9) of \cite{64}, while the former differs in sign.

Let us look at the formulas for physical observables.

The quantity $T_m(t)$ (\ref{I.1}) is defined as the following
combination of polarization operators:
\begin{eqnarray}
\nonumber
t + T_m(t) &=& \frac{16 \pi s^4}{3\bar{\alpha}} \{ \frac{c^2}{s^2}
[\Pi_Z(m^2_Z) - \Pi_W(m^2_W)] + \\
&+& \Pi_W(m^2_W) - \Pi_W(0)- \Pi_{\gamma}(m^2_Z)\} ,
\label{J.17}
\end{eqnarray}
since $\Pi_{\gamma Z}(0) =0 $ for fermion loops.

Using equations (\ref{J.7}), (\ref{J.10})--(\ref{J.12}) we
obtain
\begin{eqnarray}
\nonumber t + T_m(t) &=& t + \left(
\frac{2}{3} - \frac{8}{9} s^2 \right) \ln t - \frac{4}{3} +
\frac{32}{9} s^2 + \frac{c^2 - s^2}{3c^2} t + \\
\nonumber &+&
\frac{2}{3} (c^2 - s^2) \left( \frac{t^3}{c^6} - \frac{3t}{c^2} + 2
\right) \ln | 1 - \frac{c^2}{t} | + \frac{2(c^2 - s^2)}{3c^4} t^2 +
\\ & + & \left( \frac{2}{3} - \frac{16}{9} s^2 - \frac{2}{3}t -
\frac{32}{9} s^2t \right) F_t(t)\;\;,
\label{J.18}
\end{eqnarray}
where we have taken into account that $\Pi_{\gamma}(m^2_Z)$
cancels out with the sum of terms proportional to
$Q^2_t$ and $Q^2_b$ in $\Pi_Z(M^2_Z)$. The terms proportional to $t$
arise from $\Pi_W(0)$ and $\Pi_W(M^2_Z)$; those proportional to $t^2$
arise from $\Pi_W(m^2_W)$; the terms proportional to $\ln t$ and
the constants arise from $\Pi_W(m^2_W)$ and $\Pi_Z(m^2_Z)$;
$\ln|1 -\frac{c^2}{t}|$ corresponds to the threshold $(t\bar{b})$
in $\Pi_W(m^2_W)$, and finally, the only source of terms proportional
to $F_t(t)$ is $\Pi_Z(m^2_Z)$.

The infinities in the contribution of the doublet $(t,b)$ to the
observables must cancel each other since the introduction of an
additional fermion family into the electroweak theory does not
violate its renormalizability.

Substituting the terms proportional to $\Delta_t$ in (\ref{J.17})
and taking into account (\ref{J.16}), we obtain zero:
\begin{eqnarray}
\nonumber
\Delta_t \left\{
\frac{c^2}{s^2} \left[ \frac{\bar{\alpha}}{8\pi s^2 c^2} (2 - 3t) -
\frac{\bar{\alpha}}{2\pi c^2} - \frac{\bar{\alpha}}{4\pi s^2} \left(
1 - \frac{3t}{2c^2} \right) \right] + \right. \\
+ \left. \frac{\bar{\alpha}}{4\pi s^2} \left( 1 -
\frac{3t}{2c^2} \right) + \frac{\bar{\alpha}}{4 \pi s^2 c^2}
\frac{3}{2}t \right\} = 0 \;\;.
\label{J.19}
\end{eqnarray}

The expression for $T_A(t)$ is
\begin{equation}
t + T_A(t) = \frac{16\pi s^2 c^2}{3\bar{\alpha}} [\Pi_Z(m^2_Z) -
\Sigma'_Z(m^2_Z) - \Pi_W(0)]\;\;.
\label{J.20}
\end{equation}

Using (\ref{J.10}), (\ref{J.12}) and (\ref{J.14}), we have
\begin{eqnarray}
\nonumber
t + T_A(t) &=& t + \frac{2}{3} - \frac{8}{9} s^2 + \frac{16}{27} s^4
- F'_t + \left( \frac{32}{9} s^4 - \frac{8}{3} s^2 - \frac{1}{2}
\right) \times \\
& \times & \left( \frac{4}{3} tF_t - \frac{2(1 + 2t)}{3} F'_t
\right)\;\;.
\label{J.21}
\end{eqnarray}

The terms proportional to $\Delta_t$ obviously cancel out.
$\Pi_W(0)$ gives a contribution proportional to $t$, while all
other terms arise from the difference $\Pi_Z(m^2_Z) - \Sigma'_Z(m^2_Z)$.

Finally, we look at $T_R(t)$:
\begin{equation}
t + T_R(t) = -\frac{16\pi c^2 s^2}{ 3\bar{\alpha}}
\left\{
\frac{(c^2 - s^2)}{cs} \Pi_{Z\gamma}(m^2_Z) + \Pi_{\gamma}(m_Z^2) -
\Pi_Z(m^2_Z) + \Pi_W(0)
\right\} \;\;.
\label{J.22}
\end{equation}

The terms proportional to $Q^2_t$ and $Q^2_b$ cancel out, so
that only $\Pi_W(0)$ remain, as well as terms that do not
contain $Q^2_{t,b}$ coming from $\Pi_{Z\gamma}$ and $\Pi_Z$.
Substituting (\ref{J.10}), (\ref{J.9}) and (\ref{J.12}), we have
\begin{equation}
t + T_R(t) = t + \frac{4}{9} + \frac{2}{9} \ln t -
\frac{2}{9}(1 + 11 t) F(t)\;\;.
\label{J.23}
\end{equation}

Here $t$ comes from $\Pi_W(0)$, the term proportional to $F_t(t)$
 -- from $\Pi_Z$, and $\frac{4}{9} + \frac{2}{9}\ln t$
 -- from  $\Pi_Z$
and $\Pi_{Z\gamma}$.

Cancelation of infinities in (\ref{J.22}) follows from
\begin{equation}
\Delta_t \left\{ (c^2 - s^2) \frac{\alpha}{4 \pi c s} + c s
\left[ \frac{\alpha}{2 \pi c^2} - \frac{(2 - 3t)\alpha}{8 \pi
c^2 s^2} - \frac{3t\alpha}{8 \pi s^2 c^2} \right] \right\} = 0\;\;.
\label{J.24}
\end{equation}

\newpage

\setcounter{equation}{0}
\renewcommand{\theequation}{K.\arabic{equation}}
\begin{center}
{\large\bf Appendix K. }

\vspace{3mm}
{\Large\bf The contribution of light fermions to the self-energy
  of the vector bosons.}
\end{center}
\vspace{3mm}

The contribution of the doublet of light fermions to the polarization
operators is readily obtained using the formulas of the
preceding Appendix.

To achieve this, $\Delta_{Z,W}$ must be substituted into the
formulas of Appendix J instead of $\Delta_q +\ln(m_q /
m_{Z,W})^2$. The physical reason for the absence of terms
proportional to the logarithm of the mass of light quarks and
leptons is the infrared stability of the quantities that are
analyzed in this Appendix. In the formulas (\ref{K.4}) and (\ref{K.7})
we use the equality $Q_u -Q_d =1$. The subscripts $u$ and $d$
stand for the upper and lower components of the doublet:
\begin{equation}
\Pi_{\gamma}(0) = 0 \;\;.
\label{K.1}
\end{equation}
\begin{equation}
\Pi_{\gamma}(m_Z^2)=\frac{N_c\bar{\alpha}}{3\pi} (Q_u^2
+Q_d^2)(\Delta_Z +\frac{5}{3}) \;\;.
\label{K.2}
\end{equation}
\begin{equation}
\Pi_{\gamma Z}(0) = 0 \;\;.
\label{K.3}
\end{equation}
\begin{equation}
\Pi_{\gamma Z}(m_Z^2)=\frac{N_c\bar{\alpha}}{3cs\pi}(\Delta_Z
+\frac{5}{3})[\frac{1}{4}-(Q_u^2 +Q_d^2)s^2] \;\;.
\label{K.4}
\end{equation}
\begin{equation}
\Pi_W(0) = 0 \;\;.
\label{K.5}
\end{equation}
\begin{equation}
\Pi_W(m_W^2)=\frac{N_c\bar{\alpha}}{12\pi
s^2}(\Delta_W+\frac{5}{3});\;.
\label{K.6}
\end{equation}
\begin{equation}
\Pi_Z(m_Z^2)=\frac{N_c\bar{\alpha}}{3\pi s^2 c^2}(\Delta_Z
+\frac{5}{3})[\frac{1}{4}-\frac{s^2}{2}+ s^4(Q_u^2 +Q_d^2)] \;\;.
\label{K.7}
\end{equation}
\begin{equation}
\Pi_Z(0) = 0 \;\;.
\label{K.8}
\end{equation}
\begin{equation}
\Sigma'_Z(m_Z^2)=\frac{N_c\bar{\alpha}}{3\pi s^2 c^2}(\Delta_Z
+\frac{2}{3})[\frac{1}{4}-\frac{s^2}{2}+ s^4(Q_u^2 +Q_d^2)] \;\;.
\label{K.9}
\end{equation}

Formulas (\ref{K.1})-- \ref{K.9}) must be used for three
lepton doublets $(\nu_e, e)$, $(\nu_{\mu}, \mu)$ and $(\nu_{\tau},
\tau)$ with $N_c =1$ and two quarks doublets $(u,d)$ and $(c,s)$
with $N_c =3$.

Substituting equations (\ref{K.1}) - (\ref{K.9}) into expressions
for physical observables in terms of polarization operators
(\ref{L.16}), (\ref{L.20}) and (\ref{L.24}), we arrive at the
contributions to the constants $C_i$ owing to the
self-energies.

\newpage

\setcounter{equation}{0}
\renewcommand{\theequation}{L.\arabic{equation}}
\begin{center}
{\large\bf Appendix L.}

\vspace{3mm}

{\Large\bf The contribution of the vector and scalar bosons to the
self-energy of the vector bosons.}
\end{center}

This Appendix gives formulas for boson contributions to
polarization operators that we reproduced from \cite{64},
pp~53, 54.(There is a misprint in \cite{64}: the term
 proportional to $\Delta_W$ in the expressions for $\Pi_W(q^2)$
 must be multiplied by 1/3).

The polarization operators in \cite{64} depend on $c_W$ and $s_W$
via coupling constants and depend dynamically on the ratio $m_W/m_Z$,
which arises from Feynman integrals. We substitute everywhere
$c$ for $c_W$ (and $m_W/m_Z$) and $s$ for and $s_W$. In the
framework of the one-loop approximation, this substitution is
justified. After this substitution, we find expressions for
physical observables; ultraviolet divergences of polarization
operators cancel out in these expressions.

In the following formulas  $\Pi_i$ denotes only
 boson contributions to the corresponding polarization operator
(all calculations were performed in the 'tHooft--Feynman gauge):

\begin{eqnarray}
\Pi_{\gamma}(m^2_Z) & = & -\frac{\bar{\alpha}}{4\pi} \left\{
3\Delta_W + 2(3 + 4 c^2) \left[
1 - \sqrt{4 c^2 - 1} \arcsin \left( \frac{1}{2c}\right)
\right] \right\} =
\nonumber \\
& = & -\frac{\bar{\alpha}}{4\pi}(3\Delta_W + 1.53) \;\;.
\label{L.1}
\end{eqnarray}

\begin{equation}
\Pi_{\gamma}(0) = 0\;\;.
\label{L.2}
\end{equation}

\begin{equation}
\Pi_{\gamma Z}(0) = -\frac{\bar{\alpha}}{4\pi cs} (2c^2 \Delta_W) =
-\frac{c \bar{\alpha}}{2\pi s} \Delta_W\;\;.
\label{L.3}
\end{equation}


\begin{eqnarray}
\Pi_{\gamma Z}(m^2_Z) & = & -\frac{\bar{\alpha}}{4\pi c s} \left\{
(5c^2 + \frac{1}{6}) \Delta_W \right.
\nonumber \\
& + & \left.
2(\frac{1}{6} + \frac{13}{3} c^2 + 4 c^4) \left[ 1 - \sqrt{4
c^2 - 1} \arcsin \left( \frac{1}{2c} \right) \right] + \frac{1}{9}
\right\} =
\nonumber \\
& = & \frac{-\bar{\alpha}}{4\pi} \left( \frac{30 c^2 + 1}{6 c s}
\Delta_W + 3.76 \right) \;\;.
\label{L.4}
\end{eqnarray}

\begin{eqnarray}
\Pi_W(0) & =& \frac{\bar{\alpha}}{4\pi s^2} \left[ \left(
\frac{s^2}{c^2} - 1 \right) \Delta_W + \frac{3}{4(1 - c^2/h)} \ln
\frac{c^2}{h} - \frac{h}{8c^2} + s^2 \right.
\nonumber \\
&+& \left. \frac{s^4}{c^2} - \frac{1}{8c^2} - \frac{39}{12} + \left(
\frac{s^2}{c^2} + 3 - \frac{17}{4s^2} \right) \ln c^2 \right]
\nonumber \\
& = &
\frac{\bar{\alpha}
}{4\pi s^2} \left[ \left( \frac{s^2}{c^2} - 1 \right) \Delta_W +
\frac{3}{4(1 - c^2/h)} \ln \frac{c^2}{h} - \frac{h}{8c^2} + 0.85
\right].
\label{L.5}
\end{eqnarray}

\begin{eqnarray}
\Pi_W(m^2_W)&=&\frac{\bar{\alpha}}{4\pi s^2}
\left\{ -\left( \frac{25}{6} -
\frac{s^2}{c^2} \right) \Delta_W \right.
\nonumber \\
& +& \left[ \frac{s^4}{c^2}-\frac{c^2}{3}
\left( \frac{7}{c^2} + 17- 2\frac{s^4}
{c^4} \right)-
\frac{1}{6} \left( \frac{1}{2} + \frac{1}{c^2} -
\frac{s^4}{2c^4} \right) \right]
F_h(1/c^2) \nonumber \\
& + & \left[ \frac{c^2}{3} \left( \frac{3}{c^2} + 21 \right) -
\frac{s^4}{c^2} + \frac{1}{4}
\right] \frac{1}{s^2} \ln (\frac{1}{c^2})-
3s^2 -\frac{1}{6c^2} +
\frac{s^4}{c^2}-\frac{113}{18}  \nonumber \\
& + & \left. \left( 1-\frac{h}{3c^2} +
\frac{h^2}{12c^4}\right)
F_h(h/c^2)-1-\frac{h}{6c^2} + \frac{3h}{4(c^2-h)} \ln (h/c^2)
\right\} \nonumber \\
& =& \frac{\bar{\alpha}}{4\pi s^2} \left[ \left( \frac{s^2}{c^2}
-\frac{25}{6} \right) \Delta_W - 1.76 +
\left( 1-\frac{h}{3c^2}+\frac{h^2}{12c^4}
\right) F_h(h/c^2)-\frac{h}{6c^2} \right. \nonumber \\
& + & \left. \frac{3h\ln(h/c^2)}{4(c^2-h)} \right].
\label{L.6}
\end{eqnarray}

\begin{eqnarray}
\Pi_Z(m^2_Z) & = & \frac{\bar{\alpha}}{4\pi s^2}
\left\{ \left( 7s^2 -\frac{25}{6} +\frac{7}{6}
\frac{s^2}{c^2}
\right) \Delta_W +\frac{73}{36c^2}
-\frac{2}{9}+\frac{13}{12c^2} \ln (c^2) \right.
\nonumber \\
& + & \left( 2+\frac{1+8c^2}{6c^2}(c^2-s^2)^2 -\frac{20}{3}c^2(1+
2c^2) \right)
\nonumber \\
& \times & \left[ 1-\sqrt{4c^2-1}
\arcsin \left( \frac{1}{2c} \right) \right]
\nonumber \\
& - & \left. \frac{1}{c^2} -\frac{h}{6c^2}
+\frac{3h}{4c^2(1-h)}\ln h +
\left(\frac{1}{c^2} -
\frac{h}{3c^2} + \frac{h^2}{12c^2}\right) F_h(h)\right\}
\nonumber \\
& = &\frac{\bar{\alpha}}{4\pi s^2} \left[ \left( 7s^2-\frac{25}{6}
+\frac{7}{6}\frac{s^2}{c^2} \right) \Delta_W -
0.58-\frac{h}{6c^2}+ \frac{3h}{4c^2(1-h)}\ln h \right.
\nonumber \\
& + & \left. \left( \frac{1}{c^2}-\frac{h}{3c^2} + \frac{h^2}{12c^2}
\right) F_h(h) \right].
\label{L.7}
\end{eqnarray}

\begin{eqnarray}
\Sigma^{\prime}_Z(m^2_Z) & = & \frac{d\Sigma_Z(s)}{ds}
\mid _{s=m^2_Z} \nonumber \\
& = & \frac{\bar{\alpha}}{4\pi} \left(3-\frac{19}{6s^2} +
\frac{1}{6c^2} \right) \Delta_W \nonumber \\
& + & \frac{\bar{\alpha}}{48\pi s^2 c^2} \left\{ \left[-40c^4 +
(c^2-s^2)^2 \right] 2 \left[ 1- \sqrt{4c^2-1}\arcsin
\left( \frac{1}{2c} \right) \right] \right. \nonumber \\
& + & \left[ 12c^2 +(8c^2 +1)(c^2-s^2)^2 - 40 c^4(1+2 c^2)\right]
\nonumber \\
& \times & \left[-1+\frac{4c^2}{\sqrt{4c^2-1}}\arcsin
\left(\frac{1}{2c}\right)\right] \nonumber \\
& + & \left[1-(h-1)^2\right] F_h(h) +\left[11-2h+(1-h)^2\right]
F^{\prime}_h(h) \nonumber \\
& + & \left.\left[1-\frac{1+h}{2(h-1)}\ln h
-\frac{1}{2}\ln\frac{h}{c^4}\right] + \frac{2}{3}\left[1+(c^2 -s^2)^2
-4c^4\right]\right\} \nonumber \\
& = &
\frac{\bar{\alpha}}{4\pi}
\left( 3-\frac{19}{6s^2} +\frac{1}{6c^2} \right) \Delta_W \nonumber
\\
& + & \frac{\bar{\alpha}}{4\pi s^2c^2} \left[ \left( 1-\frac{h}{3}
+ \frac{h^2}{12} \right) F^{\prime}_h(h) + \left( \frac{h}{6}
-\frac{h^2}{12} \right) F_h(h) \right. \nonumber \\
& + & \left. \frac{h}{12(1-h)} \ln h - 1.67 \right] \;.
\label{L.8}
\end{eqnarray}

The functions $F_h(h)$ and $F'_h(h)$ are defined as

\begin{equation}
F_h(h)\equiv F(s,m_Z,m_H)\mid_{s=m^2_Z}\equiv F(1,1,h) \;\;,
\label{L.9}
\end{equation}

\begin{equation}
F^{\prime}_h(h) \equiv s\frac{dF(s,m_Z,m_H)}{ds}\mid_{s=m^2_Z} \;.
\label{L.10}
\end{equation}
Using \cite{82}, pp~88, we obtain

\begin{eqnarray}
F_h(h) & = & 1+
\left( \frac{h}{h-1}-\frac{1}{2}h \right) \ln h+
h \sqrt{1-\frac{4}{h}} \ln
\left( \sqrt{\frac{h}{4} -1} + \sqrt{\frac{h}{4}} \right)
\;\;\;, h >4 , \nonumber \\
\\
& = & 1 + \left( \frac{h}{h-1} - \frac{1}{2} h \right) \ln h - h
\sqrt{\frac{4}{h}-1} \arctan \sqrt{\frac{4}{h}-1}\;\; \;, h
\;\;<4. \nonumber
\label{L.11}
\end{eqnarray}

If $h\rightarrow\infty$,
\begin{equation}
F_h(h)\approx \frac{1}{2h}-\frac{1}{h^2}
\left( 1+\frac{4}{h^2} \right) \ln h+
\frac{5}{3h^2}+\frac{59}{12h^3} \;\;.
\label{L.12}
\end{equation}

If $h\rightarrow 0$,
\begin{equation}
F_h(h)\approx 1-\pi \sqrt{h}+(1-\frac{3}{2}\ln h)h \;\;.
\label{L.13}
\end{equation}

Finally, for $F^{\prime}_h(h)$ we have
\begin{eqnarray}
F^{\prime}_h(h)& = & -1 + \frac{h-1}{2}\ln h+(3-h)
\sqrt{\frac{h}{h-4}}\ln
\left(\sqrt{\frac{h-4}{4}}+\sqrt{\frac{h}{4}}\right) \;,\;\; h>4,
\nonumber \\
\\
& = & -1 +\frac{h-1}{2} \ln h+(3-h)
\sqrt{\frac{h}{4-h}}\arctan\sqrt{\frac{4-h}{h}} \;,\;\;\;\; h<4.
\nonumber
\label{L.14}
\end{eqnarray}
If $h\rightarrow\infty$,
\begin{equation}
F^{\prime}_h(h)\approx \frac{1}{2h}-\frac{1}{h^2}\ln(h) \;\;.
\label{L.15}
\end{equation}

All infinities in formulas (\ref{L.1})--(\ref{L.8}) are
collected into the factors $\Delta_W$ by replacing the factors
$\Delta_Z$ and $\Delta_H$ using the equation
$$
\Delta_i = \Delta_j +\ln(m_j^2 /m_i^2) \;\;.
$$

The function $H_m(h)$ is

\begin{eqnarray}
H_m(h) & = & \frac{16\pi
s^4}{3\bar{\alpha}}\left(\frac{c^2}{s^2}\left[\Pi_Z(m^2_Z)-
\Pi_W(m^2_W)\right] + \Pi_W(m^2_W) -\Pi_W(0)\right. \nonumber \\
& - & \left. \Pi_{\gamma}(m^2_Z)-2\frac{s}{c}\Pi_{\gamma Z}(0)\right)
- div H_m \;\;,
\label{L.16}
\end{eqnarray}
where $div H_m$ denotes the sum of terms, proportional to
$\Delta_W$, in polarization operators in (\ref{L.16}).
Substituting the finite parts of the formulas for polarization
operators from this Appendix, we obtain

\begin{eqnarray}
H_m(h) & = & -\frac{h}{h-1}\ln h +\frac{c^2 h}{h-c^2}\ln
\frac{h}{c^2}- \frac{s^2}{18c^2} h+\left(\frac{h^2}{9} - \frac{4h}{9}
+ \frac{4}{3} \right) F_h (h) \nonumber \\
& - & (c^2-s^2)\left(\frac{h^2}{9c^4}-\frac{4h}{9c^2}
+\frac{4}{3}\right) F_h\left(\frac{h}{c^2}\right)+ 0.50 \;\;,
\label{L.17}
\end{eqnarray}
where the term proportional to $F_h(h)$ arises from
$\Pi_Z$, and the term proportional to $F_h(h/c^2)$ arises from
$\Pi_W(m^2_W)$. The term proportional to $\ln h$ originates from
$\Pi_Z$, while $\ln (h/c^2)$ arises both from $\Pi_W(m^2_W)$ and
from $\Pi_W(0)$. The term proportional to $h$ is contained in
$\Pi_Z$, $\Pi_W(m^2_W)$ and $\Pi_W(0)$ and finally, all four
polarization operators make contribution to the constant.

Collecting the terms proportional to $\Delta_W$
in the polarization
operators in (\ref{L.16}), we get
\begin{eqnarray}
div H_m & = & \frac{16\pi s^4}{3\bar{\alpha}}\Delta_W
\left[\frac{\bar{\alpha}}{4\pi s^2}
\frac{c^2}{s^2}\left(7s^2 -\frac{25}{6} + \frac{7}{6} \frac{s^2}{c^2}
+ \frac{25}{6} - \frac{s^2}{c^2}\right) \right. \nonumber \\
& + & \left. \frac{\bar{\alpha}}{4\pi
s^2}\left(-\frac{19}{6}\right)+\frac{3\bar{\alpha}}{4\pi}+
2\frac{s}{c}\frac{c}{s}\frac{\bar{\alpha}}{2\pi}\right] =
\frac{16\pi s^4}{3\bar{\alpha}}\Delta_W\frac{\bar{\alpha}}{\pi s^2} \;\;.
\label{L.18}
\end{eqnarray}

Note that the divergent term in $D$ (see (\ref{E.8}) and
(\ref{M.26})) exactly compensates for the divergence in (\ref{L.18}),
which justifies
the subtraction of infinity in (\ref{L.16}).

The function $H_A(h)$ is expressed in terms of polarization operators
as follows:

\begin{equation}
H_A(h)=\frac{16\pi
s^2c^2}{3\bar{\alpha}}\left[\Pi_Z(m^2_Z)-\Sigma^{\prime}_Z
(m^2_Z)-\Pi_W(0)\right]-div H_A \;\;.
\label{L.19}
\end{equation}

Substituting the finite parts of the polarization operators, we
obtain
\begin{eqnarray}
H_A(h) & = & \frac{hc^2}{h-c^2}\ln\frac{h}{c^2}-\frac{8h}{9(h-1)}
\ln(h)+(\frac{4}{3}-\frac{2}{3}h+\frac{2}{9}h^2)F_h(h) \nonumber \\
& - & \left(\frac{4}{3}-\frac{4h}{9}+\frac{h^2}{9}\right)
F^{\prime}_h(h)- \frac{h}{18}+0.78 \;\;,
\label{L.20}
\end{eqnarray}
where $\ln(h/c^2)$ stems from $\Pi_W$ and $\ln h$ stems from $\Pi_Z$.
$F_h(h)$ arises both from $\Pi_Z$ and from $\Sigma^{\prime}_Z$,
while the only source of $F^{\prime}_h(h)$ is $\Sigma^{\prime}_Z$.
The term linear in $h$ is contained in $\Pi_Z$ and $\Pi_W$,
while all three polarization operators contribute to the
constant.

Adding up the divergent terms, we have
\begin{equation}
div H_A=\frac{16}{3}c^2s^2\Delta_W=\frac{16\pi s^2c^2}{3\bar{\alpha}}
\Delta_W\frac{\bar{\alpha}}{\pi} \;\;.
\label{L.21}
\end{equation}

Note that the divergent part of  $D_A$
(see (\ref{M.18}) and (\ref{M.14})) is:
\begin{equation}
div D_A=-\frac{16c^2}{3}\Delta_W=-\frac{16\pi
s^2c^2}{3\bar{\alpha}}\Delta_W \frac{\bar{\alpha}}{\pi s^2} \;\;.
\label{L.22}
\end{equation}
Vertex parts also contain ultraviolet divergences (see
Appendix M, eqs.(\ref{M.11}), (\ref{M.15})):
\begin{equation}
div \tilde{F}_A=-(\frac{16}{3}c^2s^2-\frac{16}{3}c^2)\Delta_W=\frac{16\pi
s^2c^2}
{3\bar{\alpha}}\Delta_W\frac{c^2}{s^2}\frac{\bar{\alpha}}{\pi} \;\;.
\label{L.23}
\end{equation}
The sum of terms (\ref{L.21}), (\ref{L.22}) and (\ref{L.23}) equals
zero.

Finally, the expression for $H_R(h)$ is
\begin{eqnarray}
H_R(h) & = &
-\frac{16\pi}{3\bar{\alpha}}c^2s^2\left(\frac{(c^2-s^2)}{cs}
\Pi_{Z\gamma}(m^2_Z)+\Pi_{\gamma}(m^2_Z)-\Pi_Z(m^2_Z)+
\Pi_W(0) \right. \nonumber \\
& + & \left. 2\frac{s}{c}\Pi_{\gamma Z}(0)\right)-div
H_R(h) \;\;.
\label{L.24}
\end{eqnarray}

Collecting the finite parts of the polarization operators, we
find
\begin{equation}
H_R(h)=-\frac{h}{18}+\frac{c^2}{1-c^2/h}\ln\frac{h}{c^2}+(\frac{4}{3}-
\frac{4}{9}h+\frac{1}{9}h^2)F_h(h)+
\frac{h}{1-h}\ln h+0.03 \;\;.
\label{L.25}
\end{equation}

The term proportional to $F_h(h)$ stems from $\Pi_Z$, just as
$\ln h$ does. $\Pi_W$ generates the term $\sim\ln(h/c^2)$. The
term linear in $h$ is contained both in $\Pi_Z$ and in $\Pi_W$,
and all polarization operators with the exception of $\Pi_{\gamma Z}(0)$
contribute to the constant.

Adding up the divergent parts of the polarization operators
in (\ref{L.24}), we obtain
\begin{eqnarray}
div H_R(h) & = & \Delta_W\left[\frac{2}{9}(c^2-s^2)(1+30c^2)+4c^2s^2+
\frac{4}{3}c^2\left(7s^2-\frac{25}{6}\right. \right. \nonumber \\   \\
& + & \left. \left. \frac{7}{6}\frac{s^2}{c^2}\right)+
\frac{4}{3}c^2-\frac{4}{3}s^2+\frac{16}{3}c^2s^2\right]
=\Delta_W(8c^2-\frac{16}{3} c^4) \;\;.
\label{L.26} \nonumber
\end{eqnarray}

Taking into account the divergent term in $D$, yielding
(see eqs. (\ref{E.8}), (\ref{M.14}), (\ref{M.18}))
\begin{equation}
div D_R=-\frac{16}{3}c^2\Delta_W \;\;,
\label{L.27}
\end{equation}
and in $\tilde{F}_R$ (see Appendix M, equations (M.11) and
(M.22)),
\begin{equation}
div \tilde{F}_R=(\frac{16}{3}c^4-\frac{8}{3}c^2)\Delta_W \;\;,
\label{L.28}
\end{equation}
we confirm that divergences cancel out in expressions for $R$.

\newpage

\setcounter{equation}{0}
\renewcommand{\theequation}{M.\arabic{equation}}
\begin{center}
{\large\bf Appendix M}

\vspace{3mm}

{\Large\bf The vertex parts of
${\bf F_{Af}}$ and ${\bf F_{Vf}}$ and the constants
${\bf C_i}$.}
\end{center}
\vspace{3mm}

This Appendix collects the vertex functions that form a part of
one-loop electroweak corrections to $Z\to\nu\bar{\nu}$, $Z\to l^+ l^-$,
$Z\to u\bar{u}$, $c\bar{c}$, $d\bar{d}$ and $s\bar{s}$ decays.
In the case of the $Z\to b\bar{b}$ decay, a $t$-quark can
propagate in the loop, so vertex corrections are not reducible
to numbers but are functions of $m_t$ (see Appendix N).

The finite parts of vertex functions are given in \cite{64}, pp~29,
30.  The corresponding expressions depend on $c_W(s_W)$ and
$m_W/m_Z$.  In the framework of the one-loop approximation, we
replace $c_W$ and $m_W/m_Z$ with $c$, and $s_W$ with $s$. For this
reason, while vertex functions in \cite{64} depend on $m_t$,
$m_H$ and the new physics, ours are numbers (see also Appendix L).

This Appendix also gives the infinite parts absent from \cite{64}
and required for testing whether the infinities in physical
observables do cancel out.

We begin with the $Z\to \nu\nu$ decay:
\begin{eqnarray}
F_{\nu} & \equiv & F_{V\nu} = F_{A\nu} = \nonumber \\
& = & \frac{\bar{\alpha}}{4\pi}\frac{1}{4cs} \left[ \frac{1}{4c^2 s^2}
\Lambda_2(m_Z^2, m_Z) +\frac{2s^2 -1}{2s^2} \Lambda_2 (m_Z^2,
m_W) +\frac{3c^2}{s^2}\Lambda_3(m_Z^2, m_W)\right]
\label{M.1}
\end{eqnarray}
For the decay to a pair of charged leptons or quarks we have
\begin{equation}
F_{Vf} = \frac{\bar{\alpha}}{4\pi}\left[v_f(v_f^2 +3a^2_f)\Lambda_2
(m_Z^2, m_Z) +F_L^f\right] \;\;,
\label{M.2}
\end{equation}
\begin{equation}
F_{Af} = \frac{\bar{\alpha}}{4\pi}\left[a_f(3v_f^2 +a^2_f)\Lambda_2
(m_Z^2, m_Z) +F_L^f\right] \;\;,
\label{M.3}
\end{equation}
where $a_l =a_d =-1/4 sc$, $a_u =1/4 sc$, $v_f =(T_3^f -2Q^f
s^2)/2sc$ ($T_3^l =T_3^d =-1/2$, $T_3^u =1/2$, $Q^l=-1$, $Q^d =-1/3$,
$Q^u =2/3$). The functions $F_L^f$ are
\begin{equation}
F_L^l = \frac{1}{8s^3 c}
\Lambda_2(m_Z^2, m_W) -\frac{3c}{4s^3} \Lambda_3 (m_Z^2, m_W) \;\;,
\label{M.4}
\end{equation}
\begin{equation}
F_L^u = -\frac{1-\frac{2}{3}s^2}{8s^3 c}
\Lambda_2(m_Z^2, m_W) +\frac{3c}{4s^3} \Lambda_3 (m_Z^2, m_W) \;\;,
\label{M.5}
\end{equation}
\begin{equation}
F_L^d = \frac{1-\frac{4}{3}s^2}{8s^3 c}
\Lambda_2(m_Z^2, m_W) +\frac{3c}{4s^3} \Lambda_3 (m_Z^2, m_W) \;\;.
\label{M.6}
\end{equation}

For calculating $F_{Vf}$ and $F_{Af}$ we need to determine the
values of three constants: $\Lambda_2(m_Z^2, m_W)$, $\Lambda_2(m_Z^2, m_Z)$
and $\Lambda_3(m_Z^2, m_W)$:
\begin{eqnarray}
\Lambda_2(m^2_Z, m_W)&=&-\frac{7}{2}-2c^2-(2c^2+3)\ln(c^2) \nonumber \\
&+&2(1+c^2)^2\left[\ln c^2 \ln\left(\frac{1+c^2}{c^2}\right)-
Sp\left(-\frac{1}{c^2}\right)\right] \;\;,
\label{M.7}
\end{eqnarray}
where we have used $m_W/m_Z=c$; $Sp(x)$ is the Spence function:
\begin{equation}
Sp(x)=-\int\limits^1_0\frac{dt}{t}\ln(1-xt)\;, \;\;
Sp(-1)= -\frac{\pi^2}{12} \;\;.
\label{M.8}
\end{equation}
Using (\ref{M.7}) and (\ref{M.8}), we find
\begin{equation}
\Lambda_2(m^2_Z,
m_Z)=-\frac{7}{2}-2-8 Sp(-1)\;\;.
\label{M.9}
\end{equation}
Finally,
\begin{eqnarray}
\Lambda_3(m^2_Z, m_W)&=& \frac{5}{6}-\frac{2}{3}c^2+\frac{2}{3}(2c^2+1)
\sqrt{4c^2-1} \arctan\frac{1}{\sqrt{4c^2-1}} \nonumber \\
&-&\frac{8}{3}c^2(c^2+2)\left(\arctan\frac{1}{\sqrt{4c^2-1}}
\right)^2\;\;.
\label{M.10}
\end{eqnarray}

The expressions for divergent parts of the vertex functions,
describing the coupling of the $Z$-boson to the leptons, are
\begin{equation}
div F_{\nu} =\frac{\bar{\alpha}}{8\pi}\frac{c}{s^3}\Delta_W \;\;,
div F_{Vl} =div F_{Al} = -div F_{\nu}
\label{M.11}
\end{equation}
We switch to the calculation of the constants $C_i$. We begin
with definitions. According to \cite{35},
\begin{equation}
V_{\nu}(t,h)=t+T_{\nu}(t)+H_{\nu}(h)+L_{\nu}+D_{\nu}+
\tilde{F}_{\nu} \;\;,
\label{M.12}
\end{equation}
The value of $L_{\nu}$ represents the contribution of leptons
and light quarks to the polarization operators of the vector
bosons and can be easily obtained from the formulas of Appendix K
for polarization operators:
\begin{equation}
L_{\nu}=4-8s^2 +\frac{304}{27}s^4\;\;.
\label{M.13}
\end{equation}

$D_{\nu}$ originates from the box and vertex electroweak corrections
to the $\mu$-decay \cite{17}. The expression for $D$ see in Appendix E,
equation (E.8). For $D_{\nu}$ we have
\begin{equation}
D_{\nu}=-\frac{16\pi
s^2 c^2}{3\bar{\alpha}}(D-\frac{\bar{\alpha}}{\pi s^2}\Delta_W) \;\;  .
\label{M.14}
\end{equation}
Finally,
\begin{equation}
\tilde{F}_{\nu} =\frac{128 \pi s^3 c^3}{3\bar{\alpha}} F_{\nu} \;\;.
\label{M.15}
\end{equation}
Comparing (\ref{M.12}) and (\ref{51}), we arrive at the
expressions for $C_{\nu}$ whose ingredients are now all
determined:
\begin{equation}
C_{\nu} =L_{\nu}+D_{\nu}+\tilde{F}_{\nu}
\label{M.16}
\end{equation}
Let us switch to $V_A$:
\begin{equation}
V_A(t,h)=t+T_A(t)+H_A(h)+L_A+D_A+\tilde{F}_A \;\;.
\label{M.17}
\end{equation}
The expressions for $L_A$ and $D_A$ are already there:
\begin{equation}
L_A = L_{\nu} \;\;, \;\; D_A =D_{\nu} \;\;;
\label{M.18}
\end{equation}
the formula for the vertex function is
\begin{equation}
\tilde{F}_A=-\frac{128\pi s^3 c^3}{3\bar{\alpha}}F_{Al} \;\;.
\label{M.19}
\end{equation}
Finally,
\begin{equation}
C_A=L_A+D_A+\tilde{F}_A
\label{M.20}
\end{equation}
We now move to $V_R$:
\begin{equation}
V_R(t,h) = t+T_R(t)+H_R(h)+L_R+D_R+\tilde{F}_R \;\;,
\label{M.21}
\end{equation}
where $D_R =D_{\nu}$ and $L_R =0$ since $\Pi_W(m_W^2)$ is
absent from the ratio $g_V/g_A$. \\
For $\tilde{F}_R$ we have
\begin{equation}
\tilde{F}_R = \frac{16\pi(c^2 -s^2)cs}{3\bar{\alpha}}
[-F_{Vl}+(1-4s^2)F_{Al}] \;\;,
\label{M.22}
\end{equation}
and the formula for $C_R$ is
\begin{equation}
C_R =L_R +D_R +\tilde{F}_R \;\;.
\label{M.23}
\end{equation}

We end this Appendix with formulas for $C_m$:
\begin{equation}
C_m =L_m +D_m \;\;,
\label{M.24}
\end{equation}
\begin{equation}
L_m =4(c^2 -s^2)\ln c^2 \;\;,
\label{M.25}
\end{equation}
\begin{equation}
D_m =-\frac{16\pi s^4}{3\bar{\alpha}}\left(D-\frac{\bar{\alpha}}
{\pi s^2}\Delta_W \right) \;\; .
\label{M.26}
\end{equation}

\newpage

\setcounter{equation}{0}
\renewcommand{\theequation}{N.\arabic{equation}}
\begin{center}
{\large\bf Appendix N. }

\vspace{3mm}

{\Large\bf The functions {\boldmath$\phi(t)$}
and {\boldmath$\delta\phi(t)$} in the
{\boldmath$Z\to b\bar{b}$} decay.}
\end{center}
\vspace{5mm}

For the function $\phi(t)$ we use the expansion from \cite{51}:

$$
\phi(t) = \frac{3-2s^2}{2s^2 c^2}\left\{t+c^2[2.88 ln\frac{t}{c^2}
-6.716 + \right.
$$
\begin{equation}
+\frac{1}{t}(8.368 c^2 \ln \frac{t}{c^2}-3.408 c^2)+ \frac{1}{t^2}
(9.126 c^4 \ln \frac{t}{c^2}+2.26 c^4) +
\label{N.1}
\end{equation}
$$
\left. +\frac{1}{t^3}(4.043 c^6 \ln \frac{t}{c^2} +7.41 c^6) +... \right\}
\;\;,
$$
and for $\delta\phi(t)$ we use the leading approximation
calculated in \cite{52} and \cite{50}:
\begin{equation}
\delta\phi(t,h) = \frac{3-2s^2}{2s^2c^2}\left
\{-\frac{\pi^2}{3}(\frac{\hat{\alpha}_s(m_t)}{\pi})
t +\frac{1}{16s^2c^2}(\frac{\bar{\alpha}}{\pi})t^2
\tau_b^{(2)}(\frac{h}{t})\right\} \;\;,
\label{N.2}
\end{equation}
where the function $\tau_b^{(2)}$ is tabulated in Table 4
for $m_H / m_t <4$. For $m_H / m_t >4$ we use the expansion \cite{50}
$$
\tau_b^{(2)}(\frac{h}{t}) = \frac{1}{144}[311+24\pi^2+282\ln
r+90\ln^2 r -4r(40+6\pi^2+15\ln r+18\ln^2 r) +
$$
\begin{equation}
+\frac{3}{100}r^2(24209-6000\pi^2-45420\ln r-18000\ln^2 r)],
\label{N.3}
\end{equation}
where $r=t/h$. For $m_t = 175$ GeV and $m_H = 300 $ GeV
$$
\tau_b^{(2)} = 1.245\;\;.
$$

\newpage

\setcounter{equation}{0}
\renewcommand{\theequation}{O.\arabic{equation}}
\begin{center}
{\large\bf Appendix O.}

\vspace{3mm}

{\Large\bf The {\boldmath$\delta_2 V_i$} corrections.}
\end{center}

\vspace{2mm}

The corrections $\delta_2 V_i \sim \bar{\alpha} \hat{\alpha}_s$
arise from gluon exchanges in quark electroweak loops \cite{47}
(see also \cite{65}). For two generations of light quarks
$(q = u, d, s, c)$ we have
\begin{equation}
\delta^q_2 V_m(t,h)
= 2 \cdot [\frac{4}{3}(\frac{\hat{\alpha}_s(m_Z)}{\pi})(c^2-s^2)\ln c^2] =
(\frac{\hat{\alpha}_s(m_Z)}{\pi})(-0.377)
\label{O.1}
\end{equation}
\begin{equation}
\delta^q_2 V_A(t,h) = \delta^q_2 V_\nu(t,h) = 2 \cdot
[\frac{4}{3}(\frac{\hat{\alpha}_s(m_Z)}{\pi})
(c^2-s^2 + \frac{20}{9} s^4)]=
(\frac{\hat{\alpha}_s(m_Z)}{\pi})(1.750)
\label{O.2}
\end{equation}
\begin{equation}
\delta^q_2 V_R(t,h) = 0
\label{O.3}
\end{equation}

The result of calculations for the third generation is obtained
in the form of fairly complicated functions of the $t$-quark mass:
$$
\delta^t_2 V_m(t,h) = \frac{4}{3} (\frac{\hat{\alpha}_s (m_t)}{\pi})
\{tA_1 (\frac{1}{4t})+(1-\frac{16}{3}s^2)tV_1(\frac{1}{4t})
+(\frac{1}{2}-\frac{2}{3}s^2) \ln t
$$
\begin{equation}
-4(1-\frac{s^2}{c^2}) \times tF_1(\frac{c^2}{t})-4\frac{s^2}{c^2}tF_1 (0)\}
\label{O.4}
\end{equation}
$$
\delta ^t_2 V_A (t,h) = \delta ^t_2 V_\nu (t,h) =
\frac{4}{3}(\frac{\hat{\alpha}_s (m_t)}{\pi})
\{tA_1(\frac{1}{4t})-\frac{1}{4}A^{'}_{1} (\frac{1}{4t})+
$$
\begin{equation}
+(1-\frac{8}{3}s^2)^2[tV_1 (\frac{1}{4t})-
\frac{1}{4}V^\prime _1 (\frac{1}{4t})]+(\frac{1}{2}-\frac{2}{3}s^2 +
\frac{4}{9}s^4)-4tF_1 (0)\}
\label{O.5}
\end{equation}

\begin{equation}
\delta^t_2 V_R(t,h) = \frac{4}{3} (\frac{\hat{\alpha}_s (m_t)}{\pi})\{tA_1
(\frac{1}{4t})
-\frac{5}{3}tV_1 (\frac{1}{4t})-4tF_1(0)+\frac{1}{6} \ln t\},
\label{O.6}
\end{equation}
where
\begin{equation}
\hat{\alpha}_s (m_t) =
\frac{\hat{\alpha}_s (M_Z)}
{1 +\frac{23}{12 \pi} \hat{\alpha}_s (M_Z) \ln t}
\label{O.7}
\end{equation}

Note that $\delta_2 V_i$ are independent of $m_H$. The functions
$V_1 (r)$, $A_1 (r)$ and $F_1 (x)$ have rather complex form and
were calculated in \cite{47}. Their expansions for sufficiently
small values of arguments are (we have added cubic terms to the
expansions from \cite{47}):
\begin{equation}
V_1 (r) = r[4\zeta(3)-\frac{5}{6}]+r^2 \frac{328}{81}
+r^3 \frac{1796}{25 \times 27}+...
\label{O.8}
\end{equation}
\begin{equation}
A_1 (r) = [-6\zeta (3)-3\zeta (2) + \frac{21}{4}] + r[4\zeta (3) -
\frac{49}{18}]+
r^2 \frac{689}{405} + r^3 \frac{3382}{7 \times 25 \times 27} +...
\label{O.9}
\end{equation}

$$
F_1 (x)=[-\frac{3}{2} \zeta (3) - \frac{1}{2}\zeta (2) +
\frac{23}{16}]+x[\zeta
(3)-
\frac{1}{9} \zeta (2) - \frac{25}{72}]+
$$
\begin{equation}
+x^2[\frac{1}{8}\zeta (2) + \frac{25}{3 \times 64}]+x^3
[\frac{1}{30} \zeta (2) +\frac{5}{72}]+...
\label{O.10}
\end{equation}
where $\zeta (2) = \pi ^2 /6, \zeta (3) = 1.2020569....$

Adding up the contributions (\ref{O.4})--(\ref{O.6}) and
using the expansions (\ref{O.8})--(\ref{O.10}), we obtain (up to
terms $\sim 1/t^3$):
\begin{equation}
\delta^t_2 V_m(t,h) = (\frac{\hat{\alpha}_s(m_t)}{\pi})[-2.86t + 0.46\ln t -
1.540 - \frac{0.68}{t} - \frac{0.21}{t^2}] =
\frac{\hat{\alpha}_s(m_t)}{\pi}(-11.67)
\label{O.11}
\end{equation}
\begin{equation}
\delta^t_2 V_A(t,h) = \delta^t_2 V_\nu (t,h) =
(\frac{\hat{\alpha}_s(m_t)}{\pi})[-2.86t + 0.493 - \frac{0.19}{t} -
\frac{0.05}{t^2}]=\frac{\hat{\alpha}_s(m_t)}{\pi}(-10.10)
\label{O.12}
\end{equation}
\begin{equation}
\delta^t_2
V_R(t,h) = (\frac{\hat{\alpha}_s(m_t)}{\pi})[-2.86t + 0.22 \ln t - 1.513 -
\frac{0.42}{t} - \frac{0.08 }{t^2}]
    =\frac{\hat{\alpha}_s(m_t)}{\pi}(-11.88).
\label{O.13}
\end{equation}
These formulas hold for $m_t > m_Z$. In the region $m_t < m_Z$
we either set $\delta^t_2 V_i = 0$ or make use of the massless
limit $\delta^t_2 V_i = \frac{1}{2}\delta^q_2 V_i$.
In any case this region gives negligible contribution to the
global fit.

\newpage

\setcounter{equation}{0}
\renewcommand{\theequation}{P.\arabic{equation}}
\begin{center}
{\large\bf Appendix P.}

\vspace{3mm}

{\Large\bf The {\boldmath$\delta_5 V_i$} corrections.}

\end{center}

\vspace{2mm}

In the second order of weak interactions
quadratic dependence
on the mass of the higgs boson is given by expressions \cite{66}:

\begin{equation}
\delta_5 V_m
=\frac{\bar{\alpha}}{24\pi}(\frac{m_H^2}{m_Z^2})\times \frac{0.747}{c^2}
= 0.0011
\label{P.1}
\end{equation}
\begin{equation}
\delta_5 V_A = \delta_5 V_\nu
=\frac{\bar{\alpha}}{24\pi}(\frac{m_H^2}{m_Z^2})\times \frac{1.199}{s^2}
= 0.0057
\label{P.2}
\end{equation}
\begin{equation}
\delta_5 V_R =-\frac{\bar{\alpha}}{24\pi}(\frac{m_H^2}{m_Z^2})\frac{c^2 -s^2}
{s^2 c^2}\times 0.973 = -0.0032 \;\;.
\label{P.3}
\end{equation}

The numerical evaluations above were made with $m_H = 300$ GeV.

\end{document}